\documentclass{aa}  

\usepackage{graphicx,natbib}
\usepackage{xcolor}
\usepackage{soul}
\usepackage{float}
\usepackage[percent]{overpic}
\usepackage{rotating}
\usepackage{caption}
\usepackage{placeins}
\usepackage{subcaption}
\usepackage{txfonts}
\begin{document}

   \title{Radio continuum of galaxies with H$_{2}$O
megamaser disks: \\33\,GHz VLA data}

%    \subtitle{I. Overviewing the $\kappa$-mechanism}

   \author{F. Kamali\inst{1}
          \fnmsep\thanks{Member of the International Max Planck Research School (IMPRS)
                      for Astronomy and Astrophysics at the Universities of Bonn and Cologne.},
          C. Henkel\inst{1,2}, 
          A. Brunthaler\inst{1},
          C. M. V. Impellizzeri\inst{3,4},
          K. M. Menten\inst{1},
          J. A. Braatz\inst{3}, 
          J. E. Greene\inst{5}, 
          M. J. Reid\inst{6}, 
          J. J. Condon\inst{3}, 
          K. Y. Lo\inst{3}, 
          C. Y. Kuo\inst{7},
          E. Litzinger\inst{8,9},
          M. Kadler\inst{8}
          }

   \institute{Max-Planck-Institut f\"{u}r Radioastronomie, Auf dem H\"{u}gel 69, 53121 Bonn, Germany\\              
              \email{fkamali@mpifr-bonn.mpg.de}; 
              \email{fateme.kamali28@gmail.com}
             \and
             Astron. Dept., King Abdulaziz University, P.O. Box 80203, Jeddah 21589, Saudi Arabia\\
             \and
             National Radio Astronomy Observatory, 520 Edgemont Road, Charlottesville, VA 22903, USA\\
             \and
             Joint ALMA Office, Alonso de C{\'o}rdova 3107, Vitacura, Santiago, Chile\\
             \and
             Department of Astrophysical Sciences, Princeton University, Princeton, NJ 08544, USA \\
             \and 
             Harvard-Smithsonian Center for Astrophysics, 60 Garden Street, Cambridge, MA 02138, USA\\
             \and
             Department of Physics, National Sun Yat-Sen University, No.70, Lianhai Road, Gushan Dist., Kaohsiung City 804, Taiwan (R.O.C.)\\
             \and
             Institut für Theoretische Physik und Astrophysik, Universität Würzburg, Campus Hubland Nord, Emil-Fischer-Str. 31, 97074 Würzburg, Germany\\
             \and 
             Dr. Remeis-Observatory, Erlangen Centre for Astroparticle, Physics, University of Erlangen-Nuremberg, Sternwartstr. 7, 96049 Bamberg, Germany\\
                      }
\titlerunning{Radio continuum of galaxies with $H_{2}O$
megamaser disks: 33\,GHz VLA data}

\authorrunning{Kamali et al.}

   \date{Received January 1, 2017; accepted...}

% \abstract{}{}{}{}{} 
% 5 {} token are mandatory
 
  \abstract
  % context heading (optional)
  % {} leave it empty if necessary  
   {Galaxies with H$_2$O megamaser disks are active galaxies in whose
edge-on accretion disks 22\,GHz H$_2$O maser emission has been detected. Because their 
   geometry is known, they provide a unique view into the properties of active galactic nuclei.}
   % aims
   {The goal of this work is to investigate the nuclear environment of galaxies with H$_2$O maser disks and to relate the maser and host galaxy properties to those of the
   radio continuum emission of the galaxy.}
  % methods heading (mandatory)
   {The 33\,GHz (9\,mm) radio continuum properties of 24 galaxies with reported 22 GHz H$_2$O maser emission from their disks are studied in the context of
   the multiwavelength view
   of these sources. The 29--37\,GHz Ka-band observations are made with the Jansky Very Large Array in B, CnB, or BnA configurations, achieving a resolution of $\sim$ 0.2\,-\,0.5\,arcseconds. 
%    Spectral indices were calculated using fluxes at 33\,GHz and 1.4\,GHz, the latter obtained from the NVSS and FIRST surveys. 
   Hard X-ray data from the \textsl{Swift}/BAT survey and 22\,$\mu$m 
   infrared data from WISE, 22\,GHz H$_2$O maser data and 1.4\,GHz data from NVSS and FIRST surveys are also included in the analysis.}
  % results heading (mandatory)
   {Eighty-seven percent (21 out of 24) galaxies in our sample show 33\,GHz radio continuum emission at levels of 4.5 -- 240\,$\sigma$. Five sources show 
   extended emission (deconvolved source size larger than 2.5 times the major axis of the beam), including one source with two main components and one with three main components. 
  The remaining detected 16 sources (and also some of the above-mentioned 
   targets) exhibit compact cores within the sensitivity limits. Little evidence is found for extended jets ($>$300\,pc) in most sources. Either they do not exist, or our 
   chosen frequency of 33\,GHz is too high for a detection of these supposedly steep spectrum features. 
   In only one source among those with known maser disk orientation, NGC~4388, did we find an extended jet-like feature 
   that appears to be oriented perpendicular to the H$_2$O megamaser
disk. NGC~2273 is another candidate whose radio continuum source might be elongated perpendicular 
   to the maser disk. Smaller 100--300\,pc sized jets might also be present, as is suggested by the beam-deconvolved morphology of our sources. Whenever possible, central 
   positions with accuracies of 20-280 mas are provided. 
   A correlation analysis shows that the 33\,GHz luminosity 
   weakly correlates with the infrared luminosity. The 33\,GHz luminosity is anticorrelated with the circular velocity of the galaxy. The black hole masses show stronger correlations 
   with H$_2$O maser luminosity than with 1.4\,GHz, 33\,GHz, or hard X-ray luminosities. Furthermore, the inner radii of the
disks show stronger correlations with 1.4\,GHz, 33\,GHz, 
   and hard X-ray luminosities than their outer radii, suggesting that the outer radii may be affected by disk warping, star formation, or peculiar density distributions.}
  % conclusions heading (optional), leave it empty if necessary 
   {}
   
   \keywords{Galaxies: active -- Galaxies: ISM -- Galaxies: jets -- Galaxies: nuclei - Galaxies: Seyfert -- Radio continuum: galaxies}
               
                \maketitle
%
%________________________________________________________________

\section{Introduction}\label{sec:intro}

In 1969, the 22\,GHz ($\lambda$ $\sim$ 1.4\,cm) H$_2$O maser line that is emitted from the water vapor 6$_{16}$-5$_{23}$ rotational transition was detected for the 
first time in the Milky Way \citep{cheung1969}. Since then, this interesting phenomenon has been investigated in a variety of relevant astrophysical environments,
also including sources far outside the Milky Way \citep[e.g.,][]{impellizzeri2008}.

One highlight was the discovery of high-velocity H$_2$O maser emission in the nucleus of NGC\,4258, offset by $\pm$\,1000\,km\,s$^{-1}$ from their host galaxy in
systemic velocity 
\citep[e.g.,][]{nakai1993, miyoshi1995, herrnstein1999}. This  led to the discovery of so-called disk masers, which trace subparsec edge-on Keplerian disks 
that surround supermassive black holes (SMBHs).  Later on, the Keplerian motion was used to determine the SMBH masses and direct angular diameter distances. 
The distance measurement to NGC\,4258 is not only independent of the traditional distance ladder, but is also among the most accurate in extragalactic space \citep{herrnstein1999, 
humphreys2013}. The importance of such direct geometrical extragalactic distance measurements provided the motivation to carry out surveys for finding more H$_2$O maser disks \citep[e.g.,][]{braatz2004}.

The high accuracy of distances obtained from this method can reduce uncertainties in the Hubble constant. With this idea in mind, the Megamaser Cosmology 
Project (MCP) was initiated, with the purpose of measuring the Hubble constant with 3\% accuracy. In the framework of the MCP, thousands of galaxies have been searched 
for H$_2$O megamaser\footnote{Extragalactic masers are about a million times more luminous than many Galactic masers, hence they are called megamasers.} emission. To  
date, about 160 galaxies with H$_2$O megamaser emission are known, 39 of which are disk-maser candidates\footnote{See the MCP webpage: \url{https://safe.nrao.edu/wiki/bin/view/Main/MegamaserCosmologyProject}}
\citep[32 ``clean'' disk masers, where ``clean'' means that the maser emission arises from an edge-on Keplerian disk that dominates other emission from nuclear jets
or outflows, see][]{pesce2015}.
While the number of disk masers is low, ($\sim$1\,\% of local Seyfert\,2s and low ionization nuclear emission region \citep[LINER) galaxies, see, e.g.,][and references therein]{vandenbosch2016}, 
their unique geometrical properties, such as an edge-on disk with a putative jet in the plane of the sky, provide motivation to investigate the host galaxies 
of these H$_2$O megamaser disks
in more detail in order to better understand their nuclear environment. Investigating the radio continuum of these galaxies can reveal emission from inside the maser 
disks as well as jets or outflows in the vicinity of the central black hole. By definition, these galaxies are particularly suited for studying the accretion disk--jet
paradigm 
under extremely well-defined boundary conditions, including the knowledge of distance, inclination of the accretion disk, and mass of the SMBH.

Here, we present radio continuum data observed at a frequency range centered at 33\,GHz (9\,mm wavelength) from 24 such disk-maser sources, 
obtained with the Karl Jansky Very Large Array (VLA)\footnote{The Karl 
Jansky Very Large Array (VLA) is a facility of the National Radio Astronomy Observatory (NRAO), which is operated by the associated universities, Inc., under a cooperative 
agreement with the National Science Foundation (NSF).}. The purpose of our investigations is to compare the geometry and luminosity of the parsec-scale maser disk  
with nuclear radio continuum properties, as well as to probe these galaxies for large-scale (kpc scale) radio jets. 
Resolving structure at kpc scale requires an angular resolution of 0.2-0.5 arcseconds at the distances of our sources.
To achieve this resolution,
we chose the Ka band of the higher frequency bands, which uses frequencies not too far from the 22\,GHz H$_2$O maser line (to obtain a realistic 
idea of the radio continuum distribution and intensity near the frequency of the H$_2$O maser). At the same time, this band minimizes the atmospheric
attenuation (which may be stronger at 22\,GHz). 

This paper is organized in the following way: in Sect.\,\ref{sec:sample} we introduce our sample, and in Sect.\,\ref{sec:data} we describe the data and data reduction.
In Sect.\,\ref{sec:result} the 33\,GHz continuum maps are presented, followed by analysis and discussion in Sect.\,\ref{sec:discussion}. A summary is given in 
Sect.\,\ref{sec:sum}.
% %__________________________________________________________________
\section{Sample}\label{sec:sample}

A total of 39 H$_2$O disk-maser candidates were identified, mostly by the MCP, until February 2016. We initially selected 30 H$_2$O
disk-maser 
candidates from those that were known in August 2013. Of this sample, 24 galaxies (see Table \ref{table:calibrators}) were observed,
with declination (Dec) $>$ \hbox{--30\,$^{\circ}$} and right ascension (R.A.) 
from 0$^{\rm h}$ to 17$^{\rm h}$. Four H$_2$O disk-maser candidates with Dec $<$\,--30\,$^{\circ}$  are not suitable for observation with the VLA. Two more 
northern galaxies, NGC\,4258 and NGC\,1068, have been well studied in the past \citep[e.g.,][]{herrnstein1998, gallimore2001} and were therefore also not included in our sample.

Three of the 24 observed galaxies have been monitored with the goal of measuring the Hubble constant 
(UGC\,3789: Braatz et al. 2010, Reid et al. 2013; IC\,2560: Wagner et al., in prep.; Mrk\,1419: Impellizzeri et al., in prep.),
and 13 maser disks were used to measure the SMBH mass \citep[$M_{\rm SMBH}$, see][]{kuo2011, braatz2015, gao2017}.   
As mentioned before, the distances measured by the MCP are among the most accurate measurements, but they are only available for three sources in our sample. Therefore we adopted 
the distances from the NASA$/$IPAC Extragalactic Database (NED)\footnote{\url{https://ned.ipac.caltech.edu/}} , which are sufficiently accurate for our purposes. The NED 
distances were obtained using H$_0$=70.0\,km\,s$^{-1}$\,Mpc$^{-1}$, $\Omega_{\rm matter}$=0.27, and $\Omega_{\rm vacuum}$=0.73 as cosmological parameters. 
H$_2$O maser luminosities from the literature were also rescaled to H$_0$=70.0\,km\,s$^{-1}$\,Mpc$^{-1}$ to be consistent with other luminosities in this work.

  %------------------------------------------------------------------------  
 \begin{table*}[ht]
% \begin{center}
\caption {VLA observations} \label{table:calibrators}
{\tiny
\begin{tabular}{l  c l l l c c c c} 
\hline \hline
& & & & & & & &\tabularnewline
Galaxy  & R.A. & Dec. & Bandpass & Amplitude and & R.A. & $\delta$R.A.  & Dec. & $\delta$Dec.   \tabularnewline
&  \multicolumn{2}{c}{$J$2000} & calibrator& phase calibrator&  $J$2000& & $J$2000 &  \tabularnewline
&                &  &    &    & & (arcsec) & & (arcsec)        \tabularnewline
\hline
& & & & & \tabularnewline
ESO558-G009            & 07:04:21.02  &   --21:35:19.2 & 3C147 & J0731-2341 &     07:04:21.01  & 0.06  & -21:35:19.03 & 0.05    \tabularnewline

IC\,0485/ UGC\,4156    & 08:00:19.77  &    +26:42:05.2 & 3C147 & J0748+2400 &     ...          & ...   & ...          & ...    \tabularnewline
IC\,2560               & 10:16:18.72  &   --33:33:49.7 & 3C286 & J1018-3144 &     10:16:18.71  & 0.16  & -33:33:49.60 & 0.16    \tabularnewline

J0126-0417             & 01:26:01.66  &   --04:17:56.2 & 3C48  & J0115-0127 &     01:26:01.64  & 0.04  & -04:17:56.23 & 0.04     \tabularnewline 
J0350-0127             & 03:50:00.35  &   --01:27:57.7 & 3C147 & J0339-0146 &     03:50:00.35  & 0.15  & -01:27:57.40 & 0.15    \tabularnewline
J0437+2456             & 04:37:03.69  &    +24:56:06.9 & 3C147 & J0426+2327 &      ...         &    ...&      ...     & ...      \tabularnewline
J0437+6637             & 04:37:08.26  &    +66:37:42.3 & 3C147 & J0449+6332 &     04:37:08.28  & 0.12  & +66:37:42.10 & 0.12     \tabularnewline
J0836+3327             & 08:36:22.80  &    +33:27:38.7 & 3C286 & J0827+3525 &     ...          & ...   & ...          & ...      \tabularnewline
J1658+3923             & 16:58:15.50  &    +39:23:29.3 & 3C48  & J1653+3945 &     16:58:15.54  & 0.22  & +39:23:29.27 & 0.20     \tabularnewline

Mrk\,0001/NGC\,0449    & 01:16:07.25  &    +33:05:22.4 & 3C48  & J0112+3208 &     01:16:07.20  & 0.02  & +33:05:21.75 & 0.02     \tabularnewline
Mrk\,0078              & 07:42:41.73  &    +65:10:37.5 & 3C286 & J0805+6144 &     07:42:41.73  & 0.03  & +65:10:37.39 & 0.03    \tabularnewline
Mrk\,1029              & 02:17:03.57  &    +05:17:31.4 & 3C48  & J0224+0659 &     02:17:03.57  & 0.18  & +05:17:31.15 & 0.18     \tabularnewline
Mrk\,1210/Phoenix      & 08:04:05.86  &    +05:06:49.8 & 3C147 & J0811+0146 &     08:04:05.86  & 0.03  & +05:06:49.83 & 0.03    \tabularnewline
Mrk\,1419/NGC\,2960    & 09:40:36.38  &    +03:34:37.2 & 3C147 & J0948+0022 &     09:40:36.38  & 0.14  & +03:34:37.36 & 0.13    \tabularnewline

NGC\,0591/Mrk\,1157    & 01:33:31.27  &    +35:40:05.7 & 3C48  & J0148+3854 &     01:33:31.23  & 0.10  & +35:40:05.79 & 0.11     \tabularnewline
NGC\,1194              & 03:03:49.11  &   --01:06:13.5 & 3C48  & J0312+0133 &     03:03:49.11  & 0.03  & -01:06:13.48 & 0.03     \tabularnewline
NGC\,2273              & 06:50:08.66  &    +60:50:44.9 & 3C147 & J0650+6001 &     06:50:08.69  & 0.28  & +60:50:45.10 & 0.27    \tabularnewline
NGC\,2979              & 09:43:08.65  &   --10:22:59.7 & 3C286 & J0943-0819 &     09:43:08.64  & 0.04  & -10:23:00.02 & 0.04    \tabularnewline
NGC\,3393              & 10:48:23.46  &   --25:09:43.4 & 3C286 & J1037-2934 &     10:48:23.46  & 0.03  & -25:09:43.44 & 0.03    \tabularnewline
NGC\,4388              & 12:25:46.75  &    +12:39:43.5 & 3C286 & J1218+1105 &     12:25:46.78  & 0.02  & +12:39:43.77 & 0.02     \tabularnewline
NGC\,5495              & 14:12:23.35  &   --27:06:28.9 & 3C286 & J1409-2657 &     14:12:23.35  & 0.05  & -27:06:29.14 & 0.06     \tabularnewline
NGC\,5728              & 14:42:23.90  &   --17:15:11.1 & 3C286 & J1439-1659 &     14:42:23.89  & 1.00  & -17:15:10.76 & 1.00     \tabularnewline

UGC\,3193              & 04:52:52.58  &    +03:03:25.9 & 3C147 & J0503+0203 &     04:52:52.56  & 0.11  & +03:03:25.52 & 0.29    \tabularnewline
UGC\,3789              & 07:19:30.92  &    +59:21:18.4 & 3C147 & J0728+5701 &     07:19:30.95  & 0.08  & +59:21:18.37 & 0.08    \tabularnewline
 \hline                                                                                           
\end{tabular}\par
\bigskip
\textbf{Notes}. Column 1: name of galaxy.  Column 2: $J$2000 NED right ascension. Column 3: $J$2000 NED declination. Column 4: bandpass calibrator. Column 5: amplitude and phase 
calibrators (these are the same for each object). Column 6 and 7: $J$2000 right ascension determined from our radio maps and their uncertainties.
Column 8 and 9: $J$2000 declination determined from our radio maps and their uncertainties.
}   
% \end{center}                                                        
\end{table*}
% %__________________________________________________________________
 
\section{Data and data reduction}\label{sec:data}
\subsection{33\,GHz observations and data reduction}
Our sample of disk-maser galaxies was observed  
by the VLA in Ka band (26.5-40.0 GHz) in B, CnB, or BnA configurations using a phase-referencing mode (proposal code: VLA/13B-340). 
Synthesized full-width at half-maximum (FWHM) beam sizes ranged from 0.19 to 0.50 arcseconds. 
We used a total bandwidth of 8\,GHz (4$\times$2~GHz, 
from 29\,GHz to 37\,GHz), full polarization, and 3-bit sampling. The sources were grouped into 12 sets of pairs for observation, with a total time of one hour per pair
and an integration time of $\sim$10 minutes on source per target. Table\,\ref{table:calibrators} indicates the $J$2000 coordinates, amplitude, phase, and bandpass calibrators.
The data were calibrated with the NRAO VLA calibration pipeline, using standard procedures of the Common Astronomy Software Applications (CASA)\footnote{\url{http://casa.nrao.edu/}} 
package \citep[see][]{mcmullin2007}. These standard procedures include radio frequency interference flagging, deterministic flagging (e.g., online flags, end channels),
opacity and antenna corrections, and bandpass, amplitude and phase calibrations. The majority of our sources was not bright enough for self-calibration ($\sim$1\,mJy, see 
Table\,\ref{table:properties}). However, for Mrk\,1210, phase and amplitude self-calibration could be performed, which resulted in an improvement 
of the quality of its radio map.
For imaging, the CLEAN algorithm with natural weighting was used. The latter is chosen to
maximize the chance of detecting weak compact or slightly extended sources and is hence optimal for detection projects.

\subsection{Complementary data}
For further analysis and investigation we obtained from the literature 1.4\,GHz radio data from the NVSS\footnote{NRAO VLA Sky Survey, \url{http://www.cv.nrao.edu/nvss/}} \citep{condon1998}, 1.4\,GHz data from the 
FIRST\footnote{Faint Images of the Radio Sky at Twenty-cm, \url{http://sundog.stsci.edu/}} survey \citep{becker1995}, and hard X-ray data from the \textsl{Swift}/BAT satellite in a range of 20-100 keV 
(Litzinger et al., in prep.), as well as 22\,$\mu$m (W4) infrared WISE\footnote{Wide-Field Infrared Survey Explorer (WISE). See \url{http://wise.ssl.berkeley.edu/}} data \citep{wright2010}.
Seventeen of the 24 galaxies in our sample were detected with the NVSS, 8 with FIRST, 13 with \textsl{Swift}/BAT, and all 24 with WISE. 
From radio and X-ray data, the luminosities were calculated using $L\,=\,4\,\pi\,D^2\,F$, where $F$ stands for flux and $D$ for distance.
The infrared luminosities were calculated 
using
$\log\,(L/L_{\odot})\,=\,\,(M_{\odot,\,\rm W4}-M_{\rm W4})/2.5$,  where $L_{\odot}$ is the bolometric luminosity of the Sun, $M_{\odot,\,\rm W4}$
is the absolute magnitude of the Sun in WISE W4 band, and $M_{\rm W4}$ is the absolute magnitude of our sources in this band.

%______________________________________________ 

 \begin{table*}[t]
\caption [The sample]{Sample properties. } \label{table:properties}
{\tiny
\begin{tabular}{l c c c c c l l l l l} 
\hline \hline
& & & & & & & & &\tabularnewline
Galaxy   &  Distance & 33\,GHz flux
 & rms&NVSS flux&FIRST flux & IR magnitude& hard X-ray flux&log L$_{H_2O}$  &Type of\tabularnewline
                      & (Mpc)&    (mJy)  & ($\mu$Jy/beam) &  (mJy) & (mJy) &  & ($10^{-11}$$\rm erg\,\rm s^{-1}\,\rm cm^{-2}$) & (L$_\odot$) &activity\tabularnewline
& & & & & & \tabularnewline
\hline
& & & & & & \tabularnewline
ESO558-G009           &    112.2 $\pm$ 7.9      & 0.80 $\pm$ 0.04     &  18.6   & 12.8$\pm$0.6     & ...          &  8.793                 & <0.34                         & 2.9   & $U$   \tabularnewline
IC\,0485              &    122.0 $\pm$ 8.5      & ...                 &  18.9   & ...              & 3.01          &  9.406               & 1.75 $^{+0.19}_{-0.18}$       & 3.1   & $U$   \tabularnewline
IC\,2560              &     46.4 $\pm$ 3.3      & 2.00 $\pm$ 0.10     &  17.3   & 32.0$\pm$1.7     & ...       &  9.294               & 0.41$^{+0.12}_{-0.13}$        & 2.1   & Sy2   \tabularnewline

J0126-0417            &     76.2 $\pm$ 5.4      & 0.13 $\pm$ 0.01     &   9.9   & ...              & ...       &  8.930               & <0.34                         & 2.1   & $U$    \tabularnewline 
J0350-0127            &    174.2 $\pm$ 12.2     & 0.14 $\pm$ 0.02     &  12.0   & ...              & ...       &  8.529               & 0.45$^{+0.17}_{-0.16}$        & 3.7   & $U$   \tabularnewline
J0437+2456            &     68.1 $\pm$ 4.8      & ...                 &   17.7  & ...              & ...       &  9.024           & <0.41                         & 2.3   & $U$   \tabularnewline
J0437+6637            &     52.9 $\pm$ 3.7      & 0.11 $\pm$ 0.02     &  16.8   & 3.8$\pm$0.6      & ...       &  8.934               & 0.29$\pm$0.14                 & 1.4   & $U$   \tabularnewline
J0836+3327            &    214.6 $\pm$ 15.0     & ...                 &  30.9   & ...               & 1.95         &  9.145$\pm$0.459    & <0.33                           & 3.6   & Sy2   \tabularnewline
J1658+3923            &    146.8 $\pm$ 10.3     & 0.18 $\pm$ 0.04     &  13.4   & ...              & 3.64      &  9.513               & <0.30                         & 2.9   & Sy2   \tabularnewline

Mrk\,0001             &     64.2 $\pm$ 4.5      & 3.96 $\pm$ 0.12     &  10.8   & 75.9 $\pm$ 2.3   & ...          &  9.105                 & <0.32                         & 1.9   & Sy2   \tabularnewline
Mrk\,0078             &    159.9 $\pm$ 11.2     & 1.75 $\pm$ 0.11     &  13.7   & 36.9$\pm$1.2     & ...          &  8.975                 & 0.63 $\pm$ 0.18               & 1.6   & Sy2         \tabularnewline
Mrk\,1029             &    126.0 $\pm$ 8.8      & 1.18 $\pm$ 0.14     &  17.7   & 11.9$\pm$0.6     & 8.12         &  8.799                 & <0.35                         & 2.8   & $U$         \tabularnewline
Mrk\,1210             &     61.3 $\pm$ 4.3      & 9.58 $\pm$ 0.08     &  17.9   & 114.9$\pm$3.5    & ...          &  8.523                 & 3.58$\pm$0.20                 & 2.0   & Sy2, Sy1 \tabularnewline
Mrk\,1419             &     75.3 $\pm$ 5.3      & 0.36 $\pm$ 0.06     &  16.0   & 7.4 $\pm$0.5     & 5.13         &  8.790                 & <0.34                         & 2.7   & LINER \tabularnewline

NGC\,0591             &     61.1 $\pm$ 4.3      & 1.52 $\pm$ 0.07     &  13.6   & 33.3$\pm$1.1     & ...          &  8.566                 & 0.37$^{+0.12}_{-0.11}$        & 1.5   & Sy2   \tabularnewline
NGC\,1194             &     55.4 $\pm$ 3.9      & 1.08 $\pm$ 0.04     &  13.0   & ...              &  1.49        &  8.807                 & 2.21$\pm$0.18                 & 2.8   & Sy\,1.9\tabularnewline
NGC\,2273             &     26.8 $\pm$ 1.9      & 2.69 $\pm$ 0.29     &  26.3   & 63.4$\pm$2.4     & ...          &  9.734$\pm$0.511    & 0.67 $\pm$ 0.16            & 0.9   & Sy2   \tabularnewline
NGC\,2979             &     43.8 $\pm$ 3.1      & 0.36 $\pm$ 0.04     &  14.0   & 15.7$\pm$1.0     & ...          &  9.158                 & <0.36                         & 2.2   & Sy2   \tabularnewline
NGC\,3393             &     58.6 $\pm$ 4.1      & 5.30 $\pm$ 0.36     &  18.5   & 81.5$\pm$3.3     & ...          &  8.846                 & 1.56$\pm$0.19                 & 2.7   & Sy2         \tabularnewline
NGC\,4388             &     40.8 $\pm$ 2.9      & 8.57 $\pm$ 0.43     &  17.9   & 120.4$\pm$4.7    & 45.02        &  8.814                 & 15.81$^{+0.16}_{-0.15}$       & 1.2   & Sy2         \tabularnewline
NGC\,5495             &     99.8 $\pm$ 7.0      & 0.13 $\pm$ 0.02     &  15.4   & 12.5$\pm$ 1.3    & ...          &  9.301                 & <0.40                         & 2.4   & Sy2         \tabularnewline
NGC\,5728             &     43.3 $\pm$ 3.1      & 2.27 $\pm$ 0.06     &  28.1   & 70.8$\pm$2.9     & ...          &  9.126                 & 5.20$\pm$0.20                 & 2.0   & Sy2         \tabularnewline

UGC\,3193             &     63.2 $\pm$ 4.4      & 3.95 $\pm$ 0.27     &  14.4   & 17.7$\pm$0.7     & ...          &  8.942                 & <0.39                         & 2.5   & $U$     \tabularnewline
UGC\,3789             &     47.4 $\pm$ 3.3      & 0.21 $\pm$ 0.02     &  14.9   & 17.6$\pm$ 1.0    & 14.43        & 8.326$\pm$0.142     & 0.26$^{+0.14}_{-0.13}$    & 2.7   & Sy2   \tabularnewline

\hline
                                                                                                                                                                                                                                                 
\end{tabular}\par
\bigskip
\textbf{Notes}. Column 1: name of galaxy.
Column 2: Hubble flow distances (relative to the 3\,K CMB), assuming H$_0$=70\,km\,s$^{-1}$\,Mpc$^{-1}$ (see Sect.\,2).
Column 3: 33 GHz integrated flux densities with uncertainties.
          The errors given in those cases are formal values and do not show the real level of uncertainty.  
Column 4: root mean square noise level of the clean image.
Column 5: NVSS integrated flux.
Column 6: FIRST integrated flux.
Column 7: IR (WISE, W\,4) magnitude. In our analysis, a 10\% uncertainty on the magnitude is assumed for those cases where the uncertainty is not given in the literature.
Column 8: \textsl{Swift}/BAT hard X-ray fluxes (20-100\,keV) with uncertainties (Litzinger et al., in prep). 
Column 9: logarithm of water maser luminosity (isotropy assumed) \citep[][modified for H$_0$=70\,km\,s$^{-1}$\,Mpc$^{-1}$]{zhang2012}. 
In our analysis, 10\% uncertainty on the luminosity is assumed. 
Column 10: types of nuclear activity after NED; U stands for unidentified.
}                                                                      
\end{table*}

%------------------------------------------------------------------------------------------------------------------------------ 
\section{Results}\label{sec:result}
\subsection{Sample}\label{sec:result_sample}
In this section we present the 33\,GHz radio maps for all observed sources, centered at their NED positions, as well as a table of derived source properties.
Important in this context is that our linear resolution in pc
for a 0.2$\arcsec$ beam size is close to the distance of the galaxies measured in units of Mpc,
that is, (linear resolution\,/\,pc) $\approx(D\,/\,\rm Mpc)$. 
Therefore, our beam typically covers
$\sim$100\,pc, but with large deviations to either side. This implies that even the central beam contains not only a galaxy's very center, but also the surrounding
environment, which may contain star-forming regions and supernova remnants that also contribute to activity and radio continuum emission.  

Figure\,\ref{fig:images1} presents the 33\,GHz contour maps for our sample, where the contour levels are $\pm$3,  $\pm$6, $\pm$12, $\pm$24 times
the root-mean-square noise (rms) of an emission-free field in the CLEANed image (see Table\,\ref{table:properties} for the sensitivities). 
Since our detection limit is 4.5$\sigma$, a contour level of this value is also shown in Fig.\,\ref{fig:images1} (in blue),
to facilitate distinguishing between a weakly detected source and an undetected source.
Figure\,\ref{fig:images1} clearly shows the large number of sources exhibiting a compact core without much additional structure. While our classification depends on limited signal-to-noise ratios and is 
thus sometimes debatable, we count 16 such compact sources, where ``compact'' used here and in the following denotes a beam-convolved (and deconvolved) major axis 
smaller than 2.5 times the beam major axis. One source shows two main components, another source three main components, and three other sources exhibit extended emission. We note that the 
sources with two or three main components also contain compact
cores, as do some of the extended sources.

Only three of the 24 galaxies (IC\,485, J0437+2456, and J0836+3327) were not detected at 33\,GHz at a level of 4.5\,$\sigma$ or higher. Two of these galaxies, 
IC\,485 and J0836+3327, are at relatively large distances of 122\,Mpc and 214\,Mpc, respectively. IC\,485 was detected with a flux density of 3.01\,mJy, and J0836+3327 
with 1.95\,mJy by FIRST. Adopting a mean spectral index of 0.73 obtained in this work (Sect.\,\ref{sec:sp_index}) between FIRST 1.4\,GHz and our 33\,GHz observations
\footnote{Assuming a power-law dependence for the continuum flux density given by $S$\,$\propto$\,$\nu^{-\alpha}$, see Sect.\,\ref{sec:sp_index}.}, flux densities of 
$\sim$0.30\,mJy and $\sim$0.19\,mJy are expected to be observed at 33\,GHz for IC\,485 and J0836+3327, respectively. 
These flux densities are still higher than our 4.5\,$\sigma$ detection threshold. Therefore, these sources may have spectral indices steeper than 1.0.
The third source, J0437+2456, is closer (68\,Mpc), but remained undetected by both NVSS and FIRST.

\begin{figure}[ht]
 \includegraphics[width=0.46\textwidth]{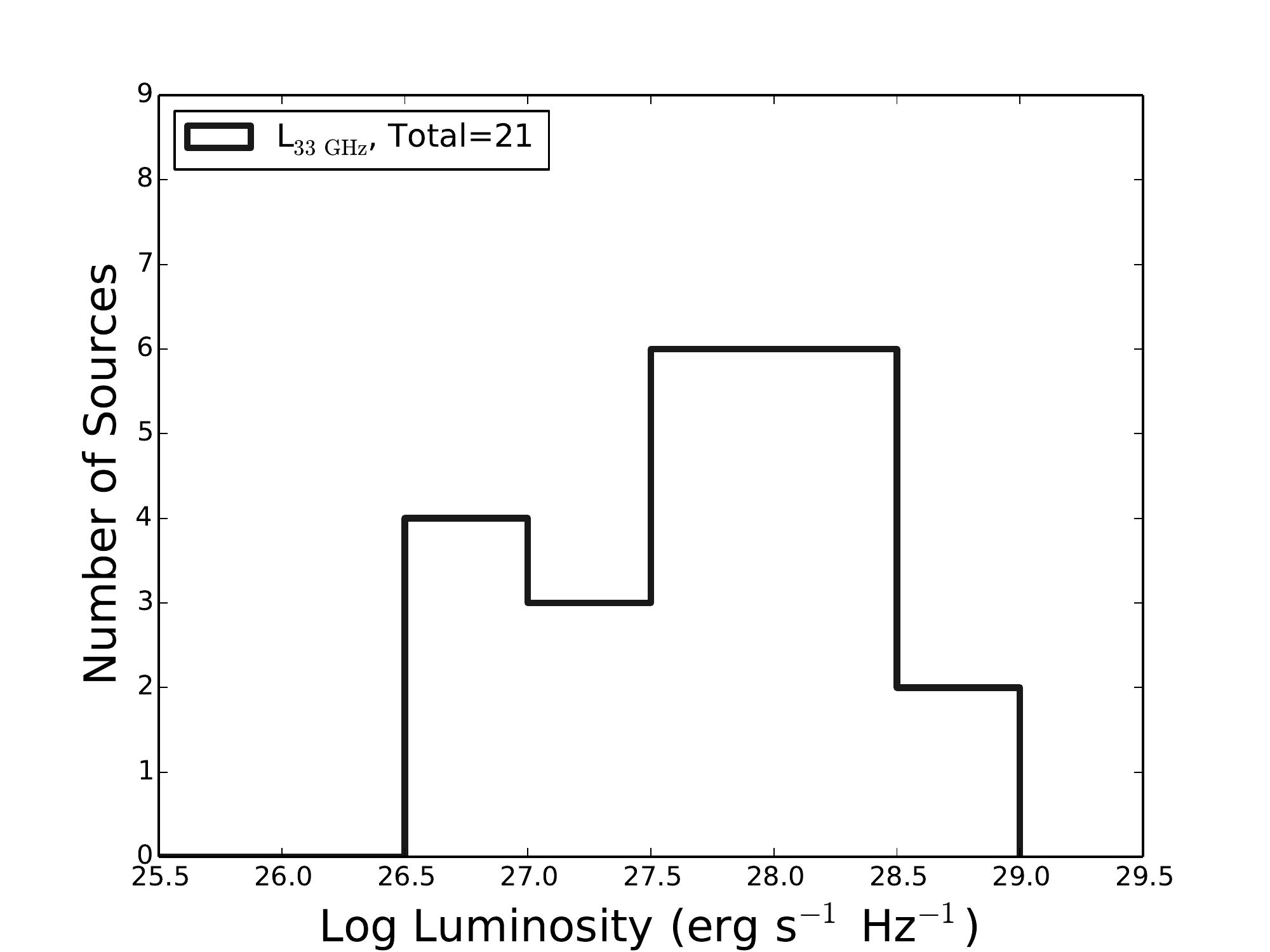}
 \caption{33\,GHz luminosity distribution for the 21 clearly detected targets.}\label{fig:lum33_hist} 
\end{figure}

With 21 galaxies detected at Ka band (29-37\,GHz), 87\% of the galaxies in our sample show radio emission at levels of 4.5\,$\sigma$ to 240\,$\sigma$.  
The 33\,GHz continuum luminosity distribution is presented in Fig.\,\ref{fig:lum33_hist}.

Furthermore, our 33\,GHz maps have improved the accuracy of the central coordinates of some galaxies in our sample. We identify the compact radio sources as the galactic 
nuclei. After fitting two-dimensional Gaussians (in the image domain) on the central component, the central position of the Gaussian fit is adopted as the new position of the center of
the respective galaxy. These new radio coordinates and their uncertainties are presented in Table\,\ref{table:calibrators}.
Our position uncertainties take into account statistical uncertainty, the uncertainty of the phase calibrator position, and systematic uncertainties. 
The systematic uncertainty was approximated from $(B^2+\Psi^2)^{1/2}\times(1/\rm SNR+1/20)$, where B is the beam size, $\Psi$ is the source major axis, and
SNR is the signal-to-noise ratio of the map \citep[for more details see][]{white1997}.
For 10 galaxies in our sample, interferometrically determined 
coordinates from the maser emission are also available \citep{kondratko2006a, kondratko2008, kuo2011, gao2017}. The central positions obtained in this work and those 
obtained from maser emission deviate on average by 121 $\pm$ 107~mas (milliarcseconds) and are mostly in agreement within the limits provided by beam size and 
signal-to-noise ratio (S/Ns).
See Table\,\ref{table:maser_33_pos} for a comparison of our determined positions and the H$_2$O maser positions.
Noteworthy deviations beyond the 3$\sigma$ level are found for ESO\,0558--G009, NGC~1194, Mrk~1419, and NGC~4388, all sources where at least slightly extended emission is
found in our continuum data. For two sources, Mrk\,1029 and NGC\,5495, our coordinates have lower uncertainties than those of the maser positions. 

In Tables\,\ref{table:properties} and \ref{table:indices_inclination}
we list the derived properties of the sources. Flux density and position angle (P.A.) have been determined 
by two-dimensional Gaussian model fitting on a region containing all the significant emission. 
For sources with more complex radio morphologies, multiple Gaussians were fitted in order to have a better estimate of the fluxes (NGC\,0591, NGC\,3393, NGC\,4388, UGC\,3193).
In what follows we mention some relevant properties of these galaxies such as their morphology, 
activity type, or other interesting features that are discussed in the literature together with some basic properties that can be extracted from our maps. Statistical properties
are discussed in Sect.\,5.
 %------------------------------------------------------------------------
 \begin{center}
\begin{figure*}[ht]
 \includegraphics[width=\textwidth]{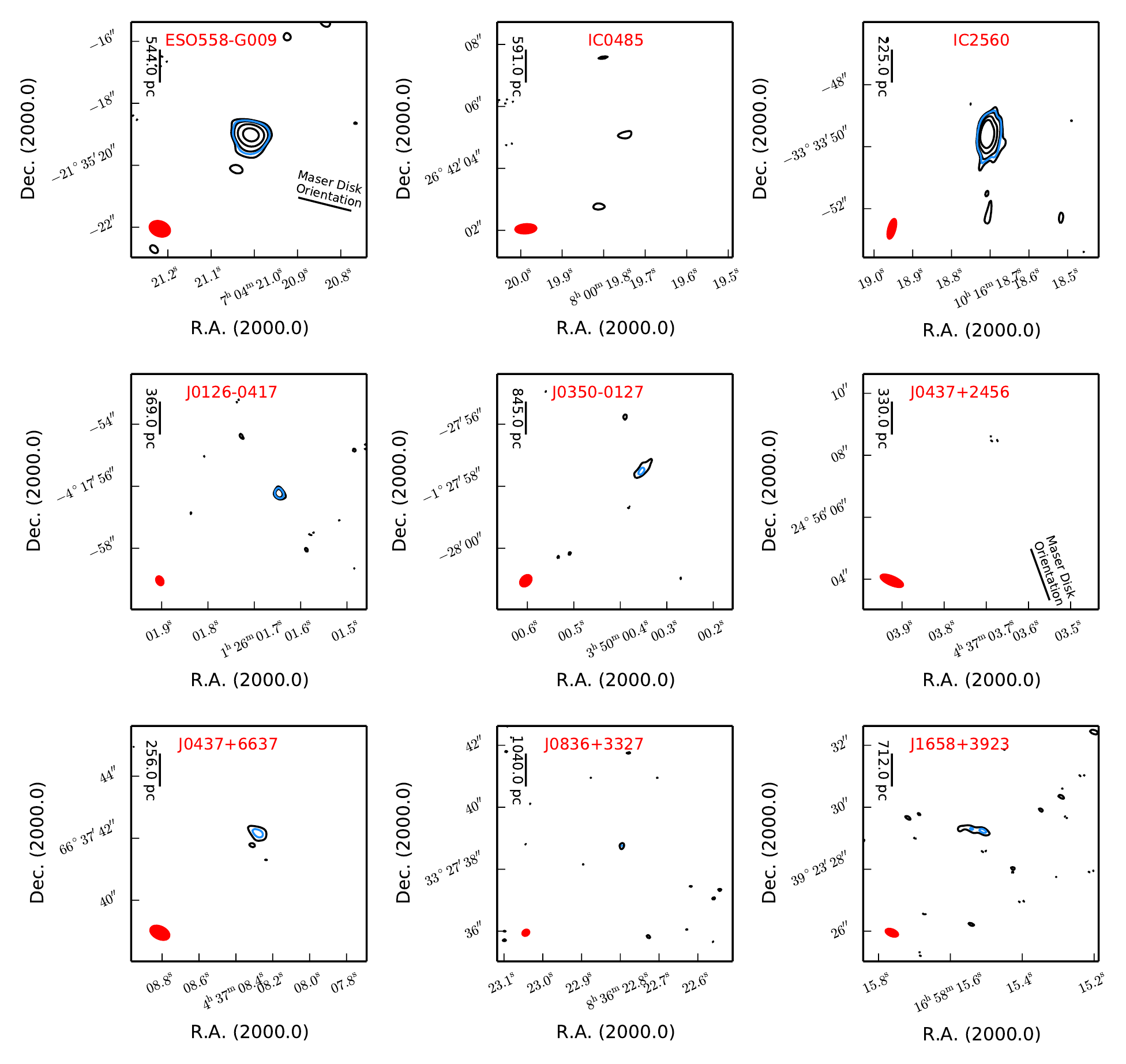}
 \caption[Contour Maps]{33\,GHz contour maps. The contour levels are $\pm$3, $\pm$6, $\pm$12, and $\pm$24 times the rms given in Table\,\ref{table:properties}.
For convenience, our detection limit of 4.5$\sigma$ is shown as a blue contour.
 The synthesized beam is shown in red in the lower left corner of each plot. When available, we also plot the orientation (not the position)
 of the H$_2$O maser disk.}\label{fig:images1}
\end{figure*}
\end{center}
 \begin{center}
\begin{figure*}[ht]
 \ContinuedFloat
%  \captionsetup{list=off,format=cont}
 \includegraphics[width=\textwidth]{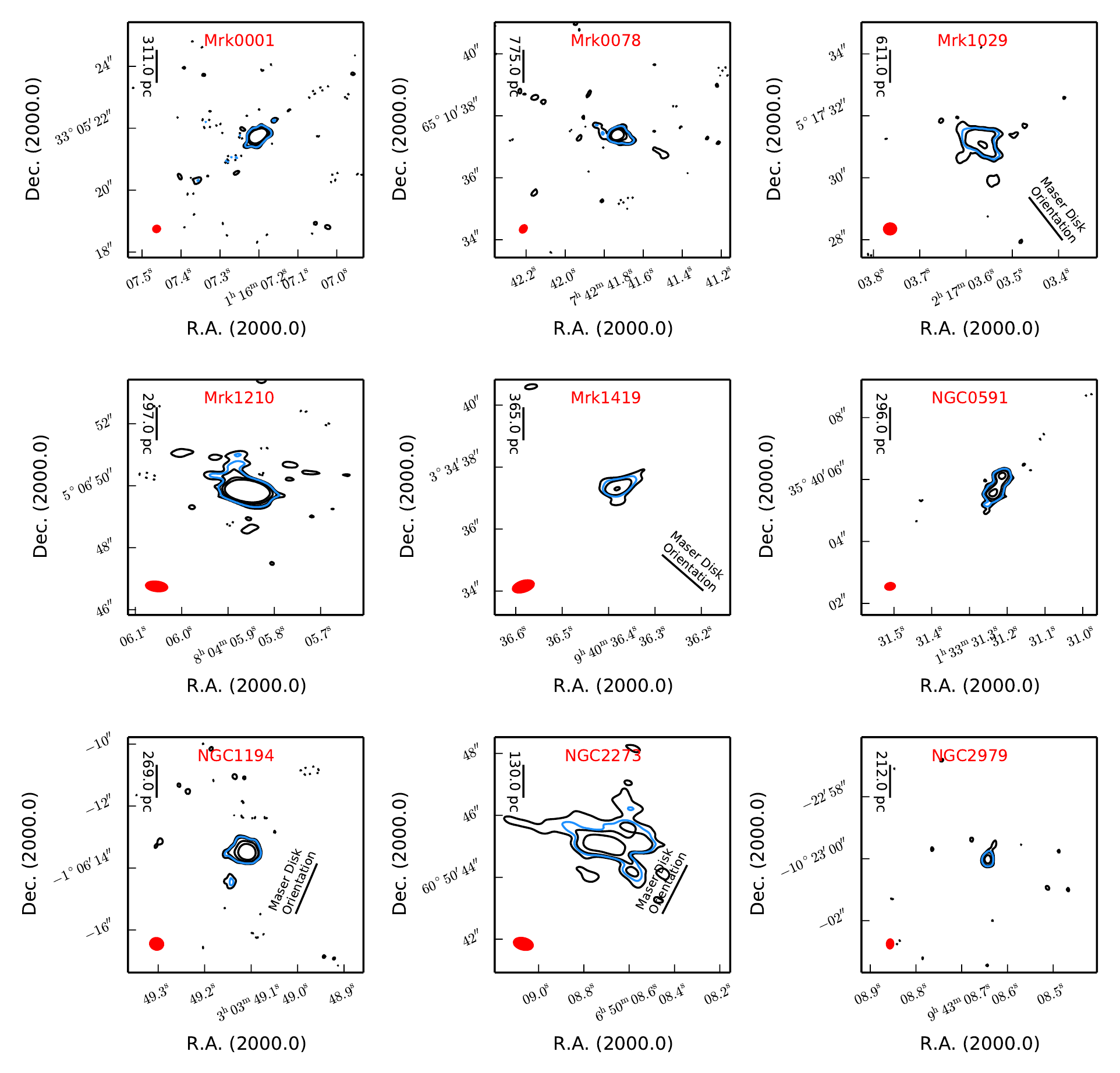}
 \caption[Contour Maps]{\emph{(cont.)}}
%  \caption[Contour Maps]{The 33\,GHz contour maps for 12 galaxies of our sample. The contour levels are $\pm$3,  $\pm$6, $\pm$12, $\pm$24 times the rms given in Table\,\ref{table:properties}.
%  The synthesized beam is shown in red in the lower left corner of each plot. If available, the orientation of the H$_2$O-maser-disk is presented as well.}\label{fig:images1}
\end{figure*}
\end{center}

 \begin{center}
\begin{figure*}[ht]
 \ContinuedFloat
%  \captionsetup{list=off,format=cont}
 \includegraphics[width=\textwidth]{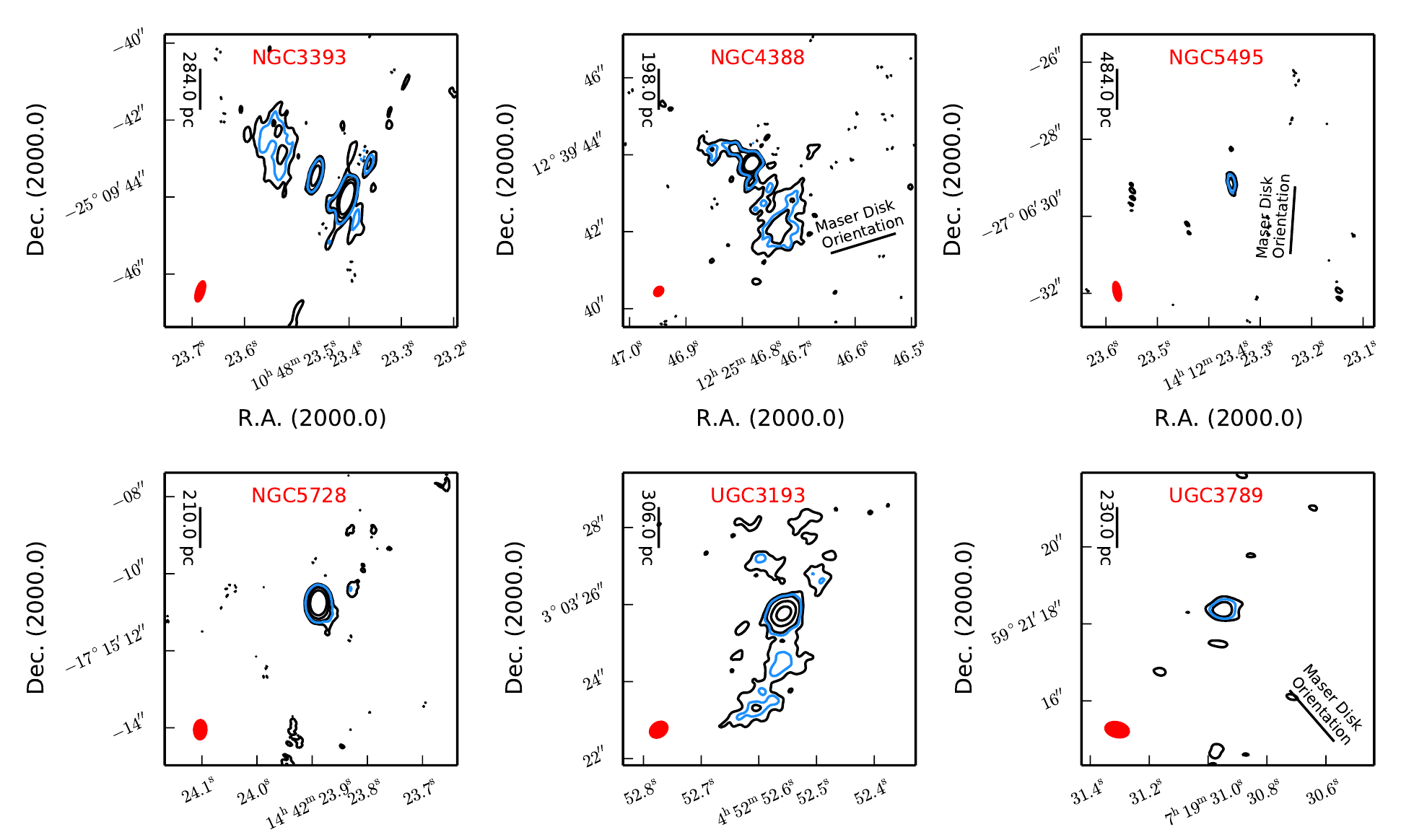}
\caption[Contour Maps]{\emph{(cont.)}}
%  \caption[Contour Maps]{The 33\,GHz contour maps for 12 galaxies of our sample. See caption of Fig\,\ref{fig:images1} for more details.}\label{fig:images2}
\end{figure*}
\end{center}
%------------------------------------------------------------------------
\begin{sidewaystable*}
\caption{H$_2$O maser positions versus our 33\,GHz positions.}\label{table:maser_33_pos}
\begin{tabular}{l l l l l || l l l l|| l l|| c}

\hline \hline
& & & & & & & & & & & \tabularnewline
Galaxy  & \multicolumn{4}{c||}{Maser} & \multicolumn{4}{c||}{33GHz} & \multicolumn{2}{c||}{Difference}& Maser\tabularnewline
\
& $J$2000 R.A. &$\delta$R.A.  & $J$2000 Dec. &$\delta$Dec.  &$J$2000 R.A.  &$\delta$R.A.  & $J$2000 Dec. &$\delta$Dec. & $\Delta$R.A. & $\Delta$Dec. & Position \tabularnewline
 &  & (\arcsec) &  & (\arcsec) &  & (\arcsec) &  & (\arcsec)& (\arcsec) & (\arcsec) &  Reference \tabularnewline
  \hline
  & & & & & & & & & & &  \tabularnewline

ESO558-G009             &                07:04:21.0113          &               0.014           &                --21:35:18.948          &               0.014           &                07:04:21.01             &               0.06            &                --21:35:19.03           &               0.05            &               0.02            &               0.08            &                1       \tabularnewline
IC\,2560                &                10:16:18.710           &               0.075           &                --33:33:49.74           &               0.01            &                10:16:18.71             &               0.16            &                --33:33:49.60           &               0.16            &               0.00            &               0.14            &                2       \tabularnewline
Mrk\,1029               &                02:17:03.566           &               0.5             &                +05:17:31.43            &               0.5             &                02:17:03.57             &               0.18            &                +05:17:31.15            &               0.18            &               0.06            &               0.28            &                1       \tabularnewline
Mrk\,1419               &                09:40:36.38370         &               0.01            &                +03:34:37.2915          &               0.01            &                09:40:36.38             &               0.14            &                +03:34:37.36            &               0.13            &               0.06            &               0.07            &                3       \tabularnewline
NGC\,1194               &                03:03:49.10864         &               0.0002          &                --01:06:13.4743 &               0.0004          &                03:03:49.11             &               0.03            &                --01:06:13.48           &               0.03            &               0.02            &               0.01            &                3       \tabularnewline
NGC\,2273               &                06:50:08.65620         &               0.01            &                +60:50:44.8979          &               0.01            &                06:50:08.69             &               0.28            &                +60:50:45.10            &               0.27            &               0.25            &               0.20            &                3       \tabularnewline
NGC\,3393               &                10:48:23.4659          &               0.001           &                --25:09:43.477          &               0.001           &                10:48:23.46             &               0.03            &                --25:09:43.44           &               0.03            &               0.08            &               0.04            &                4       \tabularnewline
NGC\,4388               &                12:25:46.77914         &               0.0004          &                +12:39:43.7516          &               0.0003          &                12:25:46.78             &               0.02            &                +12:39:43.77            &               0.02            &               0.01            &               0.02            &                3       \tabularnewline
NGC\,5495               &                14:12:23.35            &               0.3             &                --27:06:29.20           &               0.3             &                14:12:23.35             &               0.05            &                --27:06:29.14           &               0.06            &               0.00            &               0.06            &                5       \tabularnewline
UGC\,3789               &                07:19:30.9490          &               0.01            &                +59:21:18.3150          &               0.01            &                 07:19:30.95            &               0.08            &                +59:21:18.37            &               0.08            &               0.01            &               0.06            &                3       \tabularnewline

\hline \hline

\end{tabular}\par
\bigskip
\textbf{Notes}.
Column\,1: source name.
Columns\,2 and 3: $J$2000 right ascension determined from H$_2$O maser observations and their uncertainties. 
Columns\,4 and 5: $J$2000 declination determined from H$_2$O maser observations and their uncertainties. 
Columns\,6 and 7: $J$2000 right ascension determined from our radio maps and their uncertainties.
Columns\,8 and 9: $J$2000 declination determined from our radio maps and their uncertainties.
Columns\,10 and 11: right ascension and declination difference between H$_2$O maser observations and our observations.
Column\,12:  references for H$_2$O maser observations: (1) \citet{gao2017}; (2) \citet{ishihara2001}; (3) \citet{kuo2011}; (4) \citet{kondratko2008} and (5)
\citet{kondratko2006b}.
\end{sidewaystable*}
%------------------------------------------------------------------------
\subsection{Individual sources}
\textit{ESO558-G009}  is an Sbc galaxy. The MCP reported rapid intra-day variability in the maser spectrum, which was interpreted as the result of interstellar scintillation 
\citep{pesce2015}. Compact radio emission is observed in our 33\,GHz map. A two-dimensional Gaussian fit (Table~\ref{table:indices_inclination}) provides a 
formal P.A. of 155$^{\circ}\pm$33$^{\circ}$. \citet{gao2017} reported a P.A. of 256$^{\circ}\pm$2$^{\circ}$ (redshifted side, here and elsewhere) for the maser disk. While 
this might indicate that the radio continuum elongation and the maser disk are almost perpendicular (separated by 101$^{\circ}\pm$33$^{\circ}$), this needs to be verified by 
additional mesurements.

\textit{IC\,0485}  is an Sa Seyfert\,2 galaxy. Radio continuum emission was not detected at 4.5\,$\sigma$ or higher levels
(but it is tentatively detected at the 3.5\,$\sigma$ level). As mentioned before, the galaxy is at a relatively large distance of 120 Mpc.

\textit{IC\,2560}  is an SBb Seyfert\,2 galaxy. \citet{ishihara2001} reported a $J$2000 22\,GHz peak radio continuum position of RA = 10h16m18.710s$\pm$0.006s and 
Dec = --33d33m49.74s$\pm$0.$\!\!^{``}$01 ($\pm$1pc). The blueshifted maser features spatially coincide with the 22\,GHz continuum component and are interpreted as jet maser emission 
\citep{ishihara2001}. A peak and an integrated flux density of 1.8 $\pm$ 0.3 mJy/beam and 1.7$\pm$ 0.5 mJy, respectively, were reported by \citet{yamauchi2012} at 22\,GHz. 
The nearly edge-on Keplerian disk with a position angle of --46$^{\circ}$ is almost perpendicular to the large-scale galactic disk \citep{ishihara2001}. This means that 
the rotation axis of the maser disk and that of the large-scale galactic disk are nearly perpendicular. Compact radio emission is observed in our 33\,GHz radio map, where 
the peak position agrees within the uncertainties with the position of the 22\,GHz continuum peak reported by \citet{yamauchi2012}.

\textit{J0126-0417} has compact radio continuum emission detected in our 33\,GHz map.

\textit{J0350-0127} is a (within the errors) compact source detected in our 33\,GHz map; this most likely represents the nucleus.

\textit{\textup{The source} J0437+2456}   is not detected at 33\,GHz. 

\textit{\textup{The source} J0437+6637} is detected in our 33\,GHz map and is unresolved in our Gaussian deconvolution.

\textit{J0836+3327}  is a Seyfert\,2 galaxy. No radio continuum was detected at a 4.5\,$\sigma$ or higher level in our 33\,GHz map
(but it is tentatively detected at a 4\,$\sigma$ level).

\textit{J1658+3923} is an Sc Seyfert\,2 galaxy. A slightly extended radio continuum source is observed in our map.

\textit{Mrk\,0001}  is an SBa Seyfert\,2 spiral galaxy. \citet{omar2002} observed H\,{\sc i} emission from Mrk\,1 and interpreted its morphology as due to a tidal interaction 
with its nearby companion NGC~451.
The authors also reported that the line of sight toward the nucleus of Mrk\,1 is rich in both atomic and molecular gas, or in other
words, that it is heavily obscured. 
Compact 33\,GHz radio emission 
is observed in our map.

\textit{Mrk\,0078} is classified as an SB Seyfert\,2 galaxy. The narrow line region (NLR) of this galaxy is affected by a strong jet-gas interaction, and
spectroscopic observations have shown that its narrow emission lines are double peaked \citep{sargent1972,adams1973}. Based on optical and 3.6\,cm VLA images, 
\citet{whittle2004} suggested that Mrk\,78 is a post-merger system, with a highly extended asymmetric gas distribution and a nuclear dust lane. The radio nucleus 
lies within this highly obscuring dust lane \citep{whittle2004}. Slightly extended radio emission might be detected east and west of the center of our 33\,GHz map,
but the S/Ns are low, so that this source is also rated compact.

\textit{Mrk\,1029} shows extended radio emission in our 33\,GHz map with a P.A. of 59$^{\circ} \pm$11\degr. \citet{gao2017} reported a P.A. of 
218\degr$\pm$10$^{\circ}$ for the maser disk. While we consider our position angle as uncertain in spite of the low formal error, it would suggest that 
maser disk and radio continuum have approximately the same orientation (separation of 21$^{\circ} \pm$15\degr).

\textit{Mrk\,1210,} also known as the Phoenix galaxy, is an amorphously looking spiral galaxy that has been classified as both Seyfert\,1 and Seyfert\,2. 
Former studies have confirmed a double structure, a core, and a southeastern jet component \citep{middelberg2004}. Spectroscopic findings support the presence of 
a nuclear outflow instead of a hidden broad line region \citep{mazzalay2007}. A flux density of 36.28 $\pm$ 1.95 mJy at 5 GHz was reported by \citet{Xanthopoulos2010}. 
We observe compact emission in our 33\,GHz map within the uncertainties.

\textit{Mrk\,1419,} also known as NGC\,2960, is an Sa galaxy. Mrk\,1419 hosts a LINER nucleus. The P.A. of the maser disk is 49\degr , and the disk inner and 
outer radii are 0.13\,pc and 0.37\,pc, respectively \citep{kuo2011}. \citet{sun2013} reported an extension in the 20\,cm radio map at a P.A. of 125$^{\circ} \pm$ 10 
$^{\circ}$. This extension may suggest that a jet is launched from the central black hole. In our 33\,GHz map, the radio emission may show a slight extension in the 
northwest-southeast direction, with a P.A. of 117$^{\circ} \pm$84\degr. Nevertheless, we rate the source as compact.

\textit{NGC\,0591}  is an SB0/a Seyfert\,2 galaxy. Near-infrared spectral studies of stellar and gas kinematics of NGC\,591 show that the gas has two kinematic 
components: one is located inside the plane of the galaxy with similar rotation to that of stellar motion, and the other is forming an outflow that is oriented along 
the radio jet directed toward the northwest from the nucleus \citep{riffel2011}. Our 33\,GHz map shows two peaks with a separation of $\sim$\,0.66 arcsec, more 
precisely, with an R.A. separation of 360 mas and a Dec separation of 550\,mas at a position angle of about --34$^{\circ}$. The central component has a flat spectrum 
with $\alpha^{36}_{30}$=0.54$\pm$0.44 that possibly is the radio core, while the northwestern component is characterized by $\alpha^{36}_{30}$=1.39$\pm$0.63. Such 
a steep spectrum suggests the presence of a jet as the source of radio emission, which is consistent with the above-mentioned gas kinematics of NGC\,591 obtained 
at other wavelengths.

\textit{NGC\,1194}  is an S0-a Seyfert\,1.9 galaxy. Of the galaxies with a determined nuclear mass in our sample, this galaxy has the most massive nuclear core, $M_{\rm SMBH}$ = 
(6.5$\pm$0.3) $\times$ $10^{7} M_{\odot}$ \citep{kuo2011}. The position angle of the maser disk is 157\degr \ east of north and the inclination is $\sim$ 85\degr \citep{kuo2011}. 
Compact radio emission is detected in our 33\,GHz map.

\textit{NGC\,2273}  is an SBa Seyfert\,2 galaxy. The H$_2$O maser in NGC\,2273 is the least luminous of our sample with L$_{\rm H_2O}$\,< 10\,L$_\odot$. \citet{petitpas2002} 
reported a nuclear stellar bar with a P.A. of 45\degr \ that
is located approximately perpendicular to the large-scale galactic bar with a P.A. of 115\degr.
However, \citet{erwin2003} found evidence for a circumnuclear disk. \citet{greene2013} suggested the presence of an inner ring that manifests itself as a spike in the 
ellipticity profile of the inclined main galactic body at a galactocentric distance of $\sim$150\,pc. The dust lanes interior to the ring are most readily identified 
with spiral arms. In our 33\,GHz map we observe extended emission, with a P.A. of 82.8\degr $\pm$ 2.8\degr. The P.A. of the maser disk is 153$^{\circ}$ \citep{kuo2011},
which indicates that our radio continuum and the maser disk are misaligned by $\sim$70\degr.

\textit{NGC\,2979}  is an SABa Seyfert\,2 galaxy. Our 33\,GHz image shows compact radio continuum emission.

\textit{NGC\,3393} is an SBa Seyfert\,2 galaxy. \citet{fabbiano2011} reported X-ray (3-8 keV) observations of this system and provided evidence for the existence of 
a binary black hole system. However, deep Chandra imaging, combined with adaptive optics and radio imaging \citep{koss2015}, suggest a single heavily obscured 
radio-bright active galactic nucleus (AGN). VLA observations at 8.6\,GHz (A configuration) have shown a system that consists of four radio components; one at the center, one displaced to the southwest, 
and two very close components to the northeast \citep[see][]{schmitt2001}. In our 33\,GHz observations with approximately the same synthesized beam width, we see three 
components: one at the center, one toward the southwest, and one located to the northeast. That we see not two sources, like \citet{schmitt2001} in the latter case, 
may be due to the limited S/N of this component. Our spectral analysis shows $\alpha^{36}_{30}$=1.69$\pm$0.1.46 for the central component,
$\alpha^{36}_{30}$=2.66$\pm$3.2 for the northwestern component, and $\alpha^{36}_{30}$=0.84$\pm$0.29 for the southeastern component. The high uncertainty of the indices 
for the central and northwestern components do not allow us to draw a conclusion on the nature of the sources. However, an index of 0.84$\pm$0.29 for the 
southeastern component could be indicative
of radio emission from a jet.

\textit{NGC\,4388}  is an Sb edge-on Seyfert\,2 galaxy. NGC\,4388 is known to have a biconical narrow line region \citep[][Wilson et al.\,1993]{pogge1988, corbin1988}. 
The maser disk has a P.A. of 107$^{\circ}$ \citep{kuo2011}. The kiloparsec-scale disk has a P.A. of 90$^{\circ}$ and the more nuclear stellar disk has a P.A. of 75$^{\circ}$
\citep{greene2014}. \citet{damas2016} reported a southern jet with a scale of 1 kpc. Extended radio emission is observed in our 33\,GHz map with a P.A. of 
24$^{\circ}$$\pm$54$^{\circ}$ (Table\,\ref{table:indices_inclination}). The large error in the latter value is mainly caused by the bending of the jet, visible in 
Fig.\,\ref{fig:images1}. Thus maser disk and the extended radio emission seen in our data are oriented almost perpendicular to each other.

\textit{NGC\,5495}  is an SABc Seyfert\,2 galaxy. H$_2$O maser emission was detected by \citet{kondratko2006b} with an orbital velocity of $\sim$\,400 km\,s$^{-1}$. 
\citet{gao2017} reported a P.A. of 176$^{\circ}$$\pm$5$^{\circ}$ for the maser disk. Compact 33\,GHz emission is observed in our map.

\textit{NGC\,5728}   is an SBa Seyfert\,2 galaxy. Similar to NGC\,4388, this galaxy is known to have a biconical narrow line region  
\citep[][Wilson et al.\,1993]{pogge1988, corbin1988}. Our 33\,GHz map shows compact radio continuum at the central NED position.

\textit{UGC\,3193} is an SBb galaxy. Extended 33\,GHz emission, mainly oriented along an axis from north to south, is observed together with a compact core at 
the center of our map.

\textit{UGC\,3789}  is an SABa Seyfert\,2 galaxy. There is optical
evidence for a circumnuclear disk. An inner ring is visible as a spike in the 
ellipticity profile, and dust lanes are located interior to the ring, which are most readily identified with spiral arms \citep{greene2013}. The well-studied maser disk traces rotation 
speeds of up to $\sim$\,750\,km\,s$^{-1}$ \citep{braatz2008, braatz2010, reid2013}. The MCP has measured an angular-diameter distance of 49.6 $\pm$ 5.1\,Mpc for this 
galaxy \citep{reid2013}, which is compatible with the value given in Table\,\ref{table:calibrators}. The P.A. of the maser disk is 41\degr \ east of north \citep{kuo2011}.
The inner and outer radii of the disk are 0.08\,pc and 0.30\,pc, respectively. UGC\,3789 has a black hole mass of (1.4$\pm$0.05)$\times$\,$10^7$\,$M_{\odot}$ 
\citep{kuo2011}. In addition, \citet{castangia2013} classified UGC\,3789 as a Compton-thick galaxy ($N_{\rm H}$\,>\,$10^{24}$\,$cm^{-2}$). Radio emission is observed at the 
center of our 33\,GHz map. While its nominal P.A. is 117$^{\circ}$$\pm$22$^{\circ}$ (see Table \,\ref{table:indices_inclination}), the 33\,GHz radio continuum source is
too compact with respect to our resolution to discuss any potential alignments.  
%================================

% %__________________________________________________________________
\section{Analysis and discussion}\label{sec:discussion}
After reviewing the individual 33\,GHz maps, here we discuss our results in more detail. First we analyze the encountered morphologies and spectral indices obtained for 
the detected sources, and then we discuss possible correlations between various physical quantities by addressing the galaxies and in particular their nuclear regions. 
In order to do so, we applied linear fits, accounting for errros along both axes, and a Spearman rank correlation test. Slopes and intercepts of the fits, 
correlation coefficients, and $P$-values, denoting the likelihood that the parameters are unrelated, are shown in a corner of most plots. One purpose of this analysis, 
in addition to identifying correlations, is to increase the chance of disk-maser detections in future surveys.

\subsection{33\,GHz morphologies - where are the jets?}
As explained in Sect.\,\ref{sec:result_sample}, the bulk of our 33\,GHz maps contain compact cores and clearly asks for 
higher resolution studies at lower frequencies, where higher fluxes may be expected. Interestingly, two of the seven sources with $D$ $>$ 100\,Mpc (J1658+3923 and Mrk\,1029) 
show extended emission. Perhaps there is even a third, Mrk\,78. Apparently, the linear extent of the central emission can vary strongly, from $\la$50\,pc in some 
compact nearby sources (e.g., NGC\,2979) to $\sim$600\,pc in Mrk\,1029. J0350-0127, J1658+3923, and Mrk\,78 might show elongated structure, but the S/Ns 
do not permit a clear interpretation. While some of our most distant sources show structure, the most complex sources, NGC\,591, NGC\,3393, NGC\,4388, and UGC\,3193, are with 
$D$ $\la$ 65\,Mpc all relatively nearby. Each of them contains a compact core, but additional compact sources such as in NGC\,591 and NGC\,3393 indicate the presence of jets. 
This also holds for the other two objects, NGC\,4388 and UGC\,3193, where extended emission in the northeast and/or south has been observed. Nevertheless, most sources only 
exhibit a core component. 
This implies that in most of our objects jets are either shorter than a few 100\,pc, that they are still below our sensitivity limit due to a steep radio spectrum, or that they 
do not exist at all.  

\subsection{Spectral indices}\label{sec:sp_index}
Spectral indices can tell us more about the nature of observed radio emission in our sample. For instance, a steep or flat spectral index can be indicative of a jet or 
radio core, respectively. However, measuring spectral indices for our sources has its own challenges. Using archival data such as NVSS or FIRST,
will result in misestimation of spectral indices because of the different beam sizes, which cover different areas (even when
we compare the FIRST beam size with our 
33\,GHz beam size, the ratio of the covered areas is $\sim$450). Additional uncertainty in our spectral index estimation is introduced by 
the possible variability of the nuclear radio emission. Radio variability in active galaxies is a common phenomenon, especially in compact 
sources \citep{mundell2009}, which is the case for most of our targets. The variability is observed on a timescale of a few months to several years, and there are cases where 
the nuclear flux density has changed by 38\% over seven years \citep{mundell2009}.
When we consider that our observations and other observations such as NVSS or FIRST were performed in different years (NVSS observations were conducted 16 years before our 
observations and FIRST observations from 2 to 19 years before), Seyfert variability is another source of 
uncertainty in our spectral index calculations.

To exclude such errors, we can calculate the spectral indices over the wide bandwidth of the observations.
For our sample, we derived the spectral indices assuming a power-law dependence for the continuum flux density given by $S$\,$\propto$\,$\nu^{-\alpha}$,
using the 8\,GHz (29-37\,GHz) bandwidth of our observations. By splitting the data into $4\times2$\,GHz parts and
imaging each sub-band separately (with Briggs weighting and a robust parameter of 0),
we obtain the spectral indices between 30\,GHz and 36\,GHz ($\alpha^{36}_{30}$) using fluxes at the central frequencies of the sub-bands. In some cases the 
S/N was too low and the source was not detected after splitting the data. It should also be noted that the results obtained by this method have significantly 
higher errors. We also obtained $\alpha^{36}_{30}$ from spectral index maps created by CASA when using the task CLEAN with the number of Taylor coefficients set as 2. For
the sources where 1.4\,GHz NVSS flux densities \citep{condon1998} or 1.4\,GHz FIRST flux densities \citep{becker1995} were available 
(Table\,\ref{table:properties} and \ref{table:indices_inclination}), the indices were also obtained between 1.4\,GHz and 33\,GHz.

$\alpha^{36}_{30}$ obtained from CASA, based on 21 sources, 
has an unweighted  mean value of 0.68$\pm$0.28, while for the splitting method we obtained $\alpha^{36}_{30}$ for 16 sources with an unweighted mean 
value of 0.79$\pm$0.32.
For 17 sources the indices were obtained from NVSS and 33\,GHz (from now on called $\alpha^{33}_{1.4,\,NVSS}$),
with a mean (unweighted) value of 0.97$\pm$0.06.
For the seven galaxies with measured indices between FIRST and 33\,GHz ($\alpha^{33}_{1.4,\,FIRST}$), the unweighted mean value is 0.73$\pm$0.16
(errors are the standard error of the mean).
See Table\,\ref{table:indices_inclination} for spectral index measurement of the individual sources.

\subsection{Radio continuum versus H$_2$O disk-maser host galaxy properties}
\subsubsection{Correlations with optical or near-infrared morphology}
\begin{figure}[h]
\begin{center}
 \includegraphics[width=0.46\textwidth]{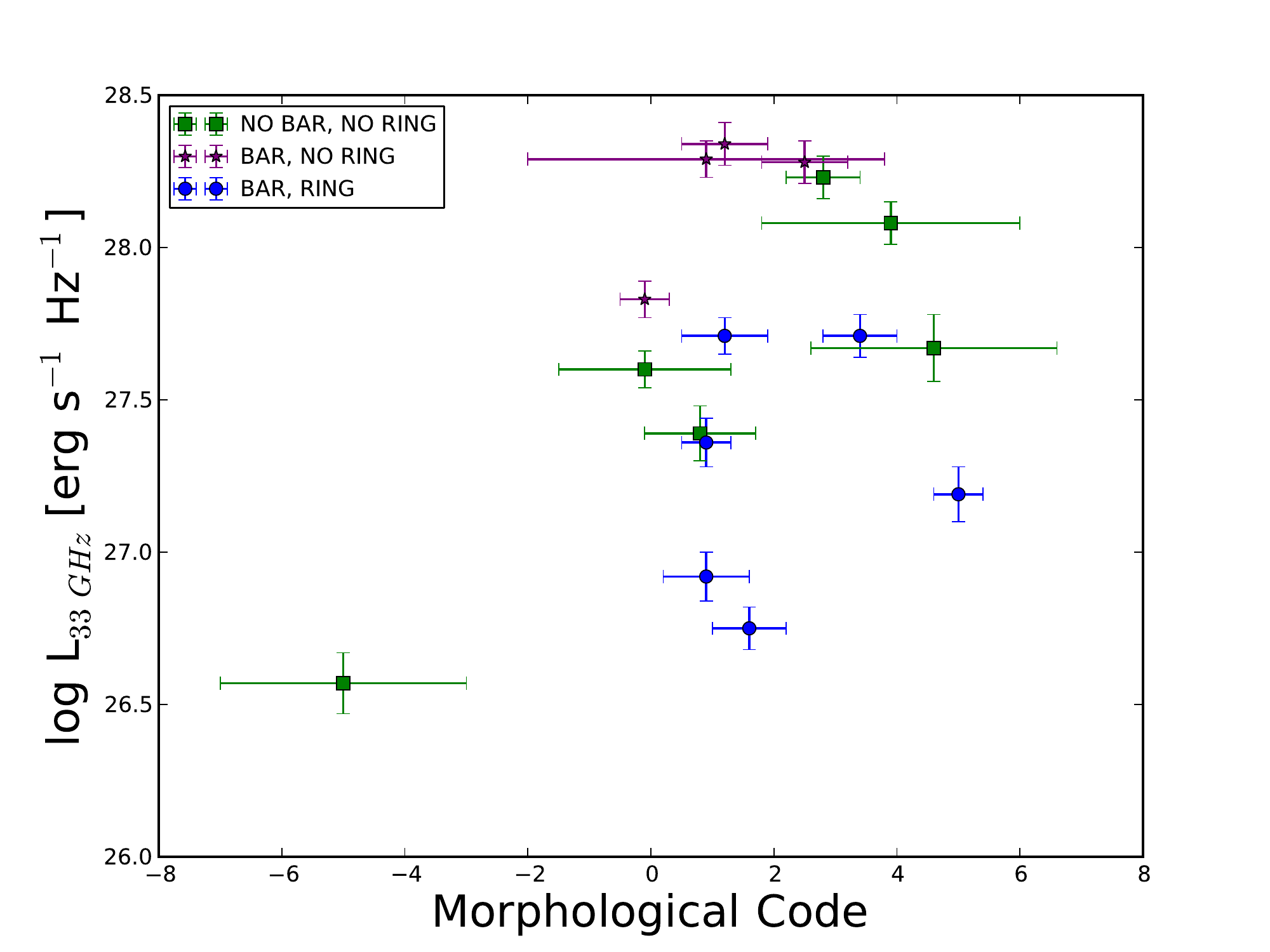}
 \includegraphics[width=0.46\textwidth]{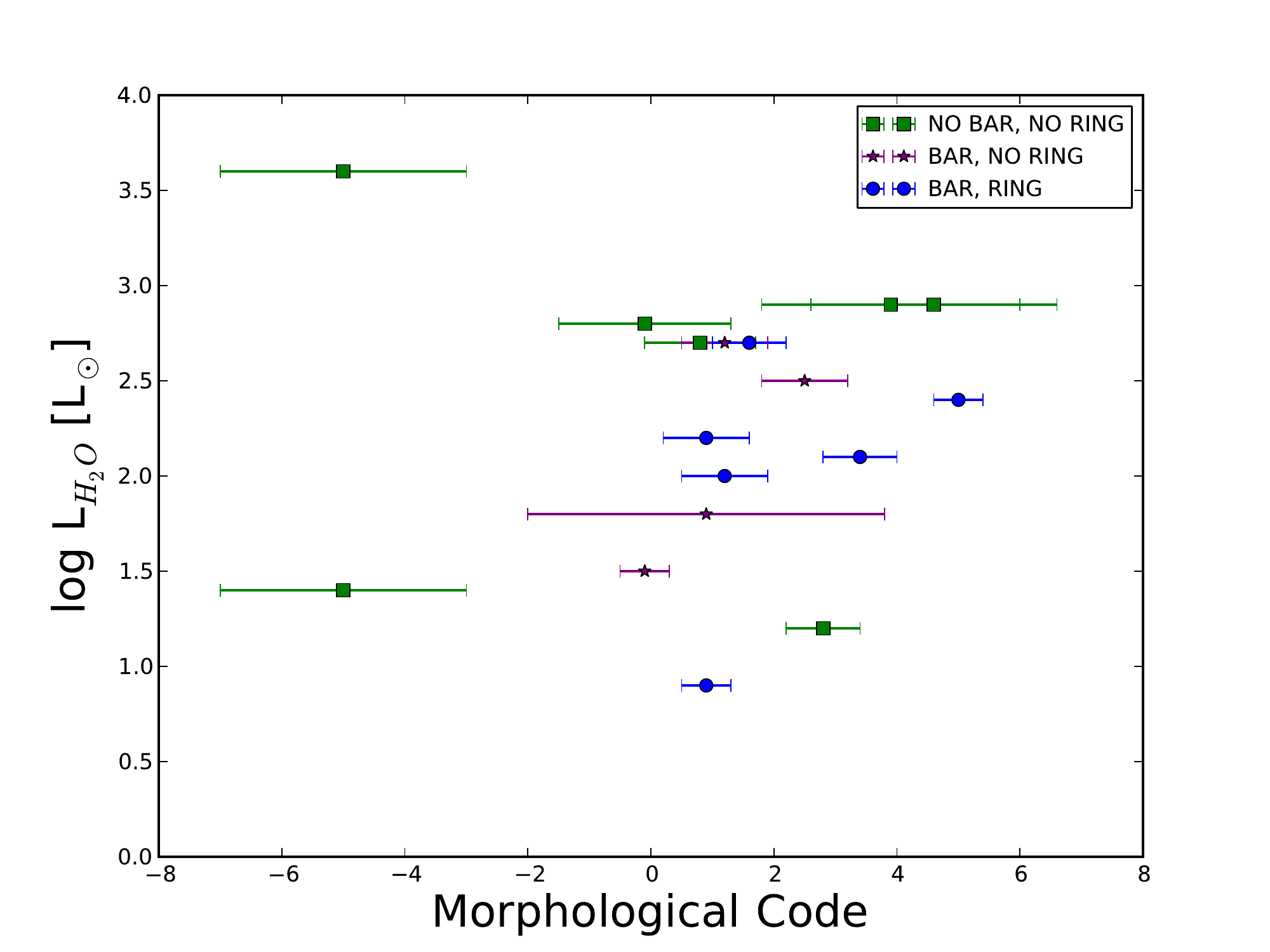}
 \caption{Top: 33\,GHz radio continuum luminosity versus morphological type. 
 Bottom: H$_2$O-maser luminosity versus morphological type.}\label{fig:morph}
\end{center}
 \end{figure}

As mentioned before, H$_2$O disk-masers are observed in Seyfert\,2s and LINERs (Table\,\ref{table:properties}), that is, in spiral galaxies. We adopted the morphological type 
in de Vaucouleur's scale from the HyperLeda\footnote{HyperLeda is a database for physics of galaxies. For more information see: \url{http://leda.univ-lyon1.fr}} database 
\citep{makarov2014}. In this scale, negative numbers represent ellipticals and positive numbers spirals, with higher positive numbers representing later-type spirals. 
The morphology of the galaxies in our sample lies between S0 and Sc, with the exception of two galaxies that are classified as ellipticals. However, accounting for uncertainties, 
these could also be S0 galaxies. Fifteen (62\%) of our galaxies are barred, and 9 (37\%) contain a ring. Figure\,\ref{fig:morph} presents 33\,GHz continuum and H$_2$O 
maser luminosities
versus the morphological type. Galaxies with both a bar and a ring in the central parts apparently have lower radio continuum luminosities than those with only a bar. 
It is well known that in the presence of a bar, gas will be driven inward by angular momentum transport. This transport drives turbulence within the gas that temporarily 
keeps it strongly gravitationally stable and prevents the onset of rapid star formation. However, at some point, the rotation curve must transition from an approximately flat to 
an approximately solid body, and the resulting reduction in shear reduces the transport rates and causes gas to build up, eventually producing a gravitationally unstable region, 
a ring, that is subject to rapid and violent star formation \citep[e.g.,][]{piner1995, krumholtz2015}. The region inside the ring might be characterized by a lower rate of inflow, 
leading to less star formation and synchrotron emission of the ambient gas as well as to a less active nucleus with a lower radio luminosity. We note, however, that our 
finding is based on small number statistics. Moreover, following the ``stuff inside stuff'' scenario \citep{hopkins2010}, there may be yet another so far undetected bar 
inside the ring channeling material closer to the center. Furthermore, not in all our sample galaxies searches for a bar or ring have been performed. Additional morphological studies \citep[see, 
e.g.,][]{greene2010, riffel2011, greene2013} would thus be highly desirable.

\subsubsection{Position angle and inclination}
\begin{figure}[h]
\begin{center}
 \includegraphics[width=0.46\textwidth]{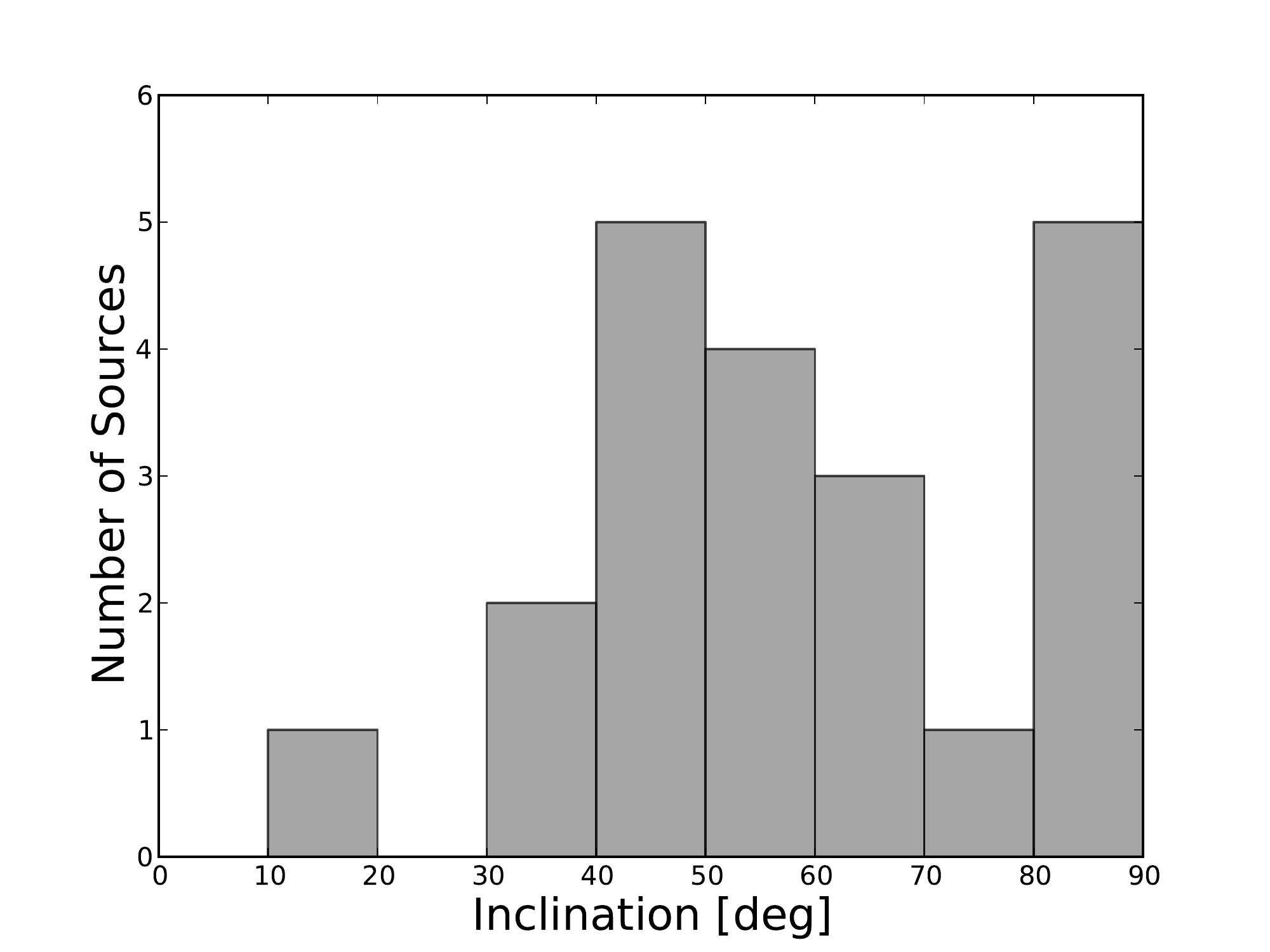}
 \caption{Distribution of the inclinations of the galactic large-scale disks.}\label{fig:inc_dist}
\end{center}
\end{figure}

For four galaxies in our sample maser disk inclinations have been reported, with values between 84$^{\circ}$ and 90$^{\circ}$ \citep{kuo2011}. This inclination 
range for the maser disk should hold for the entire sample.
In spiral galaxies, the inclination of the central region is independent of the inclination of the large-scale disk \citep{ulvestad1984, nagar1999, kinney2000}. 
Apparently, this is also the case for circumnuclear disks and bars at galactocentric distances of several 100\,pc \citep{greene2014}. Although the inclination of the 
H$_2$O maser disks of our sample is almost 90$^{\circ}$, the large-scale inclination of their host galaxies can reach any value. In Table\,\ref{table:indices_inclination} we list 
P.A. and inclination of the galaxies taken from HyperLeda. The distribution of the inclination is displayed in Fig.\,\ref{fig:inc_dist}.

\begin{figure}[h]
\begin{center}
 \includegraphics[width=0.46\textwidth]{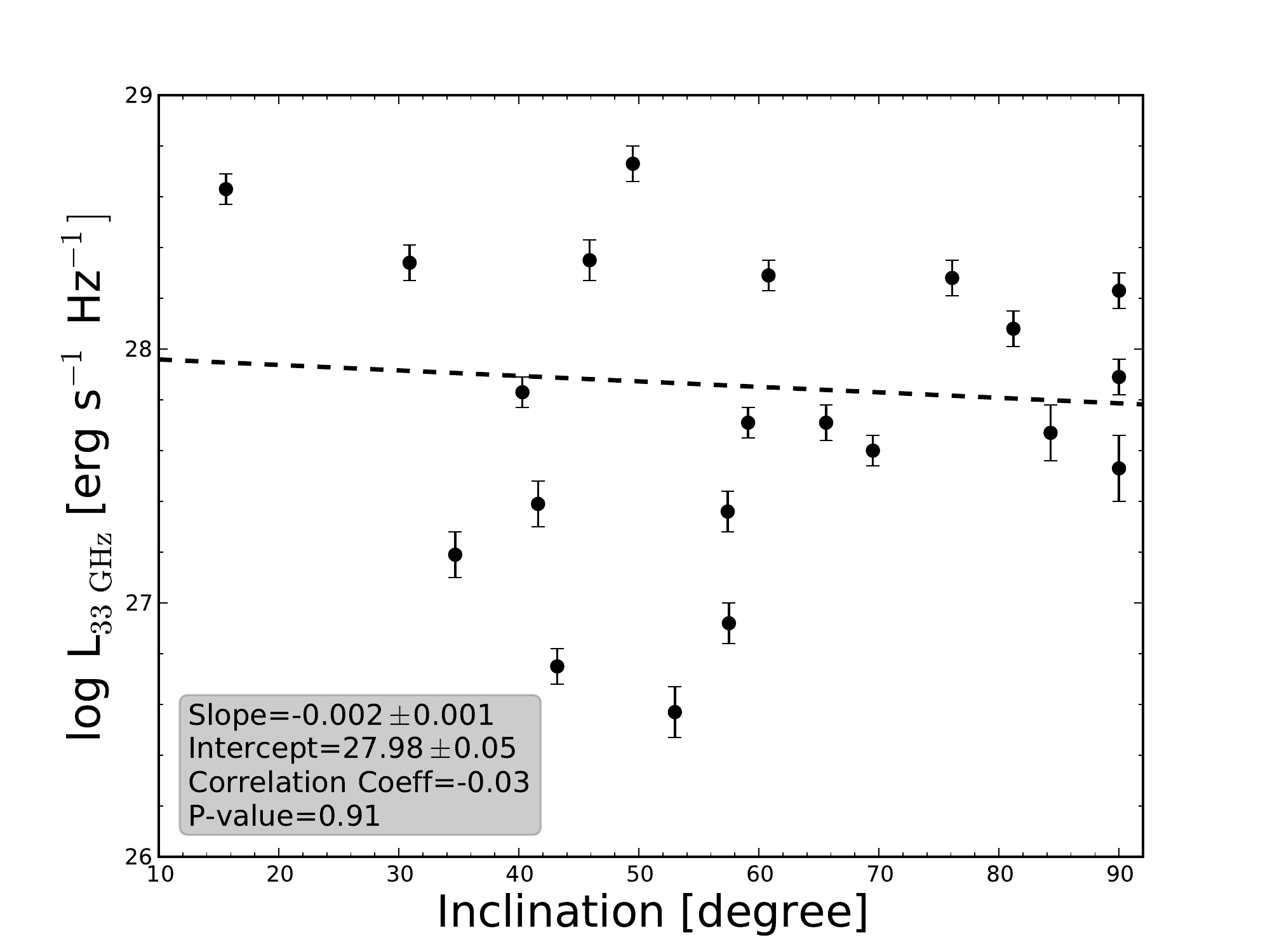}
 \includegraphics[width=0.46\textwidth]{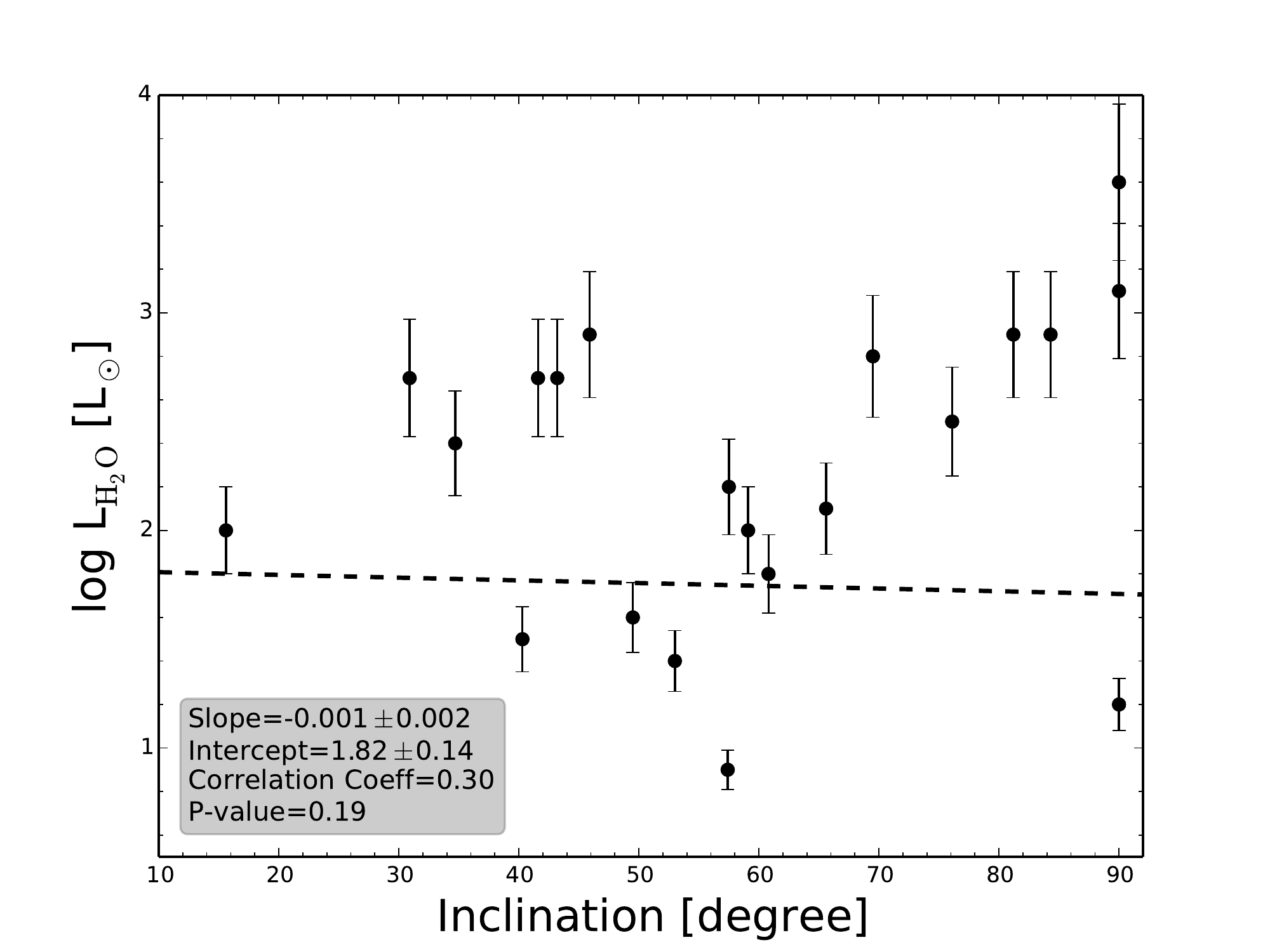}
 \caption{Top: 33\,GHz continuum luminosity versus large-scale inclination. Bottom: H$_2$O-maser 
 luminosity versus large-scale inclination of the galaxies. The dashed lines show linear fits to the data 
 points. The slope and intercept of the fits, the correlation coefficient, and the $P$-value of the Spearman test, indicating the likelihood that the two quantities are 
 unrelated, are indicated in a corner of each panel.}\label{fig:inc_lum}
\end{center}
\end{figure}

We also checked for possible relations 
between the H$_2$O maser luminosity and large-scale inclination, and for 33\,GHz continuum luminosity and large-scale inclination. As shown in Fig.\,\ref{fig:inc_lum},
no correlation is found.
Of the four galaxies of our sample in which jet-like features can be identified (NGC~591, NGC~3393, NGC~4388, and UGC~3193; see Fig.\,\ref{fig:images1}), one source has also 
been observed with high resolution by the MCP in the 22\,GHz H$_2$O line. This galaxy is NGC~4388, where \citet{kuo2011} found an edge-on disk with a P.A. (blueshifted side) 
of 107$^{\circ}$. As stated in Sect.\,\ref{sec:result_sample}, our supposed jet with a P.A. = 24$^{\circ}$ $\pm$ 54$^{\circ}$ appears to be perpendicular to the maser 
disk. Inspecting the image shown in Fig.\,\ref{fig:images1}, the formal error in this position angle, accounting for all the emission seen, may well overestimate the actual uncertainty.
The other interesting case is NGC~2273 (see again Sect.\,\ref{sec:result_sample} and Fig.\,\ref{fig:images1}), where an angle not far from 90$^{\circ}$ is also likely.  
\subsubsection{Stellar velocity dispersion and circular velocity}\label{sec:vel}
The discovery of a correlation between SMBH mass and the velocity dispersion of the bulge component of bright elliptical galaxies, together with 
similar correlations with bulge luminosity and mass, led to the widespread belief that SMBHs and bulges coevolve by regulating each other's growth 
\citep[e.g.,][]{kormendy2013}. However, by studying less luminous spiral galaxies and making use of the quite accurately determined SMBH masses 
of H$_2$O megamaser galaxies, it has become clear more recently
that the relationship between $M_{\rm SMBH}$ and velocity dispersion has a larger scatter and more structure 
than was originally thought \citep{greene2010, sun2013, greene2016}.
In particular, it seems (although still based on small number statistics) that the 
megamaser galaxies may show a different scaling relation than galaxies of similar size and Hubble type studied at other wavelengths, possibly indicating 
an observational bias in either the masing or the non-masing galaxy samples. It is thus essential to look deeper into this apparent scatter to shed more light 
onto the underlying physics that probably involves inflow, black hole growth, and subsequent feedback.
\begin{figure}
\begin{center}
 \includegraphics[width=0.46\textwidth]{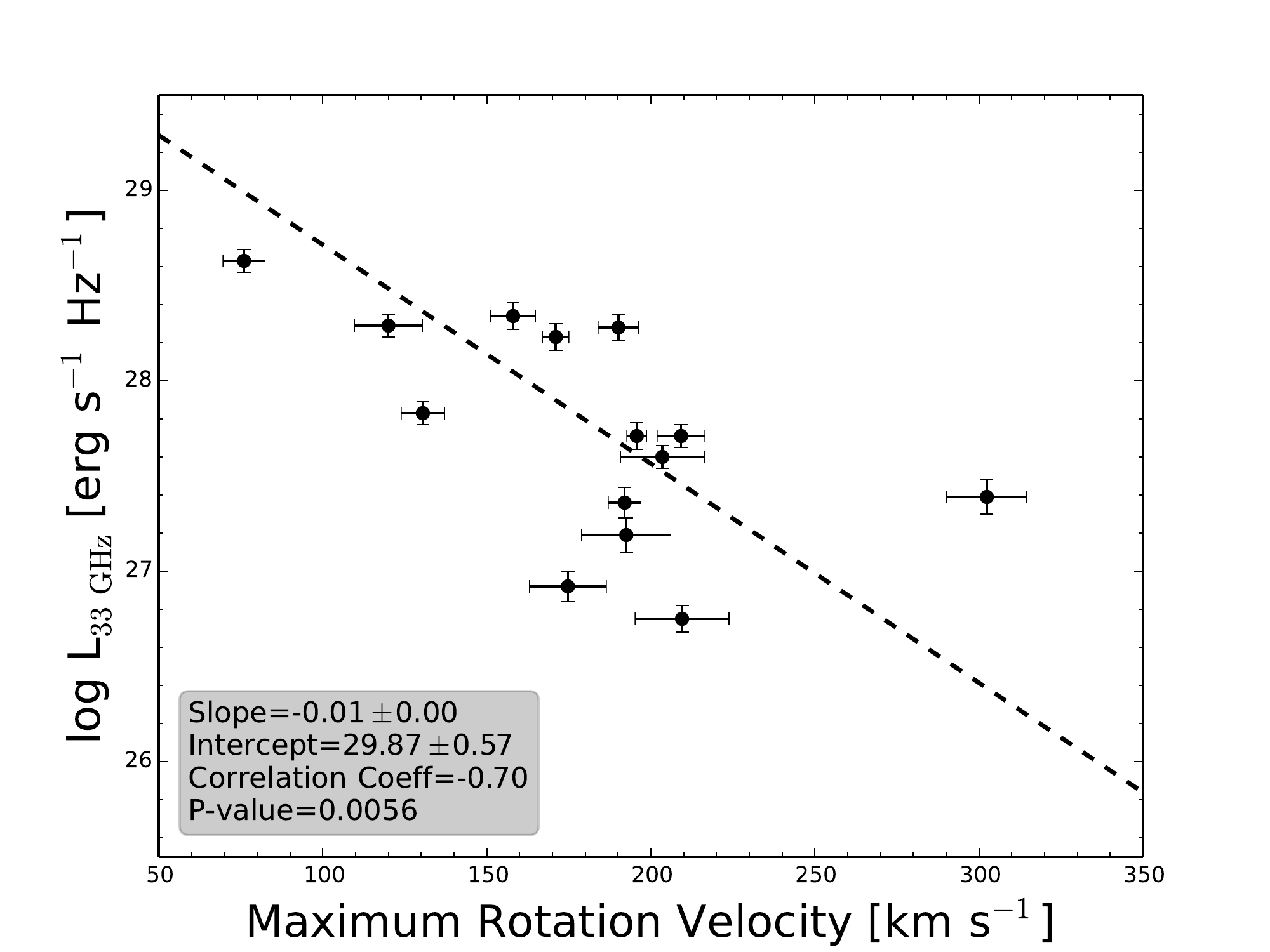}
 \includegraphics[width=0.46\textwidth]{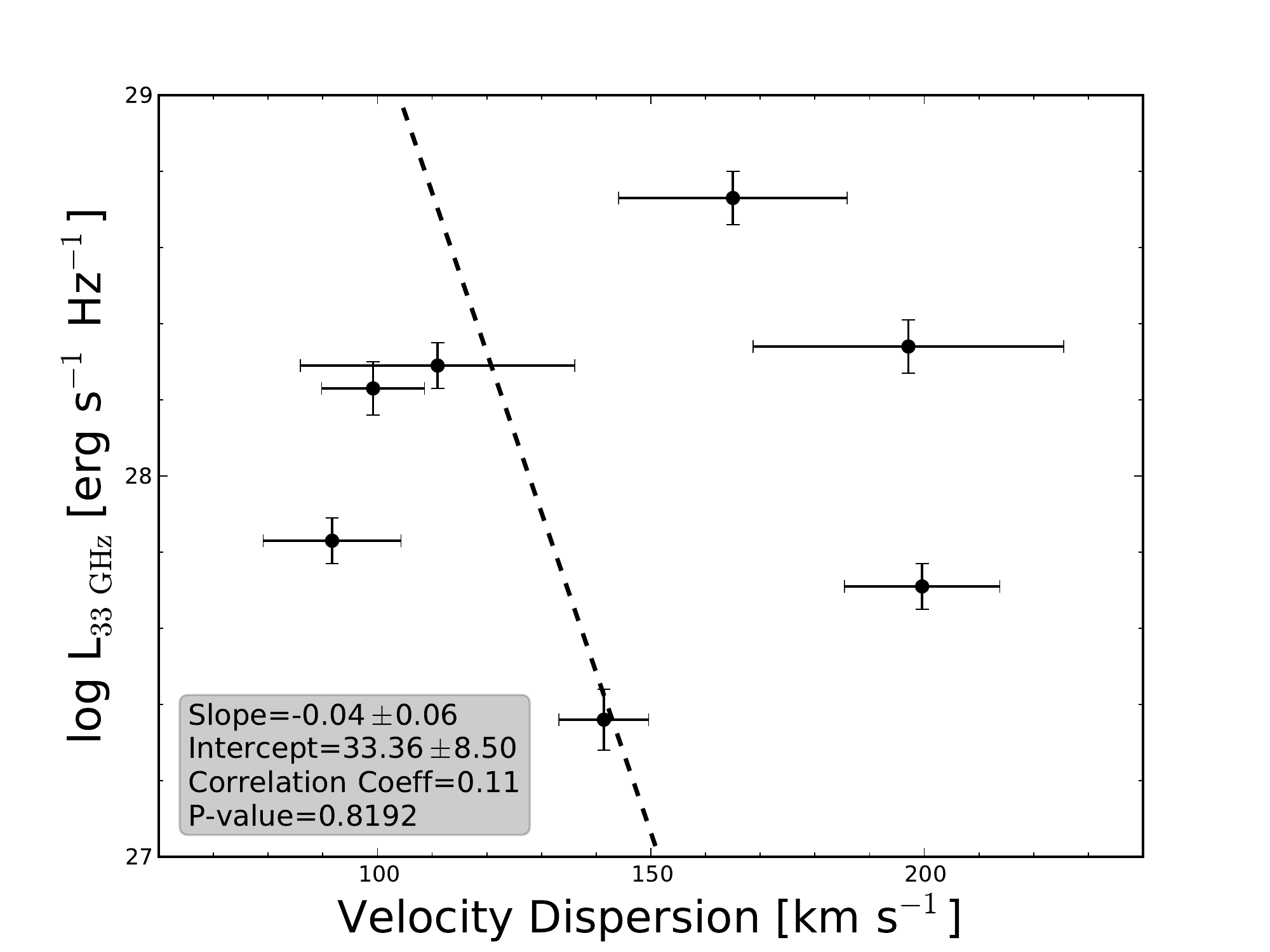}
 \caption{Top: Radio continuum versus circular velocity (from HyperLeda). Bottom: Radio 
 continuum versus central stellar velocity dispersion. ''Central'' refers to standardized aperture r$_{\rm norm}$ of 0.595 h$^{-1}$kpc,
 which is equivalent to an angular diameter of 3.4 arcsec at the distance of the Coma cluster. The lines show linear fits to the data points accounting for errors along both axes. The slope and intercept of the fit, the correlation 
 coefficient, and the $P$-value of the Spearman test (see also Fig.\,\ref{fig:inc_lum}) are shown in a corner of each plot.}\label{fig:vel_lum33}
\end{center}
\end{figure}

We searched for possible relations between the 33\,GHz luminosity, the stellar velocity dispersion $\sigma_*$ , and the maximum rotation velocity of the gas, 
$V_{\rm rot}$. As shown in the bottom panel of Fig.\,\ref{fig:vel_lum33},
$\sigma_*$ and 33\,GHz continuum luminosity are not correlated. Surprisingly, however, 
we find an anticorrelation between the 33\,GHz continuum luminosity and circular velocity of the galaxies, with a correlation coefficient of --0.70 and 
an associated $P$-value of 0.0056. 
For a better understanding of the implications, we note that the HyperLeda rotation velocities are obtained from H{\sc i} 21\,cm line data and corrected for the 
inclination of the main body of the respective galaxy. In spiral galaxies, H{\sc i} is most prominent outside the cerntral region, so that here we
compare a central parameter with 
a parameter obtained from emission originating outside of the central region. How can these be related? And why do we find an anticorrelation?
The rotation velocity of spirals tends to reach near-maximum values at $r$ $\la$ 2\,kpc \citep[e.g.,][]{hlavacek2011, chemin2015}, while our images 
typically cover 1\,kpc with most of the detected emission arising from an even smaller region.

Following \citet{safranov1960}, \citet{toomre1964} and \citet{quirk1972}, for a gaseous differentially rotating disk, the following criterion must hold:
$$
  Q \,\,\,   = \,\,\, \frac{\sigma_{\rm gas}\,\,\kappa}{\pi\,\,{\rm G}\,\,\Sigma_{\rm gas}} \,\,\, > \,\,\, 1,
$$  
with $\sigma_{\rm gas}$ denoting the velocity dispersion, $\kappa$ the epicyclic frequency, and G the gravitational constant. $\Sigma_{\rm gas}$ stands for surface density. 
Thus, a higher rotational velocity keeps the disk more stable and less prone to fragmentation and subsequent star formation, reducing radio continuum
emission that is due to star formation. More regular motion may also reduce accretion by the AGN and the luminosity of the radio core. Nevertheless,
this may be a far-fetched explanation. Although the correlation is statistically convincing, we find the result surprising, and confirmation is needed by measurements of more 
such galaxies by employing both lower and higher angular resolution to fully cover the relevant linear scales.

\subsubsection{Infrared versus radio luminosity}
\begin{figure}
\begin{center}
 \includegraphics[scale=0.45]{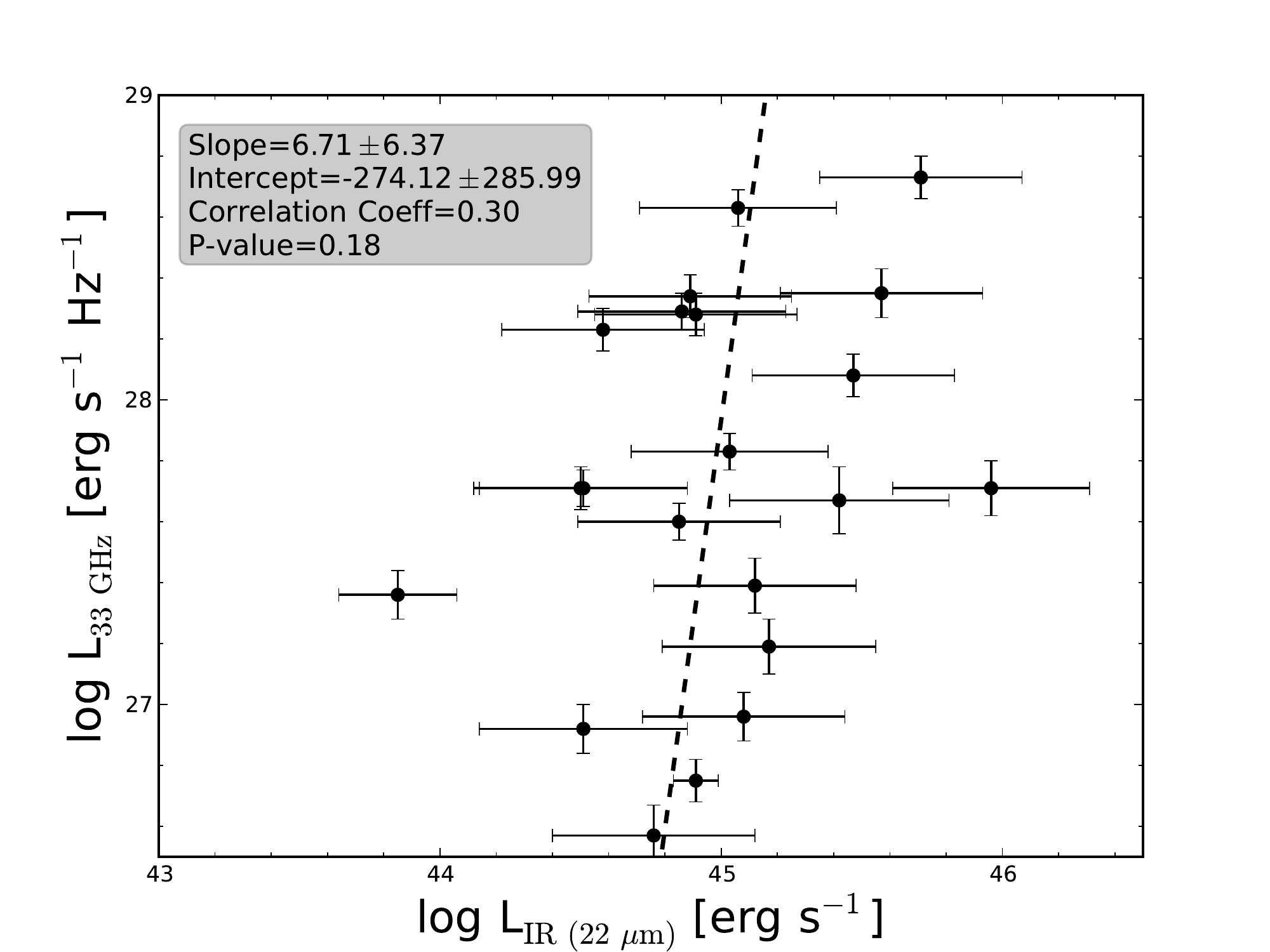}
 \caption{Radio luminosity versus infrared luminosity (WISE 22\,$\mu$m).
 The line shows a linear fit to the data points accounting for errors along both axes. For the inset in the upper left corner, see the caption to Fig.\,\ref{fig:inc_lum}.} 
 \label{fig:ir-radio}
 \end{center}
\end{figure}

As pointed out by \citet{diamond-stanic2009}, the radio luminosity can be an indicator of the AGN isotropic luminosity. On the other hand, in Seyfert\,2 galaxies,
the UV/optical
photons of the AGN are absorbed by the circumnuclear dusty torus and are re-emitted in the mid-IR. Therefore the mid-IR photometry can  also be used as an 
indicator of the AGN activity \citep[e.g.,][]{stern2012,fiore2008, georgantopoulos2008}. 
Hence we investigate the radio to mid-infrared correlation of our disk-maser sample, since both bands can be indicators of the total AGN power.
As shown in Fig.\,\ref{fig:ir-radio}, a weak correlation with correlation coefficient of 0.30 is observed.

\subsection{Radio continuum, X-ray, and IR luminosities versus maser-disk properties}
\subsubsection{H$_2$O maser luminosity}
\begin{figure*}[h]
\begin{center}
\includegraphics[width=0.46\textwidth]{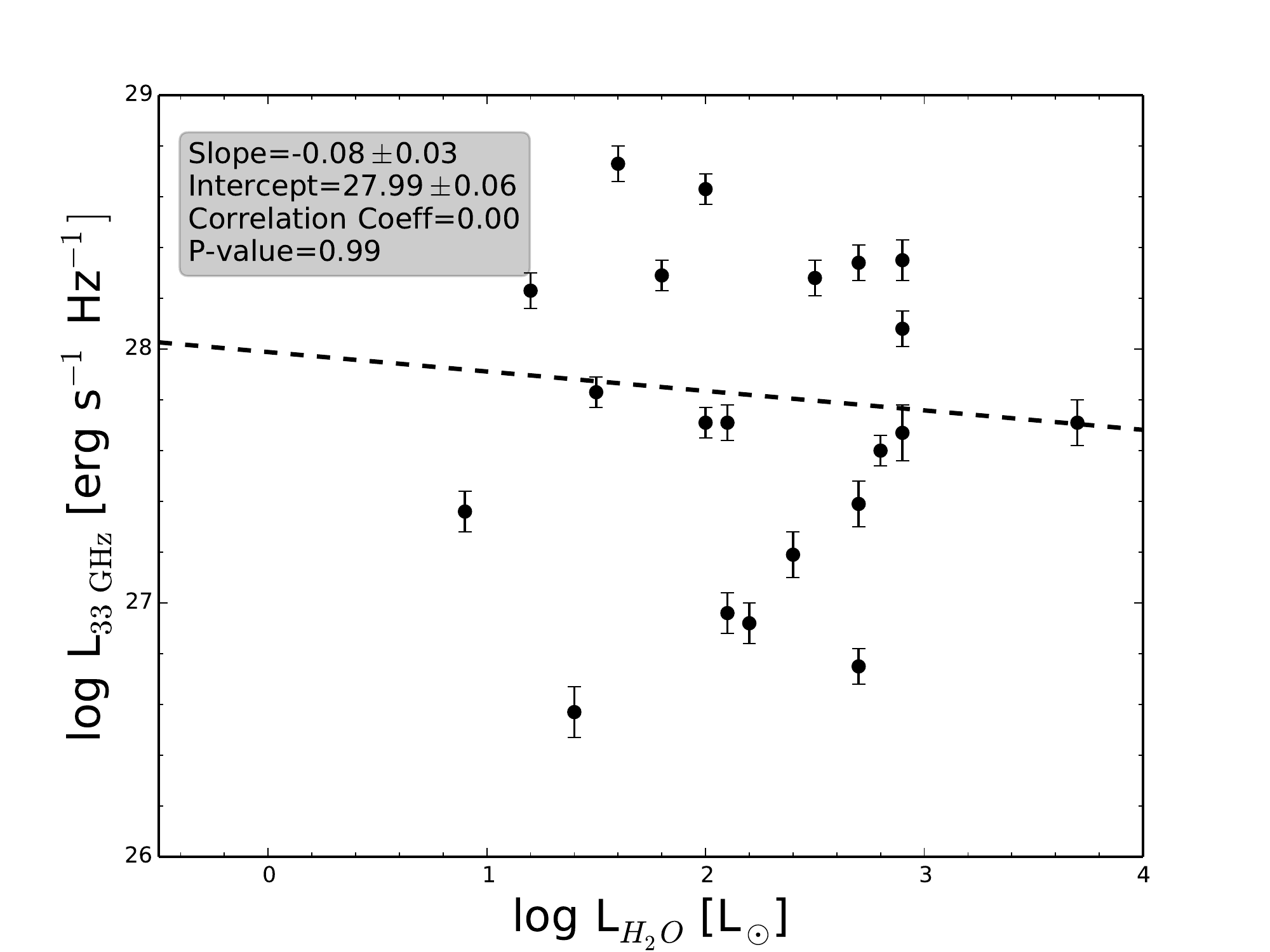}\includegraphics[width=0.46\textwidth]{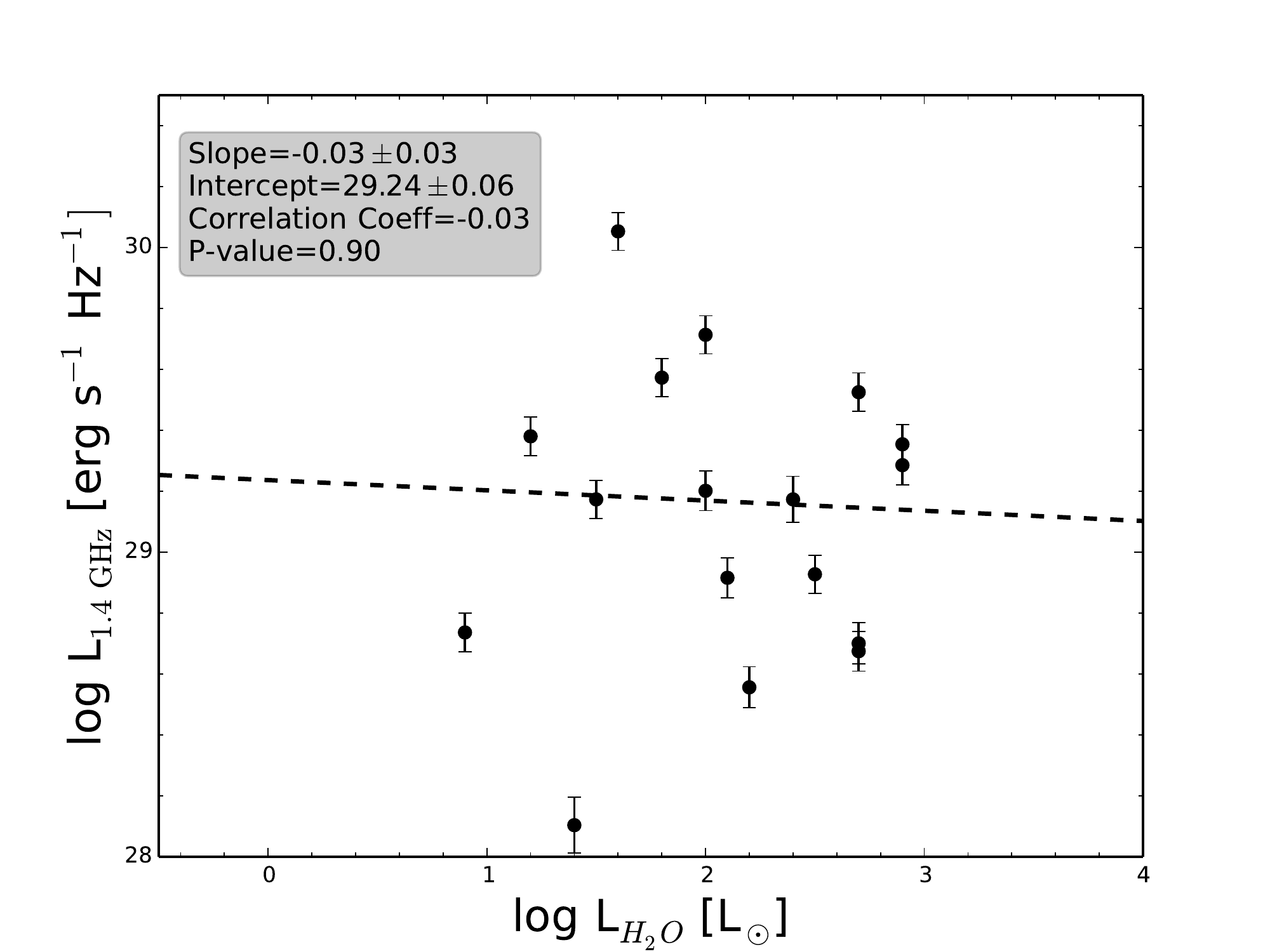}
\includegraphics[width=0.46\textwidth]{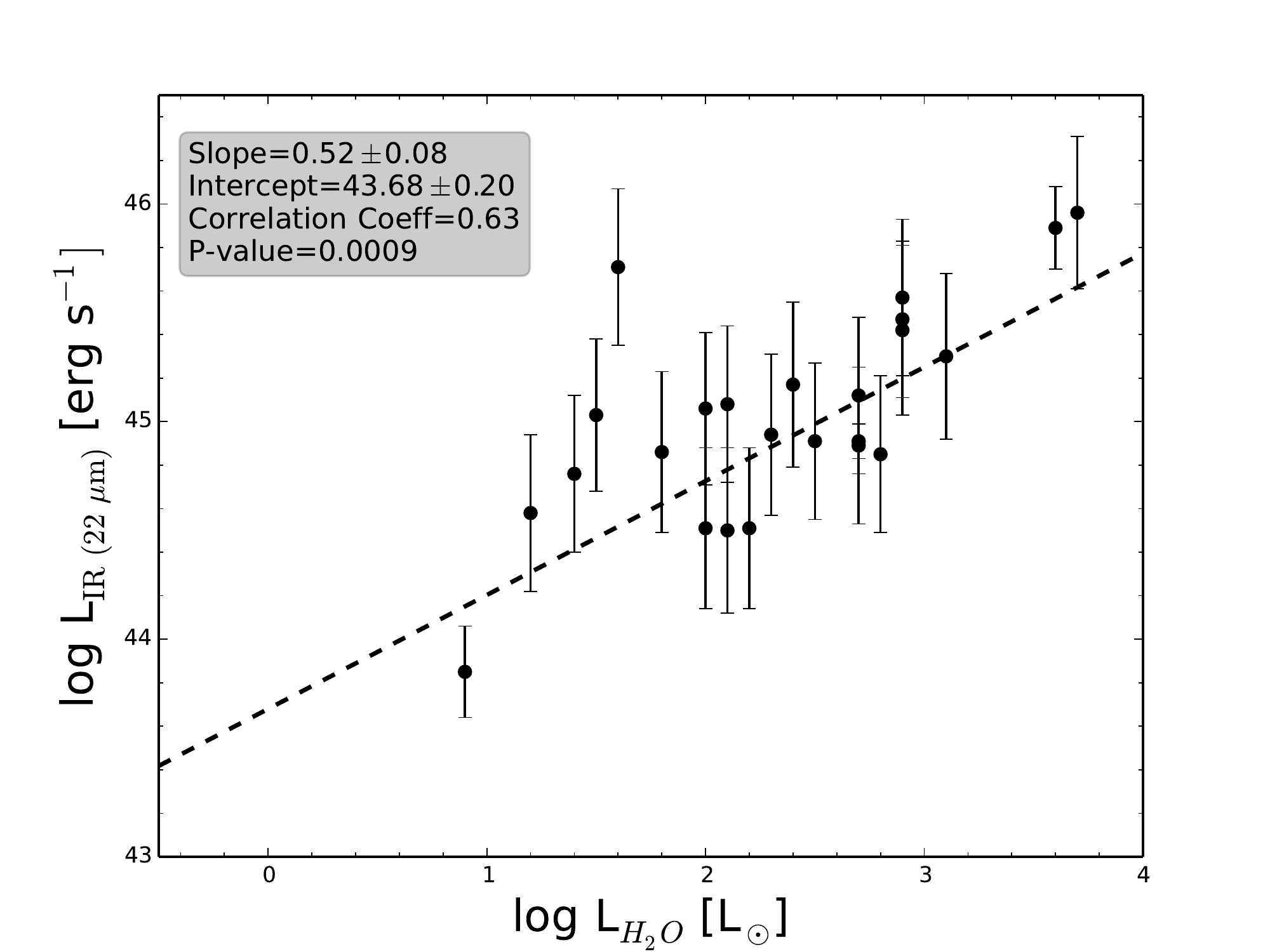}\includegraphics[width=0.46\textwidth]{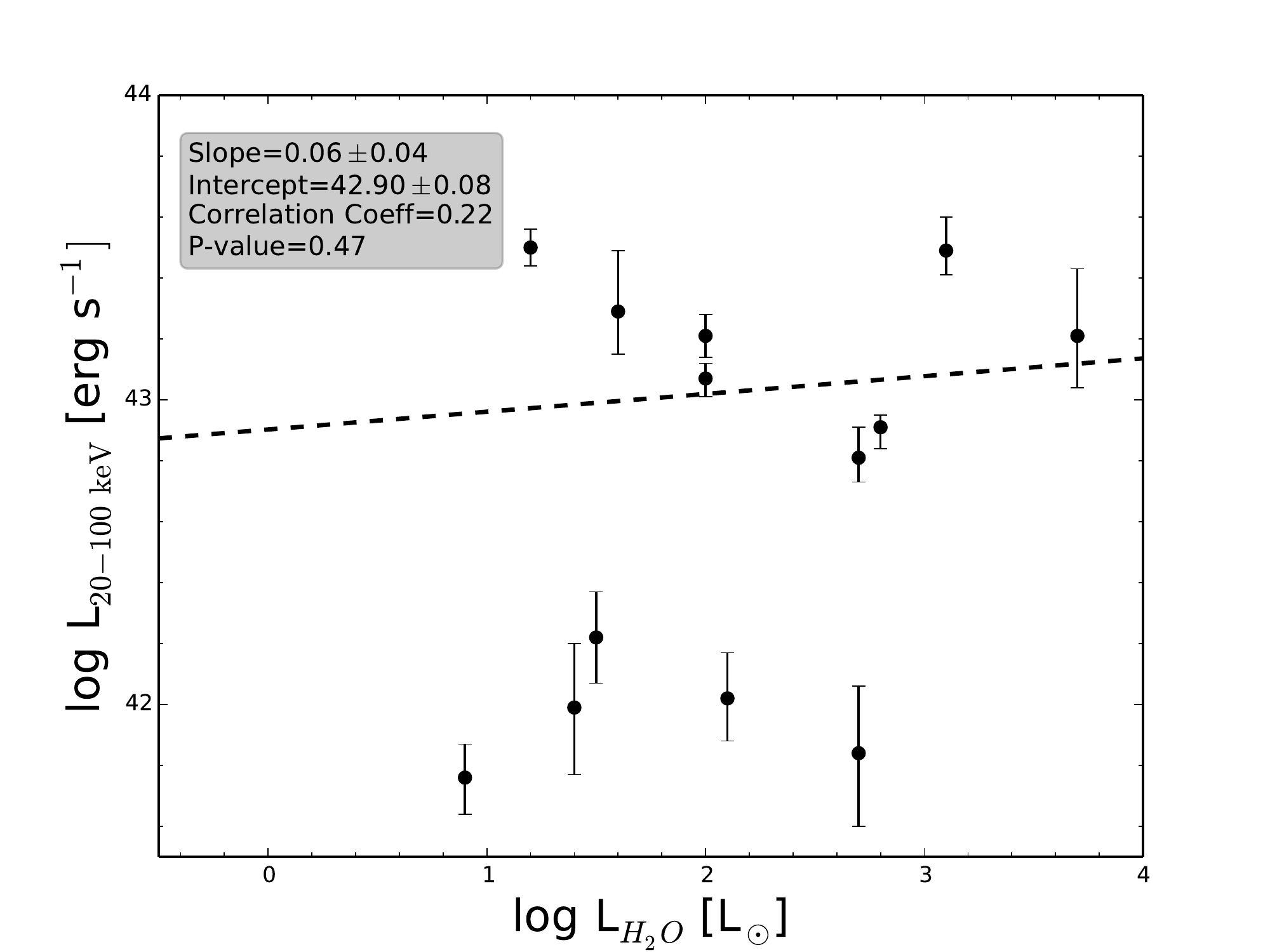}
\caption[]{\textbf{Top left:} 33\,GHz continuum luminosity versus H$_2$O maser luminosity.
\textbf{Top right:} 1.4\,GHz continuum luminosity versus H$_2$O maser luminosity.
\textbf{Bottom left:} IR luminosity versus H$_2$O maser luminosity.
\textbf{Bottom right:} Hard X-ray luminosity versus H$_2$O maser luminosity. The H$_2$O maser luminosity (isotropic emision assumed)
is on a logarithmic scale, in units of L$_\odot$. The lines show linear fits to the data points. The slope and intercept of the fit, as well as the correlation 
coefficient and the $P$-value of the Spearman test are shown in the upper left corner of each plot.}\label{fig:lum_lum}
\end{center}
\end{figure*}

Since the AGN luminosity is most likely the power source for H$_2$O megamasers \citep{lo2005}, we expect correlations between AGN luminosity and H$_2$O maser luminosity. 
For example, \citet{henkel2005} reported a correlation between far-infrared (FIR) and H$_2$O maser flux densities for maser sources that are related to either star-forming regions
or AGNs.
We also observe a weak correlation between IR 
and H$_2$O maser luminosity with a correlation coefficient of 0.63.
\citet{henkel2005} proposed that for the case of maser emission associated with AGN, the observed correlation between IR and H$_2$O maser
luminosity can be related to the existence of spatially extended dust-rich bars that fuel the central engine.
Our observed weak positive correlation between the mass of the central engine 
and the H$_2$O maser luminosity (see Sect.\,\ref{sec:bhm}) may support the suggestions of \citet{henkel2005}.
However, we do not detect such a relation between H$_2$O maser and radio (33\,GHz and 
1.4\,GHz) luminosities, or between H$_2$O-maser and hard X-ray luminosities (see Fig.\,\ref{fig:lum_lum}).

\citet{vandenbosch2016} reported a weak correlation between maser and [OIII] luminosities from studying all maser sources, while this correlation
vanishes when only the disk masers are considered. \citet{kondratko2006a} also reported a possible correlation between soft X-ray (2-10\,KeV) and 
maser luminosities, where again most of the correlation is due to the non-disk masers. More investigations to determine possible correlations
between soft X-ray (2-10 KeV) and H$_2$O maser luminosity for a sample of 14 disk-maser galaxies were carried out by \citet{castangia2013}. They 
did not find any correlation either. Following \citet{vandenbosch2016}, the disk-maser galaxies, even
though they cover a range of two orders of magnitude in isotropic maser luminosity, may be quite similar with respect to galaxy and SMBH mass.
Their properties do not provide a sufficiently
broad range of parameters, given the constant surface density predicted by \citet{neufeld1994} and the similar sizes ($\sim$0.5 pc) for most of the masing disks.
The reason that we do not see correlations between H$_2$O maser and hard X-ray or radio luminosities might be that the
small differences in the disk inclination angle and/or different support through seed photons from the nuclear continuum (which
are probably 
responsible for the strong systemic maser features in NGC~4258 \citep{herrnstein1998}) may lead to significant scatter that efficiently hides any expected 
correlation.

\subsubsection{Maser-disk radius}
An empirical upper envelope for a scaling relation between $M_{\rm SMBH}$ and the outer radii of the megamaser disks, $r_{\rm max}$ = 0.3 $\times$ $M_{\rm SMBH}$/10$^7$\,M$_{\odot}$,
was reported by \citet{wardle2012} for 12 disk maser sources. We have searched for additional relations associated with disk radii (see Table\,\ref{table:indices_inclination}), 
addressing a connection to radio, IR, and X-ray luminosity. Recently, \citet{masini2016} have searched for correlations between the ratio of inner and outer radii and X-ray 
luminosity in the 2-10\,keV band for 14 disk-maser sources. They did not find any clear correlation. 

Although it is more plausible to see correlations at high radio frequencies (where mainly the core that is directly related to the AGN may be seen) than at low frequencies (where
the possible presence of jets with steep spectra might cause the emission to be less correlated with the maser disk size), we do not find a large difference
(see Fig.\,\ref{fig:lum_radius}). 
The strongest correlations apparently belong to the radio luminosities versus the disk inner radius (correlation coefficients of 0.87 and 0.73 for 1.4\,GHz and 33\,GHz,
respectively) and disk outer radius (correlation 
coefficients of 0.40 and 0.54 for 1.4\,GHz and 33\,GHz,
respectively). Generally, as seen in Fig.\,\ref{fig:lum_radius}, the disk inner radii show stronger correlations with
luminosity than the disk outer radii.
 
This may suggest that the core has a 
greater influence at the inner than at the outer radii, where disk warping, star formation, or peculiar density distributions may reduce the direct impact of the SMBH.
However, \citet{gao2017} found a correlation between 
maser disk outer radius (and not the inner radius) and WISE IR luminosity at Band 3 ($\lambda$=12\,$\mu$m), with a Spearman rank correlation coefficient of 0.83.

Within this context, it should be noted that our radio luminosities trace the 100\,pc scale, so that any relation with properties 
of sub-pc maser disks requires follow-up investigations with higher resolution.

\begin{figure*}[h]
\begin{center}
  \includegraphics[width=0.46\textwidth]{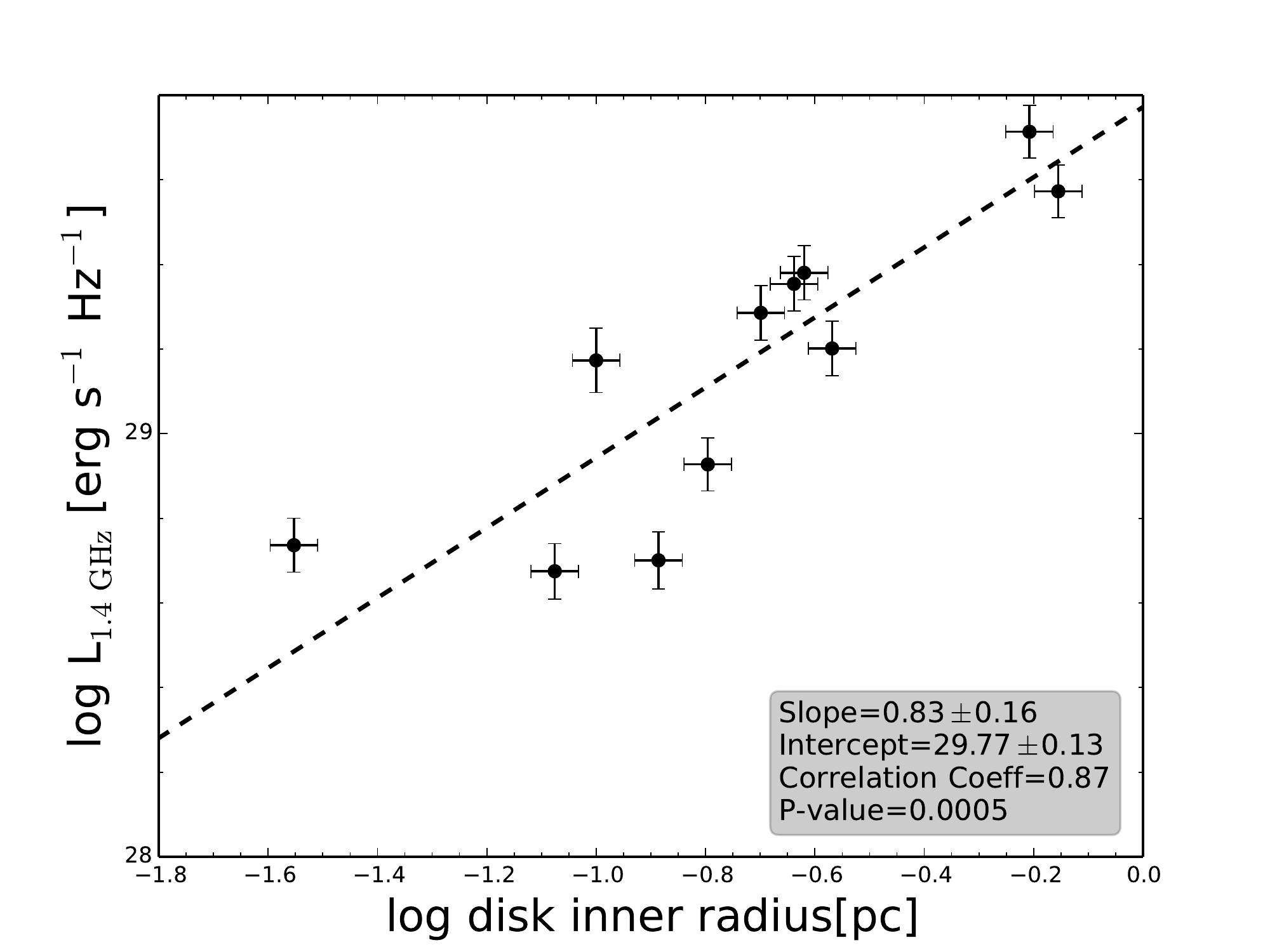}\includegraphics[width=0.46\textwidth]{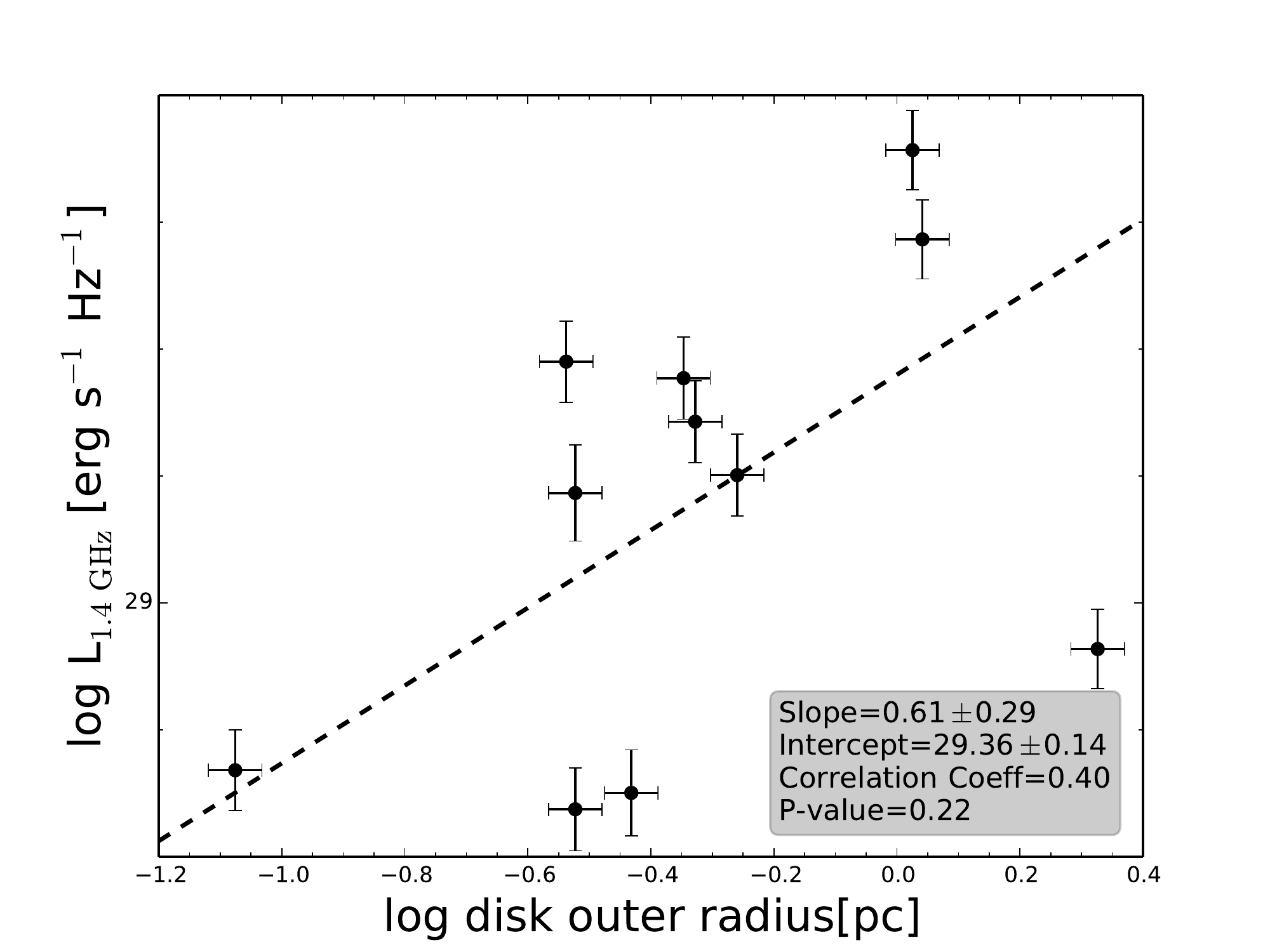}
\caption {Maser disk inner \textbf{(left)} and outer radius \textbf{(right)} versus 1.4\,GHz continuum luminosity.
The inner and outer radii are taken from from \citet{kuo2011}, \citet{braatz2015}, and \citet{gao2017}.
% In the lower plot, the red circles are upper limits for X-ray luminosities, and blue squares are the calculated \textsl{Swift}/BAT X-ray luminosities for detected sources.
 The black dashed lines are linear fits to the data.
The slope and intercept of the fit, as well as the correlation coefficient and the $P$-value of the Spearman test (see also Fig.\,\ref{fig:inc_lum}) are shown in a corner of each plot.}\label{fig:lum_radius}
\end{center}
\end{figure*}
\begin{figure*}[h]
\begin{center}
\ContinuedFloat
\includegraphics[width=0.46\textwidth]{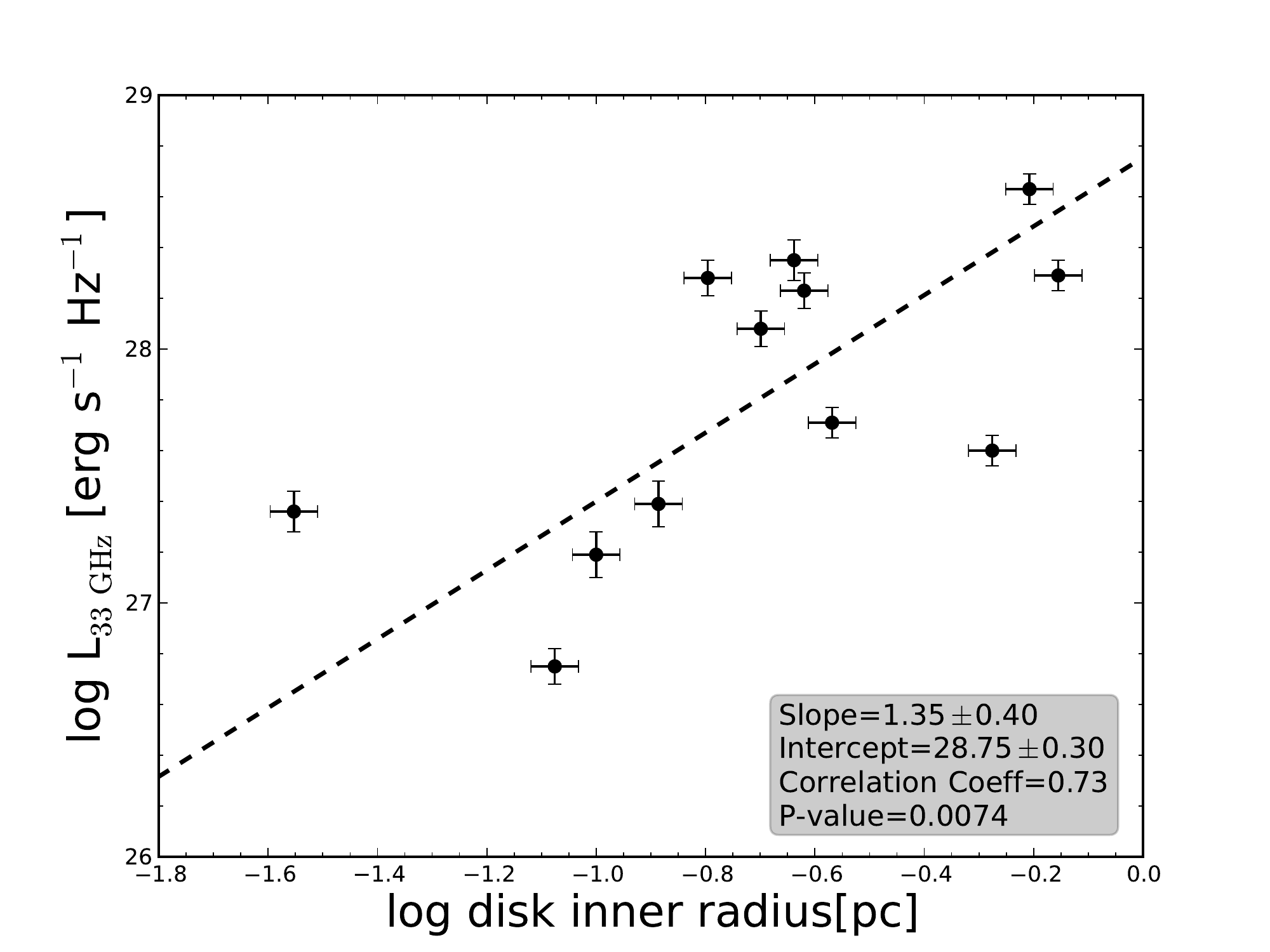}\includegraphics[width=0.46\textwidth]{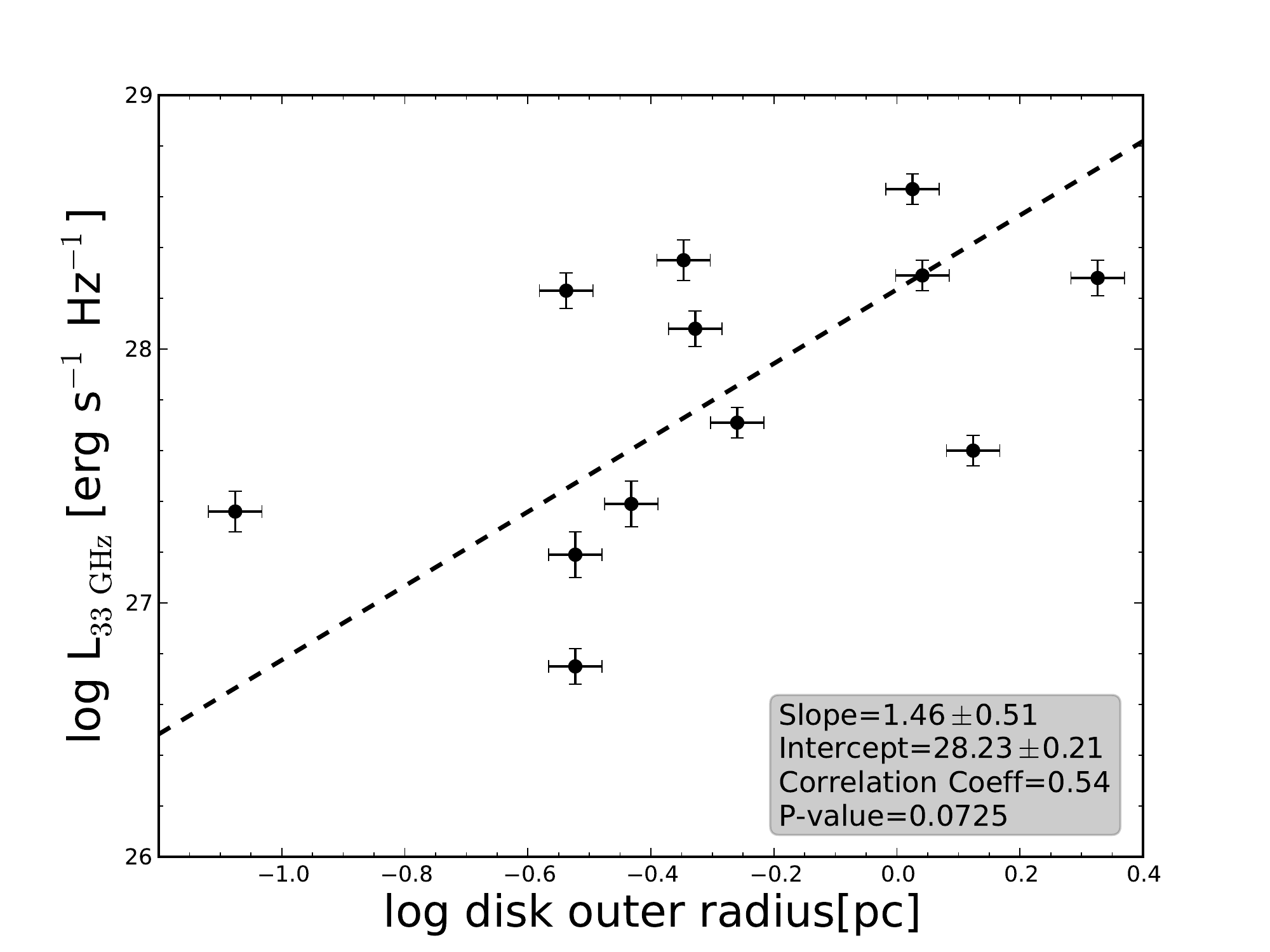}
\includegraphics[width=0.46\textwidth]{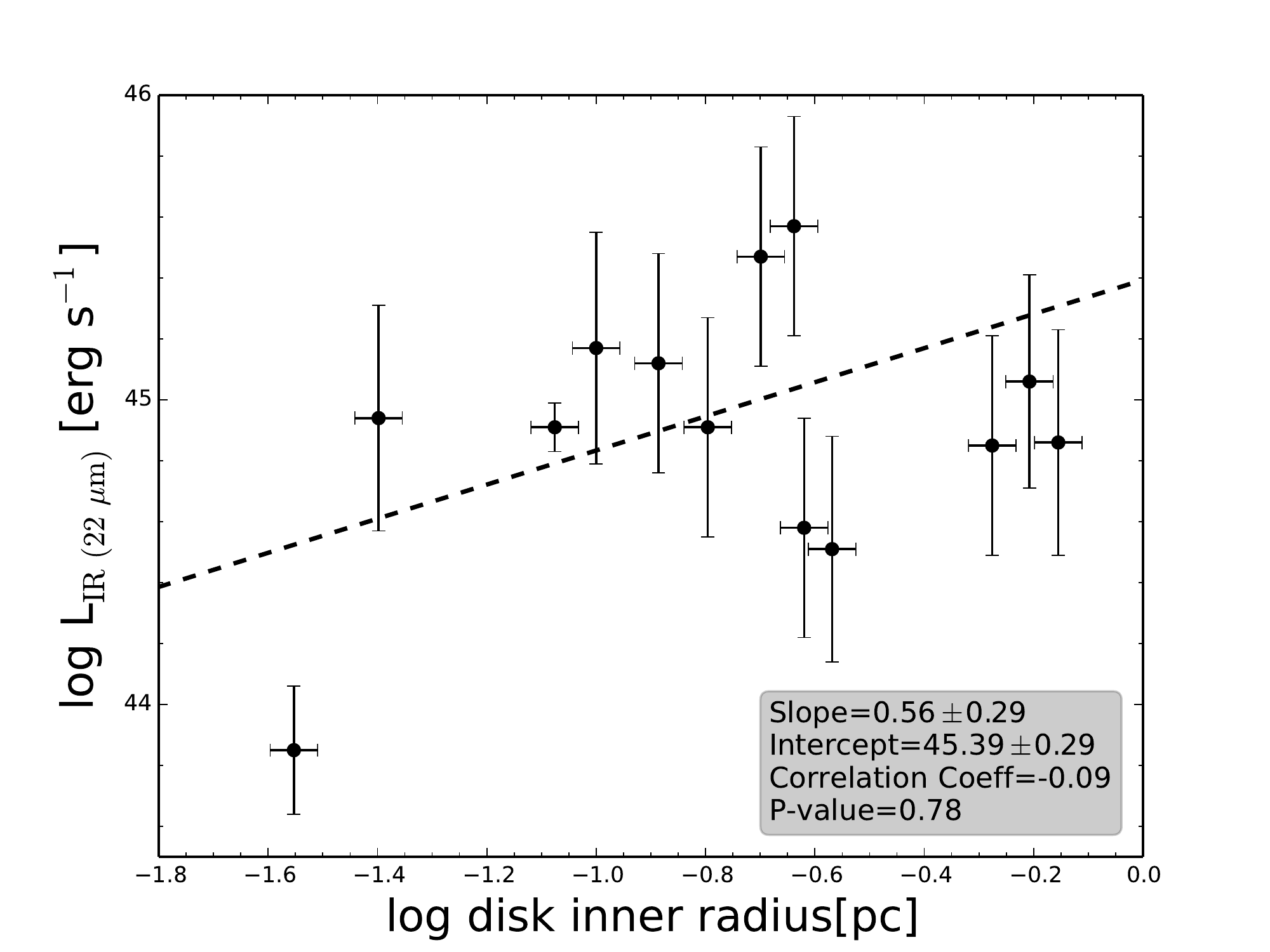}\includegraphics[width=0.46\textwidth]{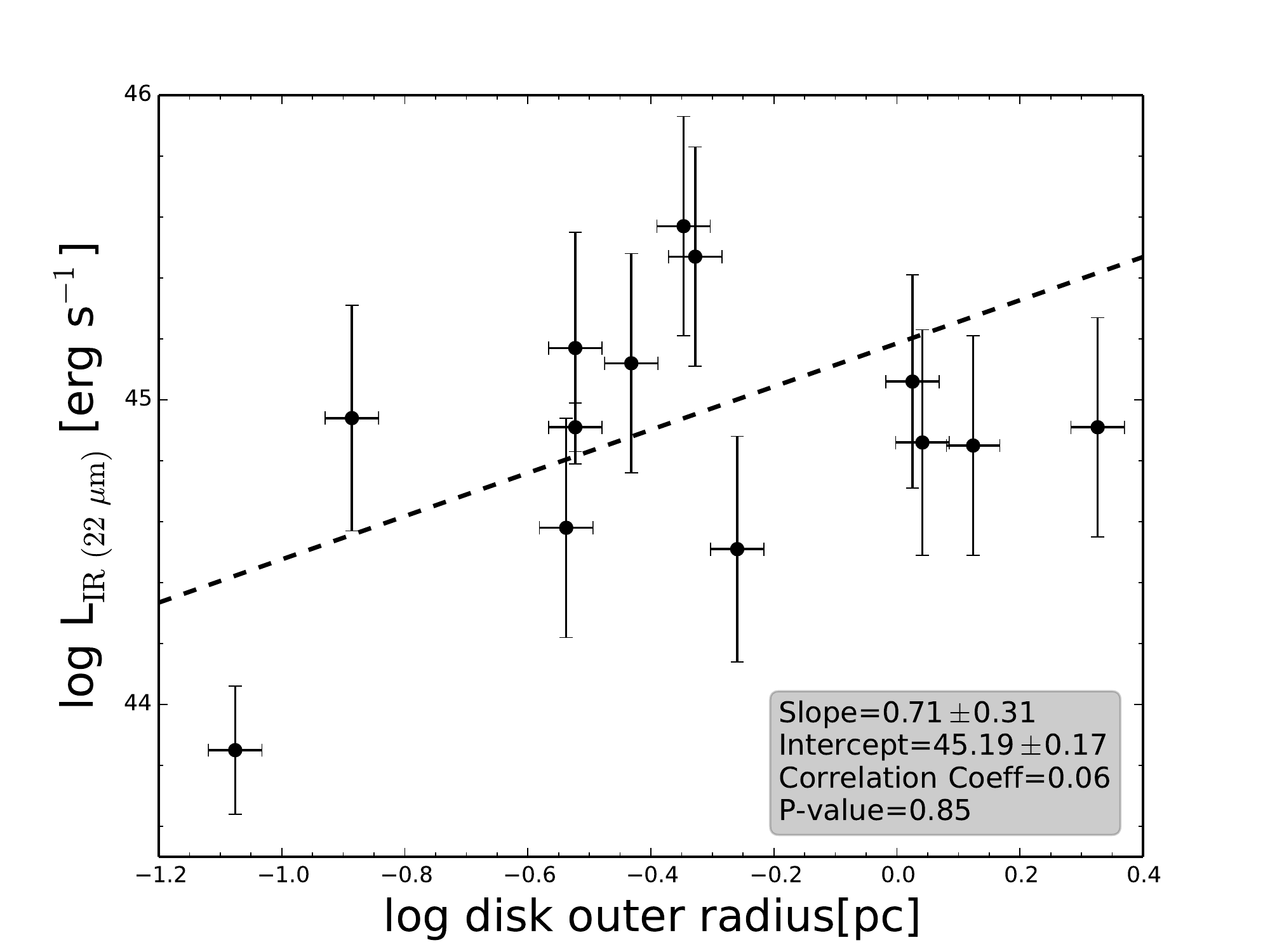}
\includegraphics[width=0.46\textwidth]{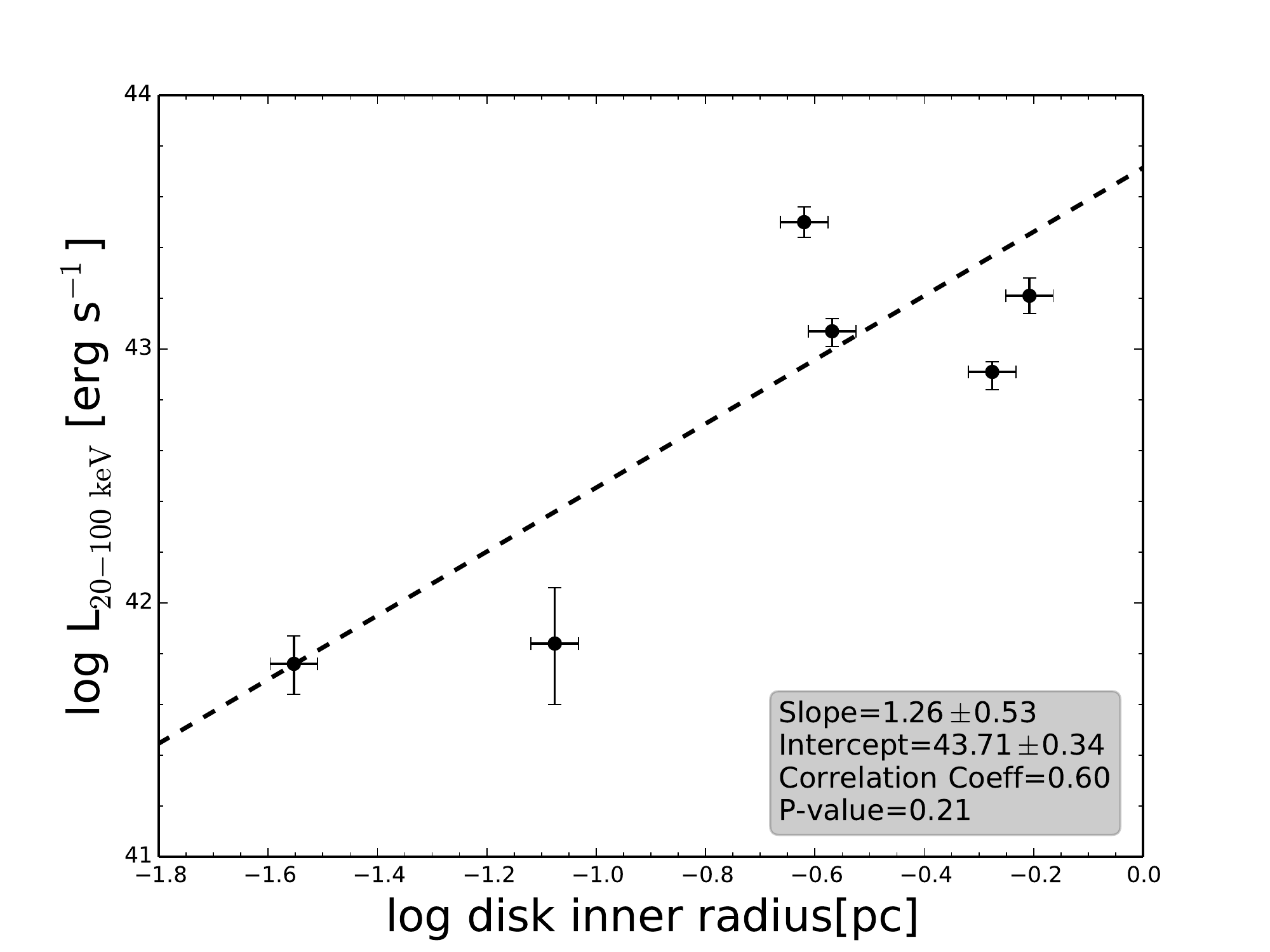}\includegraphics[width=0.46\textwidth]{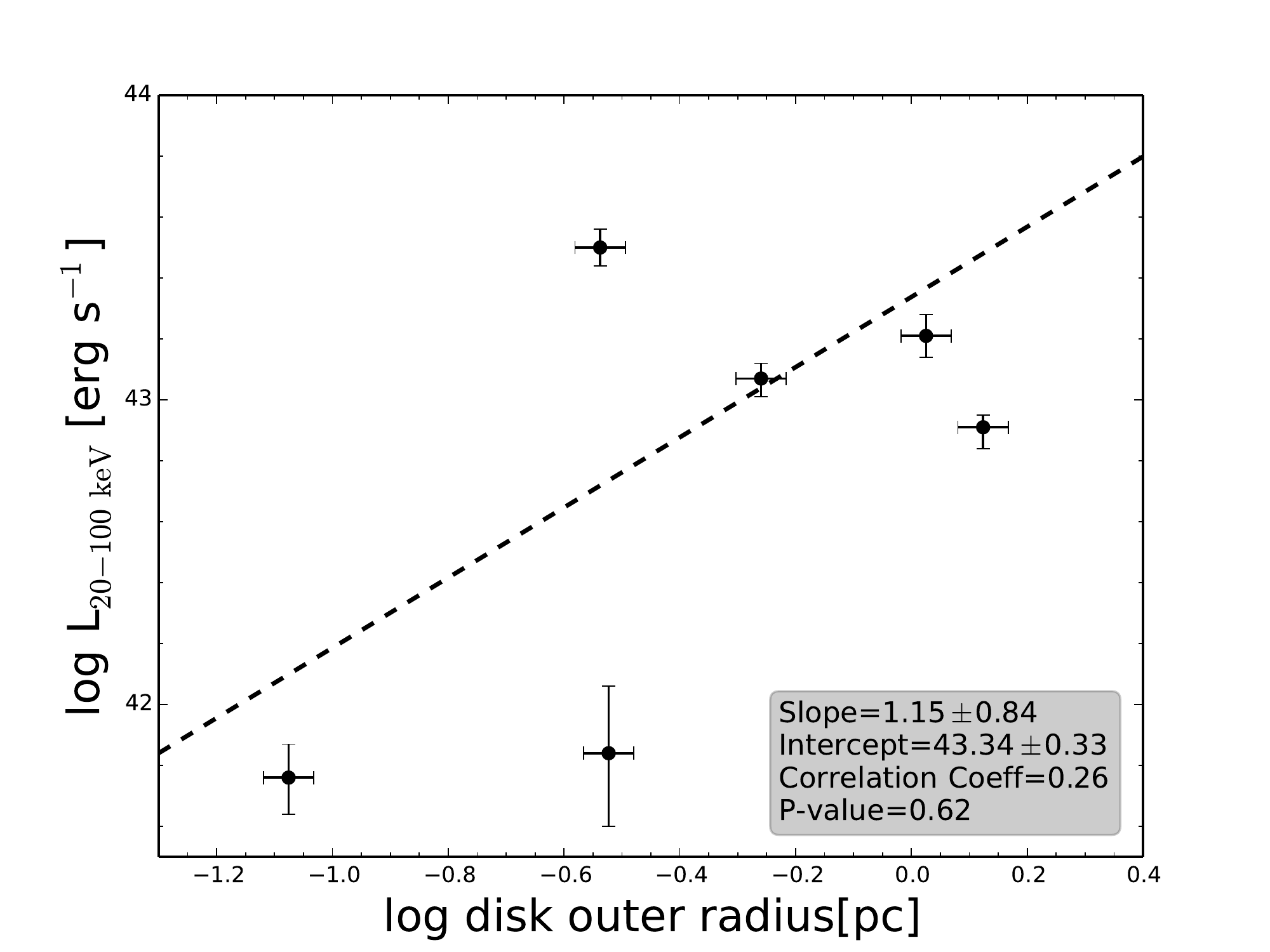}
\caption{\emph{(cont.)} 
Maser disk inner \textbf{(top left)} and outer radius \textbf{(top right)} versus 33\,GHz continuum luminosity.
Maser disk inner \textbf{(middle left)} and outer radius \textbf{(middle right)} versus IR luminosity, deduced from WISE 22\,$\mu$m data.
Maser disk inner \textbf{(bottom left)} and outer radius \textbf{(bottom right)} versus hard X-ray luminosity.}
\end{center}
\end{figure*}
\subsubsection{Black hole masses}\label{sec:bhm}
As mentioned in Sect.\,\ref{sec:vel}, relations between $M_{\rm SMBH}$ and other properties of galaxies have extensively been studied in recent decades 
\citep[e.g.,][]{kormendy2013}. Accurate measures of the black hole masses are essential to study such relations.

\begin{figure*}
\begin{center}
\includegraphics[width=0.46\textwidth]{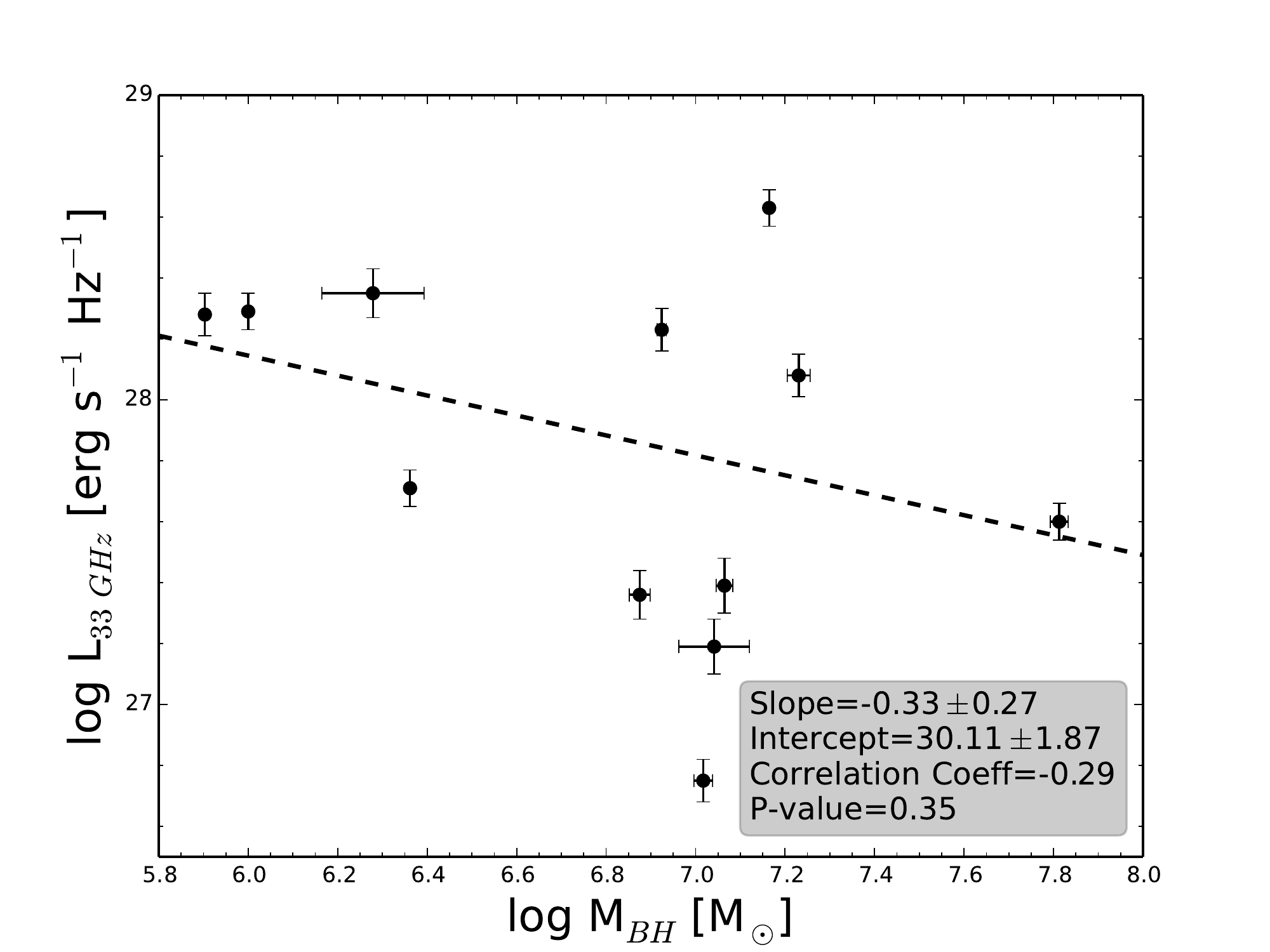}\includegraphics[width=0.46\textwidth]{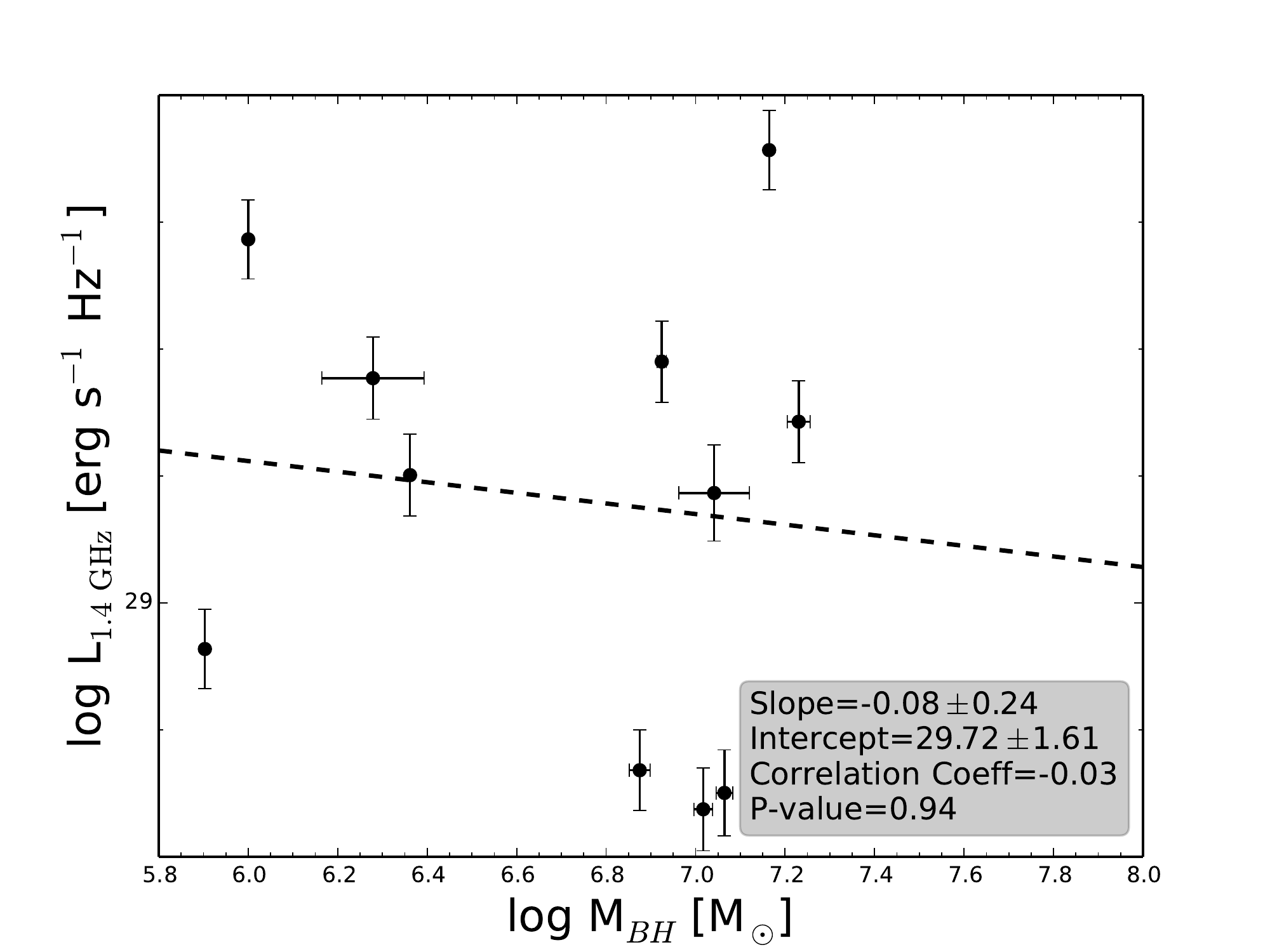}
\includegraphics[width=0.46\textwidth]{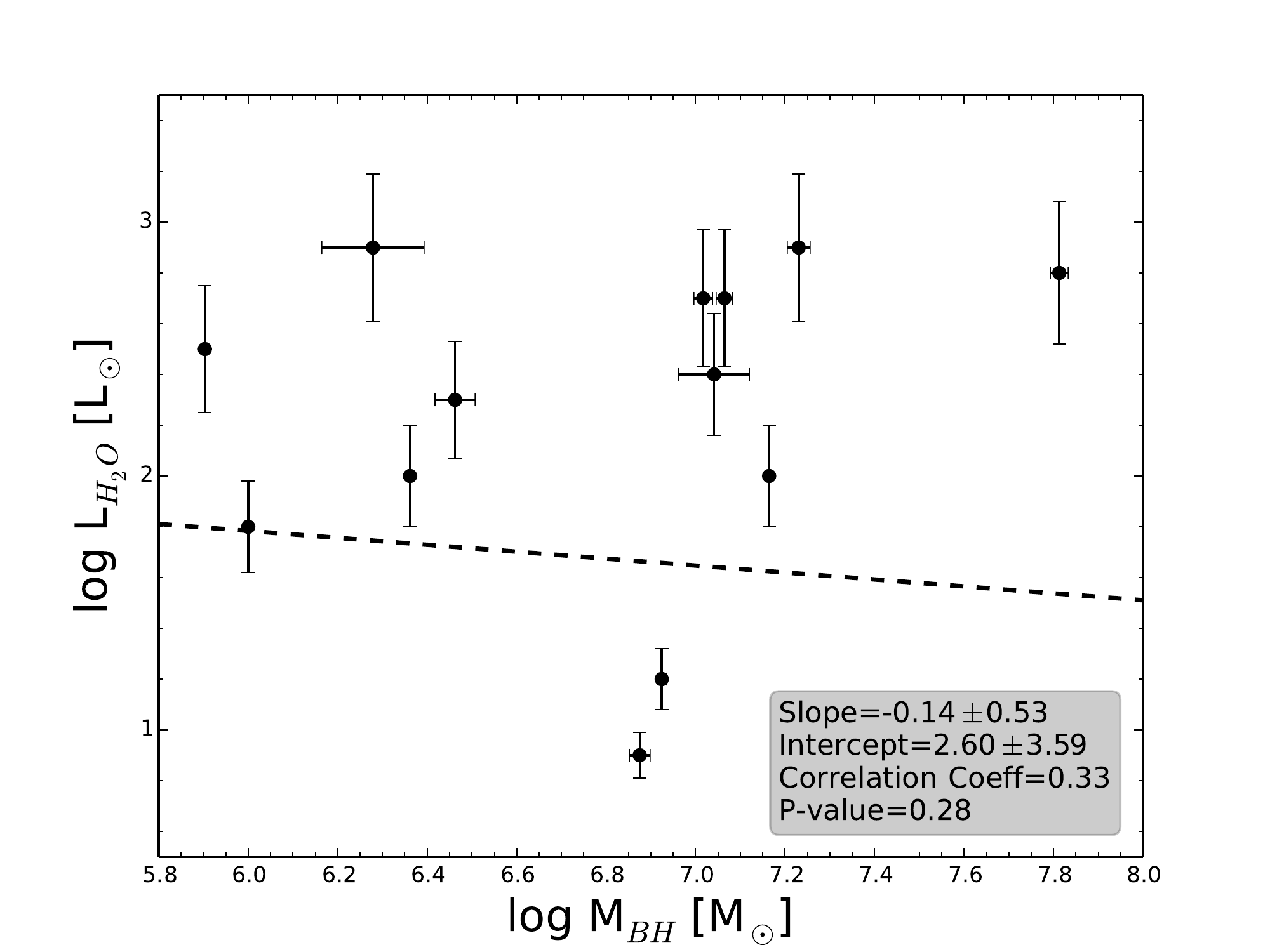}\includegraphics[width=0.46\textwidth]{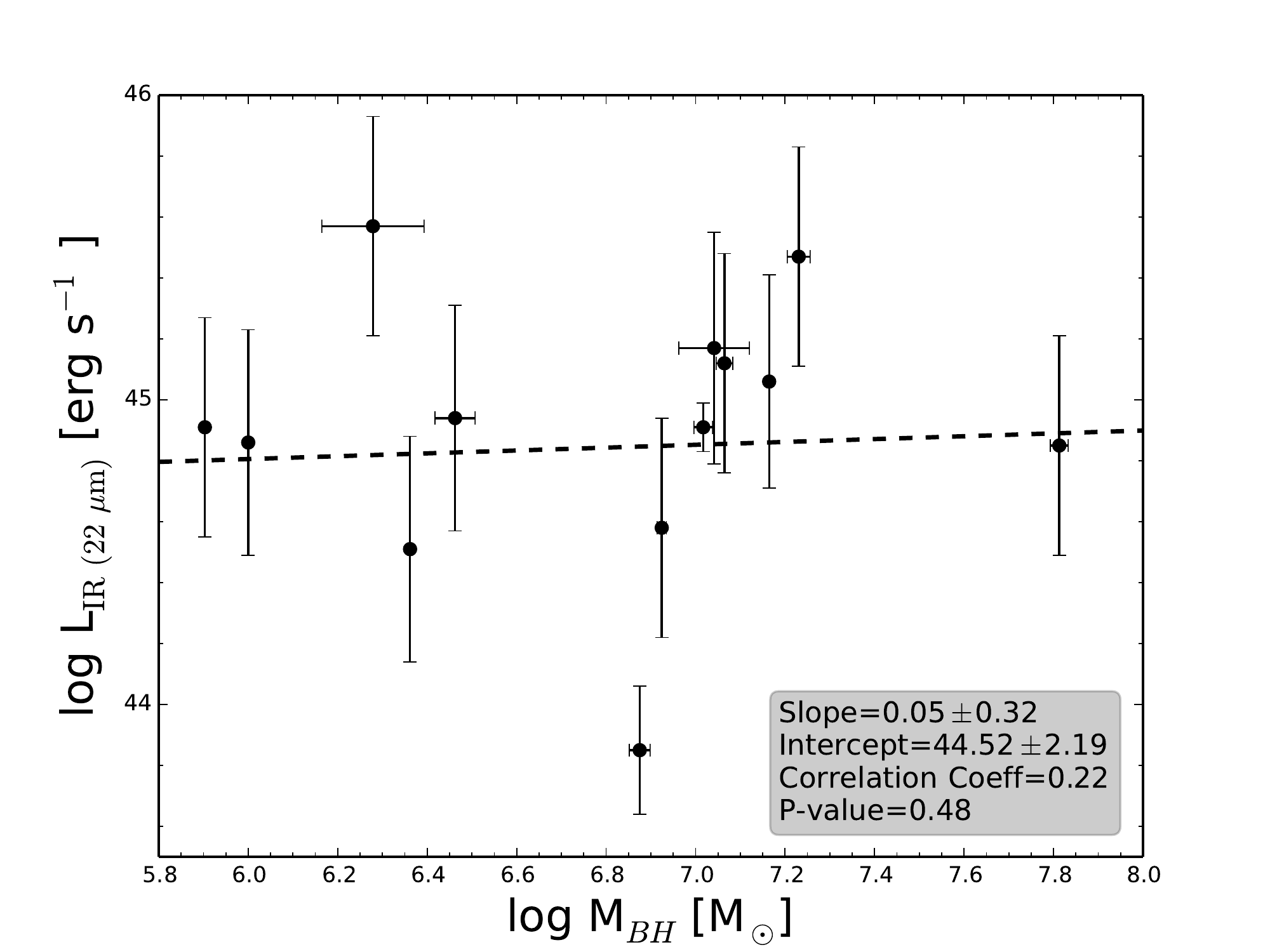}
\includegraphics[width=0.46\textwidth]{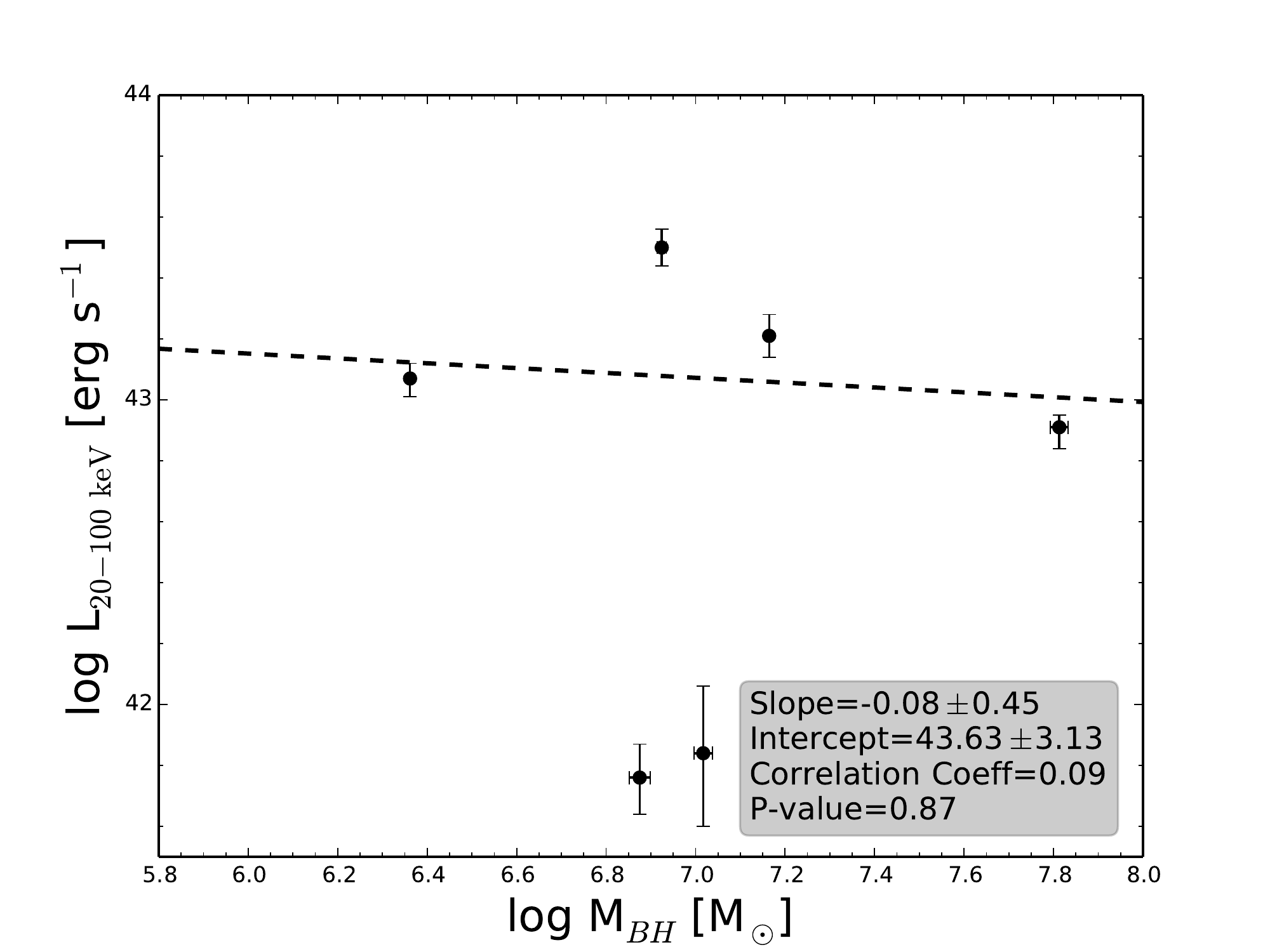}
\caption{\textbf{
Top left:} 33\,GHz continuum luminosity versus black hole mass.
\textbf{Top right:} 1.4\,GHz luminosity versus black hole mass.
\textbf{Middle left:} H$_2$O-maser luminosity versus black hole mass.
\textbf{Middle right:} IR luminosity (WISE 22\,$\mu$m data) versus black hole mass.
\textbf{Bottom:} Hard X-ray luminosity versus black hole mass.
Black hole masses are given on a logarithmic scale in units of solar mass, taken from \citet{kuo2011}, \citet{braatz2015}, and \citet{gao2017}. The lines show linear fits to the data. The slope and 
intercept of the fit, as well as the correlation coefficient and the $P$-value of the Spearman test (see also Fig.\,\ref{fig:inc_lum}) are shown in a lower 
corner of each panel.}\label{fig:bhm}
\end{center}
\end{figure*}

In nearby galaxies, VLBI observations of masers with high angular resolution provide a direct and precise measurement of the $M_{\rm SMBH}$ \citep[see][]{miyoshi1995,
herrnstein1999, humphreys2013}. The MCP has provided precise measures of the $M_{\rm SMBH}$ for 13 galaxies \citep{kuo2011, braatz2015, gao2017}. 
In Fig.\,\ref{fig:bhm} we present $M_{\rm SMBH}$ versus the continuum luminosity for these galaxies. The strongest correlation is seen for H$_2$O maser luminosity
versus $M_{\rm SMBH}$ with a correlation coefficient of 0.33. For continuum luminosities, no significant correlation is observed.
We should nevertheless keep in mind that a sample of 13 galaxies is small for statistical considerations.

%------------------------------------------------------------------------

\section{Summary}\label{sec:sum}

VLA observations of 24 mostly Seyfert\,2 galaxies with edge-on H$_2$O megamaser disks were obtained at Ka band (29--37\,GHz) to investigate the radio 
continuum emission from the center of these galaxies with $\sim$0.2-0.5\, arcsec resolution. With respect to the distance distribution of the sample galaxies,
this corresponds to a linear resolution on the order of 100\,pc. Out of 24 galaxies, 21 show radio emission at a 
4.5\,$\sigma$ or higher level, giving a detection rate of 87\%. Almost all of the galaxies 
contain compact cores with beam-convolved (and deconvolved) major axes smaller than 2.5 times the beam major axis, which even holds for most of the sources with 
more complex structure. NGC\,591 exhibits two main components,
and NGC\,3393 even three main components, which is indicative of $>$100\,pc scale radio jets. Four additional galaxies have extended structure, possibly also revealing jet-like features, but the majority shows a single compact emission region.

The radio continuum luminosity at 1.4\,GHz, 22\,GHz H$_2$O maser luminosity, infrared luminosity (22\,$\mu$m), the hard X-ray luminosity (20-100\,keV), 
optical or near-infrared morphology, position angle and inclination of the host galaxies, and their stellar velocity dispersions and circular velocities, 
maser-disk radii and black hole masses were taken into account for the further analysis of disk-maser host galaxy properties, as summarized below.

Only for one source in our sample with a jet-like 33\,GHz continuum feature (NGC\,4388) is the H$_2$O maser-disk orientation known.
The relative orientation appears to be orthogonal. NGC~2273 may show a similar scenario, but here alignments are less certain.
Galaxies with inner rings appear to exhibit less 33\,GHz emission than those without.
The maximum rotation velocity of the disk-maser host galaxies appears to be anticorrelated with 33\,GHz radio continuum luminosities, with a Spearman rank
correlation coefficient of --0.70.
The 33\,GHz radio continuum shows a weak correlation with IR luminosity at 22\,$\mu$m derived from the WISE satellite.
The H$_2$O maser luminosity does not exhibit any correlations with the radio (1.4\,GHz and 33\,GHz) or hard X-ray luminosities.
The black hole masses show stronger correlations with the H$_2$O maser luminosity than with the 1.4\,GHz, 33\,GHz, or hard X-ray luminosities.
The H$_2$O maser-disk inner radii show stronger correlations with 1.4\,GHz, 33\,GHz, and hard X-ray luminosities than the disk outer radii.
The strongest correlation is found with low-frequency 1.4\,GHz radio data.

Further studies with higher spatial resolution would be highly desirable to reveal the nature of the detected 33\,GHz continuum sources, to identify and map jets 
on scales $\la$100\,pc, and to better distinguish between radio emission arising from the AGN and that from surrounding star-forming regions. \\
%------------------------------------------------------------------------
\begin{sidewaystable*}
\caption [Position angle and inclination of the host galaxies, the spectral indices and the disk maser properties.]{Position angle and inclination of the host galaxies, the spectral indices, and the disk maser properties. } \label{table:indices_inclination}
\begin{tabular}{l ||c c c c|| c c |c| c c c c c c}
\hline \hline
& & & & & & & & & &\tabularnewline
&\multicolumn{4}{c||}{Spectral indices}&\multicolumn{2}{c|}{Galaxy scale}& This work&\multicolumn{6}{c}{Maser disk}\tabularnewline
& & & & & & & & & &\tabularnewline
Galaxy & $\alpha^{33}_{1.4\,NVSS}$ & $\alpha^{33}_{1.4\,FIRST}$ & $\alpha^{36}_{30}$ & $\alpha^{36}_{30}$ & P.A.& Inclination & P.A.& P.A. & Inclination &$R_{\rm in}$ & $R_{\rm out}$ & $M_{\rm SMBH}$ & ref \tabularnewline
& & & & &(\degr)&(\degr)&(\degr) & (\degr) & (\degr) & (pc)      & (pc)        & (10$^7$M$_{\odot}$) &\tabularnewline
\hline
& & & & & & & & & & &\tabularnewline
ESO558-G009 &  0.88 $\pm$ 0.02  & ...                &    1.11$\pm$0.94    &  0.76$\pm$0.22  &     8.0    &    81.2  & 155$\pm$33    &  256   & 98     &  0.20  &  0.47  & 1.7  &  1  \tabularnewline
                                                                               
IC\,0485    &  ...              & ...                &   ...               & ...             &    153.9   &    90.0  & ...           & ...    &  ...    & ...    &  ...   & ...   &  ... \tabularnewline
IC\,2560    &  0.88$\pm$ 0.02   & ...                &   1.31$\pm$0.88     & 0.88$\pm$0.21   &    45.7    &    65.6  & 25$\pm$11     & ...    &  ...    & ...    &  ...   & ...   &  ... \tabularnewline

J0126-0417  &  ...              & ...                &    -2.06$\pm$2.58   & 0.87$\pm$2.34   &    100.0   &    ...   & 146$\pm$86    & ...    &  ...    & ...    &  ...   & ...   &  ... \tabularnewline   
J0350-0127  &  ...              & ...                &  ...                & -2.63$\pm$4.32  &    155.0   &    ...   & 134.6$\pm$8.1 & ...    &  ...    & ...    &  ...   & ...   &  ... \tabularnewline         
J0437+2456  &  ...              & ...                &  ...                &  ...            &    ...     &    ...   & ...           &  20    &  81     & 0.04   &  0.13  & 0.29  &  1   \tabularnewline   
J0437+6637  & 1.12 $\pm$ 0.08   & ...                &  ...                & -0.52$\pm$1.58  &    117.0   &    53.0  & ...           & ...    &  ...    & ...    &  ...   & ...   &  ... \tabularnewline   
J0836+3327  & ...               & ...                &  ...                & ...             &    51.6    &    57.7  & ...           & ...    &  ...    & ...    &  ...   & ...   &  ... \tabularnewline   
J1658+3923  & ...               & 0.95$\pm$0.07      &  ...                &  2.36$\pm$2.77  &    104.3   &    90.0  &  81.4$\pm$4.7 & ...    &  ...    & ...    &  ...   & ...   &  ... \tabularnewline    

Mrk\,0001   & 0.93 $\pm$ 0.01   & ...                &   1.06$\pm$0.30     & 0.93$\pm$0.06   &    77.5    &    60.8  & 119.6$\pm$2.1 & ...    &  ...    &  0.70  & 1.10   & 1.0   &  2   \tabularnewline 
Mrk\,0078   & 0.96 $\pm$ 0.02   & ...                &    0.82$\pm$0.58    & 1.01$\pm$0.17   &    87.0    &    49.5  & 81$\pm$10     & ...    &  ...    & ...    &  ...   & ...   &  ... \tabularnewline
Mrk\,1029   & 0.73 $\pm$ 0.04   & 0.61$\pm$0.04      &    ...              & 0.72$\pm$2.37   &    73.5    &    45.9  & 59$\pm$11     &  218   &  79     &  0.23  & 0.45   & 0.19  &  1   \tabularnewline 
Mrk\,1210   & 0.79 $\pm$ 0.01   & ...                &    0.03$\pm$0.08    & -0.12$\pm$0.93  &    160.0   &    15.6  & 179$\pm$83     & ...    &  ...    &  0.62  & 1.06   & 14.6  &  2   \tabularnewline
Mrk\,1419   & 0.96 $\pm$ 0.06   & 0.84$\pm$0.05      &    3.0 $\pm$2.72    & -0.67$\pm$1.92  &    40.2    &    41.5  & 117$\pm$84    &  229  & 89      &  0.13  &  0.37  & 1.16  &  3   \tabularnewline

NGC\,0591   & 0.98 $\pm$ 0.02   & ...                &   2.47$\pm$1.49     & 1.31$\pm$1.60   &    0.2     &    40.3  & 153.3$\pm$2.8 & ...    &  ...    & ...    &  ...   & ...   &  ... \tabularnewline     
NGC\,1194   & ...               & 0.10$\pm$0.03      &   -0.20$\pm$0.63    & -0.25$\pm$0.27  &    139.3   &    71.1  & 75$\pm$82     &  157   &  85     &  0.54  & 1.33   & 6.5   &  3   \tabularnewline 
NGC\,2273   & 1.00 $\pm$ 0.04   & ...                &    0.55$\pm$1.13    & 1.86$\pm$1.15   &    63.3    &    57.4  & 82.8$\pm$2.8  & 153    &  84     &  0.028 & 0.084  & 0.75  &  3   \tabularnewline
NGC\,2979   & 1.19 $\pm$ 0.04   & ...                &    0.91$\pm$1.94    & 0.65$\pm$0.92   &    21.0    &    57.5  & 138$\pm$51    & ...    &   ...   & ...    &  ...   & ...   &  ... \tabularnewline
NGC\,3393   & 0.87 $\pm$ 0.03   & ...                &    1.02$\pm$0.26    & 2.10$\pm$3.18   &    12.8    &    30.9  & 165$\pm$50    & ...    &   ...   & ...    &  ...   & ...   &  ... \tabularnewline 
NGC\,4388   & 0.84 $\pm$ 0.02   & 0.52$\pm$ 0.02     &    -0.10$\pm$0.22   & 3.36$\pm$3.58  &    91.1    &    90.0  & 24$\pm$54     &  107   &  ...    &  0.24  & 0.29   & 0.84  &  3  \tabularnewline
NGC\,5495   & 1.44 $\pm$ 0.06   & ...                &    ...              & 0.43$\pm$1.38   &    44.5    &    34.7  & 167$\pm$24    &  176   &  95     &  0.10  & 0.30   & 1.1   &  1 \tabularnewline
NGC\,5728   & 1.09 $\pm$ 0.02   & ...                &    -1.26$\pm$0.37   & -1.53$\pm$0.17  &    14.5    &    59.1  & ...           & ...    &  ...    &  0.27  & 0.55   & 2.3   &  2 \tabularnewline

UGC\,3193   & 0.47 $\pm$ 0.02   & ...                &    2.08$\pm$1.12    & 0.96$\pm$1.92   &    177.4   &    76.1  & 167.2$\pm$4    & ...    &  ...    &  0.16  & 2.12   & 0.8   &  2 \tabularnewline
UGC\,3789   & 1.40 $\pm$ 0.04   & 1.34$\pm$0.03      &    1.89$\pm$4.26    & 2.83$\pm$1.76   &    166.0   &    43.2  & 117$\pm$22    &  41    &  >88    &  0.084 & 0.30   & 1.04  &  3 \tabularnewline
\hline
\end{tabular}\par
\bigskip
\textbf{Notes}.
Column\,1: source name.
Column\,2: spectral indices between NVSS 1.4\,GHz and 33\,GHz, with uncertainties. 
Column\,3: spectral indices between FIRST 1.4\,GHz and 33\,GHz, with uncertainties. 
Columns\,4 and 5: spectral indices between 30\,GHz and 36\,GHz from our 8\,GHz bandwidth observations, first by splitting the data and second by setting the Taylor 
                  coefficients as two in CASA (see Sect.\,\ref{sec:sp_index}).
Column\,6 and 7: large scale position angle and inclination (from HyperLeda database).
Column\,8: position angle of the 33\,GHz continuum component with formal errors, obtained from Gaussian fits to a region containing all the significant emission.
Columns\,9 and 10: maser disk position angle and inclination.
Columns\,11 and 12: maser disk inner and outer radii in pc.
Column\,13: black hole mass.
Column\,14: references for disk maser radii and $M_{\rm SMBH}$: (1) \citet{gao2017}; (2) \citet{proposal2015}; (3) \citet{kuo2011}.
\end{sidewaystable*}
%------------------------------------------------------------------------

\begin{acknowledgements}
F.K. would like to thank Sergio Dzib Quijano for his useful comments on the data reduction process. This work made use of the NASA/IPAC extragalactic Database (NED), which 
is operated by the Jet Propulsion Laboratory, California Institute of Technology, under contract with NASA. We further acknowledge the usage of the HyperLeda database 
(http://leda.univ-lyon1.fr) and the SAO/NASA ADS Astronomy Abstract Service (http://adsabs.harvard.edu) and Kapteyn Package \citep{KapteynPackage}.
      \\
      \\
\end{acknowledgements}

% WARNING
%-------------------------------------------------------------------
% Please note that we have included the references to the file aa.dem in
% order to compile it, but we ask you to:
%
% - use BibTeX with the regular commands:
%   \bibliographystyle{aa} % style aa.bst
%   \bibliography{Yourfile} % your references Yourfile.bib
%
% - join the .bib files when you upload your source files
%-------------------------------------------------------------------

\bibliography{biblio}

\begin{thebibliography}{75}
\expandafter\ifx\csname natexlab\endcsname\relax\def\natexlab#1{#1}\fi

\bibitem[{{Adams}(1973)}]{adams1973}
{Adams}, T.~F. 1973, \apj, 179, 417

\bibitem[{{Becker} {et~al.}(1995){Becker}, {White}, \& {Helfand}}]{becker1995}
{Becker}, R.~H., {White}, R.~L., \& {Helfand}, D.~J. 1995, \apj, 450, 559

\bibitem[{{Braatz} {et~al.}(2015{\natexlab{a}}){Braatz}, {Condon},
  {Constantin}, {Gao}, {Greene}, {Hao}, {Henkel}, {Impellizzeri}, {Kuo},
  {Litzinger}, {Lo}, {Pesce}, {Reid}, {Wagner}, \& {Zhao}}]{braatz2015}
{Braatz}, J., {Condon}, J., {Constantin}, A., {et~al.} 2015{\natexlab{a}}, IAU
  General Assembly, 22, 2255730

\bibitem[{{Braatz} {et~al.}(2015{\natexlab{b}}){Braatz}, {Condon},
  {Constantin}, {Gao}, {Greene}, {Hao}, {Henkel}, {Impellizzeri}, {Kuo},
  {Litzinger}, {Lo}, {Pesce}, {Reid}, {Wagner}, \& {Zhao}}]{proposal2015}
{Braatz}, J., {Condon}, J., {Constantin}, A., {et~al.} 2015{\natexlab{b}}, IAU
  General Assembly, 22, 2255730

\bibitem[{{Braatz} \& {Gugliucci}(2008)}]{braatz2008}
{Braatz}, J.~A. \& {Gugliucci}, N.~E. 2008, \apj, 678, 96

\bibitem[{{Braatz} {et~al.}(2004){Braatz}, {Henkel}, {Greenhill}, {Moran}, \&
  {Wilson}}]{braatz2004}
{Braatz}, J.~A., {Henkel}, C., {Greenhill}, L.~J., {Moran}, J.~M., \& {Wilson},
  A.~S. 2004, \apjl, 617, L29

\bibitem[{{Braatz} {et~al.}(2010){Braatz}, {Reid}, {Humphreys}, {Henkel},
  {Condon}, \& {Lo}}]{braatz2010}
{Braatz}, J.~A., {Reid}, M.~J., {Humphreys}, E.~M.~L., {et~al.} 2010, \apj,
  718, 657

\bibitem[{{Castangia} {et~al.}(2013){Castangia}, {Panessa}, {Henkel}, {Kadler},
  \& {Tarchi}}]{castangia2013}
{Castangia}, P., {Panessa}, F., {Henkel}, C., {Kadler}, M., \& {Tarchi}, A.
  2013, \mnras, 436, 3388

\bibitem[{{Chemin} {et~al.}(2015){Chemin}, {Renaud}, \&
  {Soubiran}}]{chemin2015}
{Chemin}, L., {Renaud}, F., \& {Soubiran}, C. 2015, \aap, 578, A14

\bibitem[{{Cheung} {et~al.}(1969){Cheung}, {Rank}, {Townes}, {Thornton}, \&
  {Welch}}]{cheung1969}
{Cheung}, A.~C., {Rank}, D.~M., {Townes}, C.~H., {Thornton}, D.~D., \& {Welch},
  W.~J. 1969, \nat, 221, 626

\bibitem[{{Condon} {et~al.}(1998){Condon}, {Cotton}, {Greisen}, {Yin},
  {Perley}, {Taylor}, \& {Broderick}}]{condon1998}
{Condon}, J.~J., {Cotton}, W.~D., {Greisen}, E.~W., {et~al.} 1998, \aj, 115,
  1693

\bibitem[{{Corbin} {et~al.}(1988){Corbin}, {Baldwin}, \& {Wilson}}]{corbin1988}
{Corbin}, M.~R., {Baldwin}, J.~A., \& {Wilson}, A.~S. 1988, \apj, 334, 584

\bibitem[{{Damas-Segovia} {et~al.}(2016){Damas-Segovia}, {Beck}, {Vollmer},
  {Wiegert}, {Krause}, {Irwin}, {We{\.z}gowiec}, {Li}, {Dettmar}, {English}, \&
  {Wang}}]{damas2016}
{Damas-Segovia}, A., {Beck}, R., {Vollmer}, B., {et~al.} 2016, \apj, 824, 30

\bibitem[{{Diamond-Stanic} {et~al.}(2009){Diamond-Stanic}, {Rieke}, \&
  {Rigby}}]{diamond-stanic2009}
{Diamond-Stanic}, A.~M., {Rieke}, G.~H., \& {Rigby}, J.~R. 2009, \apj, 698, 623

\bibitem[{{Erwin} \& {Sparke}(2003)}]{erwin2003}
{Erwin}, P. \& {Sparke}, L.~S. 2003, \apjs, 146, 299

\bibitem[{{Fabbiano} {et~al.}(2011){Fabbiano}, {Wang}, {Elvis}, \&
  {Risaliti}}]{fabbiano2011}
{Fabbiano}, G., {Wang}, J., {Elvis}, M., \& {Risaliti}, G. 2011, \nat, 477, 431

\bibitem[{{Fiore} {et~al.}(2008){Fiore}, {Grazian}, {Santini}, {Puccetti},
  {Brusa}, {Feruglio}, {Fontana}, {Giallongo}, {Comastri}, {Gruppioni},
  {Pozzi}, {Zamorani}, \& {Vignali}}]{fiore2008}
{Fiore}, F., {Grazian}, A., {Santini}, P., {et~al.} 2008, \apj, 672, 94

\bibitem[{{Gallimore} {et~al.}(2001){Gallimore}, {Henkel}, {Baum}, {Glass},
  {Claussen}, {Prieto}, \& {Von Kap-herr}}]{gallimore2001}
{Gallimore}, J.~F., {Henkel}, C., {Baum}, S.~A., {et~al.} 2001, \apj, 556, 694

\bibitem[{{Gao} {et~al.}(2017){Gao}, {Braatz}, {Reid}, {Condon}, {Greene},
  {Henkel}, {Impellizzeri}, {Lo}, {Kuo}, {Pesce}, {Wagner}, \&
  {Zhao}}]{gao2017}
{Gao}, F., {Braatz}, J.~A., {Reid}, M.~J., {et~al.} 2017, \apj, 834, 52

\bibitem[{{Georgantopoulos} {et~al.}(2008){Georgantopoulos}, {Georgakakis},
  {Rowan-Robinson}, \& {Rovilos}}]{georgantopoulos2008}
{Georgantopoulos}, I., {Georgakakis}, A., {Rowan-Robinson}, M., \& {Rovilos},
  E. 2008, \aap, 484, 671

\bibitem[{{Greene} {et~al.}(2010){Greene}, {Peng}, {Kim}, {Kuo}, {Braatz},
  {Impellizzeri}, {Condon}, {Lo}, {Henkel}, \& {Reid}}]{greene2010}
{Greene}, J.~E., {Peng}, C.~Y., {Kim}, M., {et~al.} 2010, \apj, 721, 26

\bibitem[{{Greene} {et~al.}(2013){Greene}, {Seth}, {den Brok}, {Braatz},
  {Henkel}, {Sun}, {Peng}, {Kuo}, {Impellizzeri}, \& {Lo}}]{greene2013}
{Greene}, J.~E., {Seth}, A., {den Brok}, M., {et~al.} 2013, \apj, 771, 121

\bibitem[{{Greene} {et~al.}(2016){Greene}, {Seth}, {Kim}, {L{\"a}sker},
  {Goulding}, {Gao}, {Braatz}, {Henkel}, {Condon}, {Lo}, \&
  {Zhao}}]{greene2016}
{Greene}, J.~E., {Seth}, A., {Kim}, M., {et~al.} 2016, \apjl, 826, L32

\bibitem[{{Greene} {et~al.}(2014){Greene}, {Seth}, {Lyubenova}, {Walsh}, {van
  de Ven}, \& {L{\"a}sker}}]{greene2014}
{Greene}, J.~E., {Seth}, A., {Lyubenova}, M., {et~al.} 2014, \apj, 788, 145

\bibitem[{{Henkel} {et~al.}(2005){Henkel}, {Peck}, {Tarchi}, {Nagar}, {Braatz},
  {Castangia}, \& {Moscadelli}}]{henkel2005}
{Henkel}, C., {Peck}, A.~B., {Tarchi}, A., {et~al.} 2005, \aap, 436, 75

\bibitem[{{Herrnstein} {et~al.}(1998){Herrnstein}, {Greenhill}, {Moran},
  {Diamond}, {Inoue}, {Nakai}, \& {Miyoshi}}]{herrnstein1998}
{Herrnstein}, J.~R., {Greenhill}, L.~J., {Moran}, J.~M., {et~al.} 1998, \apjl,
  497, L69

\bibitem[{{Herrnstein} {et~al.}(1999){Herrnstein}, {Moran}, {Greenhill},
  {Diamond}, {Inoue}, {Nakai}, {Miyoshi}, {Henkel}, \&
  {Riess}}]{herrnstein1999}
{Herrnstein}, J.~R., {Moran}, J.~M., {Greenhill}, L.~J., {et~al.} 1999, \nat,
  400, 539

\bibitem[{{Hlavacek-Larrondo} {et~al.}(2011){Hlavacek-Larrondo}, {Carignan},
  {Daigle}, {de Denus-Baillargeon}, {Marcelin}, {Epinat}, \&
  {Hernandez}}]{hlavacek2011}
{Hlavacek-Larrondo}, J., {Carignan}, C., {Daigle}, O., {et~al.} 2011, \mnras,
  411, 71

\bibitem[{{Hopkins} \& {Quataert}(2010)}]{hopkins2010}
{Hopkins}, P.~F. \& {Quataert}, E. 2010, \mnras, 407, 1529

\bibitem[{{Humphreys} {et~al.}(2013){Humphreys}, {Reid}, {Moran}, {Greenhill},
  \& {Argon}}]{humphreys2013}
{Humphreys}, E.~M.~L., {Reid}, M.~J., {Moran}, J.~M., {Greenhill}, L.~J., \&
  {Argon}, A.~L. 2013, \apj, 775, 13

\bibitem[{{Impellizzeri} {et~al.}(2008){Impellizzeri}, {McKean}, {Castangia},
  {Roy}, {Henkel}, {Brunthaler}, \& {Wucknitz}}]{impellizzeri2008}
{Impellizzeri}, C.~M.~V., {McKean}, J.~P., {Castangia}, P., {et~al.} 2008,
  \nat, 456, 927

\bibitem[{{Ishihara} {et~al.}(2001){Ishihara}, {Nakai}, {Iyomoto}, {Makishima},
  {Diamond}, \& {Hall}}]{ishihara2001}
{Ishihara}, Y., {Nakai}, N., {Iyomoto}, N., {et~al.} 2001, \pasj, 53, 215

\bibitem[{{Kinney} {et~al.}(2000){Kinney}, {Schmitt}, {Clarke}, {Pringle},
  {Ulvestad}, \& {Antonucci}}]{kinney2000}
{Kinney}, A.~L., {Schmitt}, H.~R., {Clarke}, C.~J., {et~al.} 2000, \apj, 537,
  152

\bibitem[{{Kondratko} {et~al.}(2006{\natexlab{a}}){Kondratko}, {Greenhill}, \&
  {Moran}}]{kondratko2006a}
{Kondratko}, P.~T., {Greenhill}, L.~J., \& {Moran}, J.~M. 2006{\natexlab{a}},
  \apj, 652, 136

\bibitem[{{Kondratko} {et~al.}(2008){Kondratko}, {Greenhill}, \&
  {Moran}}]{kondratko2008}
{Kondratko}, P.~T., {Greenhill}, L.~J., \& {Moran}, J.~M. 2008, \apj, 678, 87

\bibitem[{{Kondratko} {et~al.}(2006{\natexlab{b}}){Kondratko}, {Greenhill},
  {Moran}, {Lovell}, {Kuiper}, {Jauncey}, {Cameron}, {G{\'o}mez},
  {Garc{\'{\i}}a-Mir{\'o}}, {Moll}, {de Gregorio-Monsalvo}, \&
  {Jim{\'e}nez-Bail{\'o}n}}]{kondratko2006b}
{Kondratko}, P.~T., {Greenhill}, L.~J., {Moran}, J.~M., {et~al.}
  2006{\natexlab{b}}, \apj, 638, 100

\bibitem[{{Kormendy} \& {Ho}(2013)}]{kormendy2013}
{Kormendy}, J. \& {Ho}, L.~C. 2013, \araa, 51, 511

\bibitem[{{Koss} {et~al.}(2015){Koss}, {Romero-Ca{\~n}izales}, {Baronchelli},
  {Teng}, {Balokovi{\'c}}, {Puccetti}, {Bauer}, {Ar{\'e}valo}, {Assef},
  {Ballantyne}, {Brandt}, {Brightman}, {Comastri}, {Gandhi}, {Harrison}, {Luo},
  {Schawinski}, {Stern}, \& {Treister}}]{koss2015}
{Koss}, M.~J., {Romero-Ca{\~n}izales}, C., {Baronchelli}, L., {et~al.} 2015,
  \apj, 807, 149

\bibitem[{{Krumholz} \& {Kruijssen}(2015)}]{krumholtz2015}
{Krumholz}, M.~R. \& {Kruijssen}, J.~M.~D. 2015, \mnras, 453, 739

\bibitem[{{Kuo} {et~al.}(2011){Kuo}, {Braatz}, {Condon}, {Impellizzeri}, {Lo},
  {Zaw}, {Schenker}, {Henkel}, {Reid}, \& {Greene}}]{kuo2011}
{Kuo}, C.~Y., {Braatz}, J.~A., {Condon}, J.~J., {et~al.} 2011, \apj, 727, 20

\bibitem[{{Lo}(2005)}]{lo2005}
{Lo}, K.~Y. 2005, \araa, 43, 625

\bibitem[{{Makarov} {et~al.}(2014){Makarov}, {Prugniel}, {Terekhova},
  {Courtois}, \& {Vauglin}}]{makarov2014}
{Makarov}, D., {Prugniel}, P., {Terekhova}, N., {Courtois}, H., \& {Vauglin},
  I. 2014, \aap, 570, A13

\bibitem[{{Masini} {et~al.}(2016){Masini}, {Comastri}, {Balokovi{\'c}}, {Zaw},
  {Puccetti}, {Ballantyne}, {Bauer}, {Boggs}, {Brandt}, {Brightman},
  {Christensen}, {Craig}, {Gandhi}, {Hailey}, {Harrison}, {Koss}, {Madejski},
  {Ricci}, {Rivers}, {Stern}, \& {Zhang}}]{masini2016}
{Masini}, A., {Comastri}, A., {Balokovi{\'c}}, M., {et~al.} 2016, \aap, 589,
  A59

\bibitem[{{Mazzalay} \& {Rodr{\'{\i}}guez-Ardila}(2007)}]{mazzalay2007}
{Mazzalay}, X. \& {Rodr{\'{\i}}guez-Ardila}, A. 2007, \aap, 463, 445

\bibitem[{{McMullin} {et~al.}(2007){McMullin}, {Waters}, {Schiebel}, {Young},
  \& {Golap}}]{mcmullin2007}
{McMullin}, J.~P., {Waters}, B., {Schiebel}, D., {Young}, W., \& {Golap}, K.
  2007, in Astronomical Society of the Pacific Conference Series, Vol. 376,
  Astronomical Data Analysis Software and Systems XVI, ed. R.~A. {Shaw},
  F.~{Hill}, \& D.~J. {Bell}, 127

\bibitem[{{Middelberg} {et~al.}(2004){Middelberg}, {Roy}, {Nagar}, {Krichbaum},
  {Norris}, {Wilson}, {Falcke}, {Colbert}, {Witzel}, \&
  {Fricke}}]{middelberg2004}
{Middelberg}, E., {Roy}, A.~L., {Nagar}, N.~M., {et~al.} 2004, \aap, 417, 925

\bibitem[{{Miyoshi} {et~al.}(1995){Miyoshi}, {Moran}, {Herrnstein},
  {Greenhill}, {Nakai}, {Diamond}, \& {Inoue}}]{miyoshi1995}
{Miyoshi}, M., {Moran}, J., {Herrnstein}, J., {et~al.} 1995, \nat, 373, 127

\bibitem[{{Mundell} {et~al.}(2009){Mundell}, {Ferruit}, {Nagar}, \&
  {Wilson}}]{mundell2009}
{Mundell}, C.~G., {Ferruit}, P., {Nagar}, N., \& {Wilson}, A.~S. 2009, \apj,
  703, 802

\bibitem[{{Nagar} \& {Wilson}(1999)}]{nagar1999}
{Nagar}, N.~M. \& {Wilson}, A.~S. 1999, \apj, 516, 97

\bibitem[{{Nakai} {et~al.}(1993){Nakai}, {Inoue}, \& {Miyoshi}}]{nakai1993}
{Nakai}, N., {Inoue}, M., \& {Miyoshi}, M. 1993, \nat, 361, 45

\bibitem[{{Neufeld} {et~al.}(1994){Neufeld}, {Maloney}, \&
  {Conger}}]{neufeld1994}
{Neufeld}, D.~A., {Maloney}, P.~R., \& {Conger}, S. 1994, \apjl, 436, L.127

\bibitem[{{Omar} {et~al.}(2002){Omar}, {Dwarakanath}, {Rupen}, \&
  {Anantharamaiah}}]{omar2002}
{Omar}, A., {Dwarakanath}, K.~S., {Rupen}, M., \& {Anantharamaiah}, K.~R. 2002,
  \aap, 394, 405

\bibitem[{{Pesce} {et~al.}(2015){Pesce}, {Braatz}, {Condon}, {Gao}, {Henkel},
  {Litzinger}, {Lo}, \& {Reid}}]{pesce2015}
{Pesce}, D.~W., {Braatz}, J.~A., {Condon}, J.~J., {et~al.} 2015, \apj, 810, 65

\bibitem[{{Petitpas} \& {Wilson}(2002)}]{petitpas2002}
{Petitpas}, G.~R. \& {Wilson}, C.~D. 2002, \apj, 575, 814

\bibitem[{{Piner} {et~al.}(1995){Piner}, {Stone}, \& {Teuben}}]{piner1995}
{Piner}, B.~G., {Stone}, J.~M., \& {Teuben}, P.~J. 1995, \apj, 449, 508

\bibitem[{{Pogge}(1988)}]{pogge1988}
{Pogge}, R.~W. 1988, \apj, 332, 702

\bibitem[{{Quirk}(1972)}]{quirk1972}
{Quirk}, W.~J. 1972, \apjl, 176, L9

\bibitem[{{Reid} {et~al.}(2013){Reid}, {Braatz}, {Condon}, {Lo}, {Kuo},
  {Impellizzeri}, \& {Henkel}}]{reid2013}
{Reid}, M.~J., {Braatz}, J.~A., {Condon}, J.~J., {et~al.} 2013, \apj, 767, 154

\bibitem[{{Riffel} \& {Storchi-Bergmann}(2011)}]{riffel2011}
{Riffel}, R.~A. \& {Storchi-Bergmann}, T. 2011, \mnras, 417, 2752

\bibitem[{{Safronov}(1960)}]{safranov1960}
{Safronov}, V.~S. 1960, Annales d'Astrophysique, 23, 979

\bibitem[{{Sargent}(1972)}]{sargent1972}
{Sargent}, W.~L.~W. 1972, \apj, 173, 7

\bibitem[{{Schmitt} {et~al.}(2001){Schmitt}, {Ulvestad}, {Antonucci}, \&
  {Kinney}}]{schmitt2001}
{Schmitt}, H.~R., {Ulvestad}, J.~S., {Antonucci}, R.~R.~J., \& {Kinney}, A.~L.
  2001, \apjs, 132, 199

\bibitem[{{Stern} {et~al.}(2012){Stern}, {Assef}, {Benford}, {Blain}, {Cutri},
  {Dey}, {Eisenhardt}, {Griffith}, {Jarrett}, {Lake}, {Masci}, {Petty},
  {Stanford}, {Tsai}, {Wright}, {Yan}, {Harrison}, \& {Madsen}}]{stern2012}
{Stern}, D., {Assef}, R.~J., {Benford}, D.~J., {et~al.} 2012, \apj, 753, 30

\bibitem[{{Sun} {et~al.}(2013){Sun}, {Greene}, {Impellizzeri}, {Kuo}, {Braatz},
  \& {Tuttle}}]{sun2013}
{Sun}, A.-L., {Greene}, J.~E., {Impellizzeri}, C.~M.~V., {et~al.} 2013, \apj,
  778, 47

\bibitem[{{Terlouw} \& {Vogelaar}(2015)}]{KapteynPackage}
{Terlouw}, J.~P. \& {Vogelaar}, M.~G.~R. 2015, {Kapteyn Package, version 2.3},
  {Kapteyn Astronomical Institute}, Groningen, available from
  \url{http://www.astro.rug.nl/software/kapteyn/}

\bibitem[{{Toomre}(1964)}]{toomre1964}
{Toomre}, A. 1964, \apj, 139, 1217

\bibitem[{{Ulvestad} \& {Wilson}(1984)}]{ulvestad1984}
{Ulvestad}, J.~S. \& {Wilson}, A.~S. 1984, \apj, 285, 439

\bibitem[{{Van den Bosch} {et~al.}(2016){Van den Bosch}, {Greene}, {Braatz},
  {Constantin}, \& {Kuo}}]{vandenbosch2016}
{Van den Bosch}, R.~C.~E., {Greene}, J.~E., {Braatz}, J.~A., {Constantin}, A.,
  \& {Kuo}, C.-Y. 2016, \apj, 819, 11

\bibitem[{{Wardle} \& {Yusef-Zadeh}(2012)}]{wardle2012}
{Wardle}, M. \& {Yusef-Zadeh}, F. 2012, \apjl, 750, L38

\bibitem[{{White} {et~al.}(1997){White}, {Becker}, {Helfand}, \&
  {Gregg}}]{white1997}
{White}, R.~L., {Becker}, R.~H., {Helfand}, D.~J., \& {Gregg}, M.~D. 1997,
  \apj, 475, 479

\bibitem[{{Whittle} \& {Wilson}(2004)}]{whittle2004}
{Whittle}, M. \& {Wilson}, A.~S. 2004, \aj, 127, 606

\bibitem[{{Wright} {et~al.}(2010){Wright}, {Eisenhardt}, {Mainzer}, {Ressler},
  {Cutri}, {Jarrett}, {Kirkpatrick}, {Padgett}, {McMillan}, {Skrutskie},
  {Stanford}, {Cohen}, {Walker}, {Mather}, {Leisawitz}, {Gautier}, {McLean},
  {Benford}, {Lonsdale}, {Blain}, {Mendez}, {Irace}, {Duval}, {Liu}, {Royer},
  {Heinrichsen}, {Howard}, {Shannon}, {Kendall}, {Walsh}, {Larsen}, {Cardon},
  {Schick}, {Schwalm}, {Abid}, {Fabinsky}, {Naes}, \& {Tsai}}]{wright2010}
{Wright}, E.~L., {Eisenhardt}, P.~R.~M., {Mainzer}, A.~K., {et~al.} 2010, \aj,
  140, 1868

\bibitem[{{Xanthopoulos} {et~al.}(2010){Xanthopoulos}, {Thean}, {Pedlar}, \&
  {Richards}}]{Xanthopoulos2010}
{Xanthopoulos}, E., {Thean}, A.~H.~C., {Pedlar}, A., \& {Richards}, A.~M.~S.
  2010, \mnras, 404, 1966

\bibitem[{{Yamauchi} {et~al.}(2012){Yamauchi}, {Nakai}, {Ishihara}, {Diamond},
  \& {Sato}}]{yamauchi2012}
{Yamauchi}, A., {Nakai}, N., {Ishihara}, Y., {Diamond}, P., \& {Sato}, N. 2012,
  \pasj, 64 [\eprint[arXiv]{1207.6820}]

\bibitem[{{Zhang} {et~al.}(2012){Zhang}, {Henkel}, {Guo}, \&
  {Wang}}]{zhang2012}
{Zhang}, J.~S., {Henkel}, C., {Guo}, Q., \& {Wang}, J. 2012, \aap, 538, A152

\end{thebibliography}

\end{document}